\documentclass[twocolumn,pra,aps,superscriptaddress,longbibliography,floatfix,showpacs]{revtex4-2}

\usepackage{amssymb,amsmath,amsfonts}
\usepackage{epsfig,graphicx}
\usepackage{hyperref}
\usepackage{color}

\usepackage{epsfig,graphicx}
\usepackage{mathtools}
\usepackage{amsbsy}
\usepackage{amstext}
 
\usepackage[dvipsnames]{xcolor}
\hypersetup{colorlinks=true,citecolor=NavyBlue,filecolor=black,linkcolor=NavyBlue,urlcolor=black}

\makeatletter

\newcommand\ket[1]{\left|#1\right\rangle}
\newcommand\bra[1]{\left\langle #1 \right|}
\newcommand\tr{\operatorname{tr}}

\usepackage{amsmath}
\usepackage{graphicx}
\usepackage{multirow}
\usepackage{float}
\usepackage{amssymb}
\usepackage{txfonts,color}
\usepackage{dsfont}

\makeatother

\begin{document}

\title{Low resource entanglement classification from neural network interpretability}

\author{A. Garc{\'i}a-Velo}
\affiliation{IMDEA Networks, Avda. Mar Mediterr{\'a}neo, 22, 28918 Legan{\'e}s, Spain}
\author{R. Puebla}
\affiliation{Departamento de F{\'i}sica, Universidad Carlos III de Madrid, Avda. de la Universidad 30, 28911 Legan{\'e}s, Spain}
\author{Y. Ban}
\affiliation{Instituto de Ciencia de Materiales de Madrid (CSIC), Cantoblanco, E-28049 Madrid, Spain}
\author{E. Torrontegui}
\affiliation{Departamento de F{\'i}sica, Universidad Carlos III de Madrid, Avda. de la Universidad 30, 28911 Legan{\'e}s, Spain}
\author{M. Paraschiv}
\affiliation{IMDEA Networks, Avda. Mar Mediterr{\'a}neo, 22, 28918 Legan{\'e}s, Spain}


\begin{abstract}
Entanglement is a central resource in quantum information and quantum technologies, yet its characterization remains challenging due to both theoretical complexity and measurement requirements. Machine learning has emerged as a promising alternative, enabling entanglement characterization from incomplete measurement data, however model interpretability remains a challenge. In this work, we introduce a unified and interpretable framework for SLOCC entanglement classification of two- and three-qubit states, encompassing both pure and mixed states. We train dense and convolutional neural networks on Pauli-measurement outcomes, provide design guidelines for each architecture, and systematically compare their performance across types of states. To interpret the models, we compute Shapley values to quantify the contribution of each measurement, analyze measurement-importance patterns across different systems, and use these insights to guide a measurement-reduction scheme. Accuracy–versus–measurement curves and comparisons with analytical entanglement criteria demonstrate the minimal resources required for reliable classification and highlight both the capabilities and limitations of Shapley-based interpretability when using machine learning models for entanglement detection and classification.
\end{abstract}

\maketitle

\section{Introduction}

Entanglement is the most essential resource in modern quantum information~\cite{Horodecki_2009} and quantum technologies~\cite{Dowling_2003}. From quantum computation~\cite{Raussendorf_2001, Raussendorf_2003, Briegel_2009} and quantum communication~\cite{Bennett_1992, Acin_2005, Curty_2004} to quantum metrology~\cite{Paris_09, Giovannetti_2011}, most applications that surpass classical performance leverage entanglement for information processing~\cite{Nielsen_2010}. Despite its central role, the characterization of entanglement remains a major open challenge~\cite{Friis_2019}, due both to its mathematical structure and to the experimental resources required for its detection.

If no prior knowledge of a quantum state is available, entanglement characterization generally depends on full quantum state tomography. Once the state is reconstructed, information theoretical methods can be applied to detect, classify, or quantify the entanglement present in the system ~\cite{Jamiołkowski_1972, Choi_1975, Horodecki_1996, Horodecki_2009, Guhne_2007, Guhne_2009, Chruściński_2014}. Aside from the mathematical difficulty of treating entanglement, tomography requires a number of measurements that grows exponentially with system size~\cite{Gurvits_2003, Ioannou_2007, Blume-Kohout_2010}, although partial prior knowledge can reduce this overhead~\cite{Guhne_2007, Guhne_2009, Cramer_2010, Bowles_2020}. This motivates the search for alternative strategies that avoid full state reconstruction while still providing reliable entanglement identification.

A broad range of methods has been proposed to relax the demand for complete state characterization, and instead infer entanglement directly from partial information \cite{Horodecki_2009, Guhne_2009, Friis_2019}. Modern approaches increasingly exploit machine learning. On the one hand, machine learning has been used for approximate tomography based on incomplete measurement sets~\cite{Torlai_2018, Xin_2019, Schmale_2022}, achieving performance improvements over traditional estimators. On the other hand, neural networks have been trained to detect~\cite{Lu_2018, Roik_2021, Asif_2023, Urena_2024, Travnicek_2024}, classify~\cite{Hiesmayr_2021, Rizvi_2022}, and quantify~\cite{Dominik_2023, Feng_2024, Pan_2024, Balaz_2025} entanglement, frequently using only incomplete or locally accessible data.

While machine learning offers powerful predictive performance, its lack of interpretability poses conceptual challenges \cite{molnar2025}. This is especially important in fundamental sciences, where explanatory insight—rather than black-box prediction—is essential. While machine learning techniques can effectively address challenging problems, they often increase algorithmic complexity, which in turn complicates the interpretation of their behavour. By loosing interpretability, we limit our trust in the solution, and restrict the ability to generalize them to new physical settings.

This challenge has led to a growing interest in interpretable machine learning~\cite{molnar2025}. Among the most theoretically sound and widely adopted approaches are methods based on Shapley values~\cite{Shapley_1953}, a concept derived from cooperative game theory and used as a feature-importance analysis tool, such as the SHAP method~\cite{Strumbelj_2014}. This method assigns numerical values to the input features that quantify their contribution to the final neural network output.

Shapley-based interpretability methods have recently emerged as a valuable tool for elucidating machine-learning models employed in entanglement characterization. In particular, they have been applied to entanglement detection protocols based on collective multi-copy measurements, where Shapley values were used to quantify the relative contribution of individual measurement outcomes to the detection task~\cite{Travnicek_2024}. Related approaches have been developed for the estimation of quantum correlations, including entanglement negativity and Bell nonlocality inferred from singlet-projection measurements~\cite{Tulewicz_2025}. In these settings, Shapley analyses enabled the ranking of measurement configurations and the construction of reduced measurement schemes that retain high accuracy while significantly lowering experimental complexity. Further applications to entanglement quantification have been demonstrated using measurements, or measurement-derived observables, on paradigmatic families of states such as Werner and Horodecki states~\cite{Feng_2024}.

Beyond entanglement-focused tasks, Shapley values and related explainability techniques have also found use in a broader range of quantum-physical problems. Notably, they have been employed to investigate the microscopic origins of quantum decoherence in solution-phase bond-breaking reactions, identifying which solvent descriptors most strongly influence decoherence dynamics~\cite{Mei_2024}. In a different direction, Shapley-based methods have been adapted to parameterized quantum circuits, where they provide a systematic way to assess the importance of individual gates or groups of gates in quantum machine-learning applications~\cite{heese_2025}. Collectively, these studies underscore the potential of Shapley-based explanations to yield nontrivial physical insight into complex, data-driven quantum models.

As a summary, in the context of machine learning applied to quantum entanglement, and the use Shapley values for its interpretability, existing works have established three key facts: (i) neural networks can learn entanglement properties from incomplete measurement data; (ii) neural networks can reconstruct or approximate quantum states from local observables; and (iii) Shapley values can provide meaningful interpretability for machine-learning models applied to quantum mechanics. However, a unified, systematic analysis that combines entanglement classification, diverse families of quantum states, and a thorough Shapley-value study of measurement importance has yet to be established.

In this work, we introduce a unified and interpretable machine-learning framework for SLOCC entanglement classification across multiple scenarios. We generate two- and three-qubit states—both pure and mixed—and encode them through their Pauli-measurement outcomes, providing a common representation for all scenarios. Dense and convolutional neural networks are trained on these incomplete measurement sets, and we give explicit guidelines for designing both architectures in the pure- and mixed-state regimes. We further compare both architectures to assess how their performance depends on the analyzed physical state.

To interpret the learned models, we compute Shapley values that quantify the contribution of each Pauli observable to the classification. Using these Shapley profiles, we perform systematic measurement-reduction studies and obtain accuracy–versus–measurement curves that reveal the minimal resources needed to achieve given performance thresholds. We additionally benchmark our results against analytical entanglement criteria and known witnesses.

With respect to the existing literature, we present several contributions. (i) A unified treatment of two- and three-qubit pure and mixed states, which enables a direct comparison of measurement-importance patterns and minimal measurement requirements across system sizes, purity levels, and restricted families of three-qubit mixed states. (ii) A systematic Shapley analysis for measurement reduction, calibrated against analytic criteria. This provides two fundamental insights, namely that (a) Shapley can fail to recover minimal sufficient statistics in highly redundant pure-state regimes while (b) it correctly identifies meaningful measurement blocks in less redundant mixed-state regimes. (iii) We compare convolutional and fully connected architectures in terms of both classification performance and Shapley stability. Our results show that, on the one hand, convolutional networks can be advantageous for pure-state classification but produce Shapley profiles that are dependent on input feature order and less interpretable. On the other hand, fully connected networks yield more stable, permutation-invariant importance rankings, especially for mixed states. This explicit link between architecture choice and the reliability of Shapley in quantum tasks is not, to our knowledge, discussed in previous work. (iv) Finally, for each scenario we provide explicit accuracy-vs-number-of-measurements curves that reveal the minimal resources needed to achieve given performance thresholds. Altogether, our analysis highlights both the strengths and the limitations of Shapley-based interpretability tools in machine-learning models for entanglement classification.

The paper is organized as follows. In Section~\ref{sect.:pre.}, we introduce the theoretical foundations underlying our machine-learning approach to entanglement classification. This includes quantum state tomography (Sect.~\ref{Sect.:repr.}), SLOCC entanglement classes (Sect.~\ref{sect.:SLOCC}), and Shapley values (Sect.~\ref{sect.:pre._shapley}). Section~\ref{sect.:method} presents our methodology for training and interpreting neural-network classifiers. We first describe the dataset generation (Sect.~\ref{sect.:dataset}), then discuss the design, selection, and benchmarking of network architectures (Sect.~\ref{sect.:model}), and finally explain the implementation of the SHAP interpretability framework (Sect.~\ref{sect.:shap}). In Sect.~\ref{sect.:results}, we present our main findings: the classification performance of the neural networks and the relative importance of individual measurements, validated through measurement-reduction studies. These analyses are performed separately for pure states (Sect.~\ref{sect.:results_pure}) and mixed states (Sect.~\ref{sect.:results_mixed}). Concluding remarks are provided in Sect.~\ref{sect.:conclusion}.

\section{Preliminaries}\label{sect.:pre.}

This section establishes the theoretical and technical framework necessary for our machine-learning approach to entanglement classification. The state representation (Sect.~\ref{Sect.:repr.}) details how we encode quantum states into feature vectors via Pauli tomography. The SLOCC entanglement classes (Sect.~\ref{sect.:SLOCC}) define the classification labels for two and three qubits. Finally, we introduce SHAP method (Sect.~\ref{sect.:pre._shapley}) that provides the interpretability framework we use to extract physical insight from the trained neural network predictions.

\subsection{State representation} \label{Sect.:repr.}

We consider isolated $N$-qubit systems, with $N = 2,3$. These are described by Hilbert spaces $\mathcal H$ of dimension $2^N$. By choosing an appropriate basis, the standard identification $\mathcal H \cong (\mathbb C^2)^{\otimes N} \cong \mathbb C^{2^N}$ is established. To extract information from states and use it as input to our neural network, we employ projective measurements. Given an observable $A$, its ideal expectation value with respect to a state $\varrho$ is
\begin{equation} \label{eq.:meas.}
    \langle A, \varrho\rangle \equiv \langle A\rangle = \tr(A\varrho) = \sum_{i=1}^{2^N} a_i \, p_A(i \mid \varrho) \, ,
\end{equation}

where $\{a_i\}_{i=1}^{2^N}$ are the eigenvalues of $A$~\cite{Nielsen_2010}. Each measurement probes the state from a distinct perspective, motivating the question of which observables are most informative for entanglement classification and how many measurements are required for this task. A complete description of any $N$-qubit state is provided by the Pauli decomposition
\begin{equation} \label{eq.:Pauli}
    \varrho = \frac{1}{2^N} \sum_{i_1,\ldots,i_N = 0}^3 T_{i_1 \ldots i_N} \, \sigma_{i_1} \otimes \cdots \otimes \sigma_{i_N} \, ,
\end{equation}

where $\sigma_0 = \mathbb I_2$, $\{\sigma_i\}_{i=1}^3$ are usual Pauli matrices, and 
$T_{i_1 \ldots i_N} = \langle \sigma_{i_1}\otimes\cdots\otimes\sigma_{i_N} \rangle$, with $T_{0\ldots 0}=1$~\cite{Nielsen_1999}. The correlation tensor $T := [T_{i_1\ldots i_N}]$ can be estimated experimentally, often by full state tomography protocols \cite{Blume-Kohout_2010, Cramer_2010}, and we will use it as input to the network. For convenience, we index tensor components lexicographically,
\begin{equation}\label{eq.:lex_order}
    (i_1,\ldots,i_N)\ \mapsto\ j = \sum_{n=1}^{N} 4^{\,n-1} i_n \, .
\end{equation}

A generic $N$-qubit density matrix requires $4^{N+1}$ real parameters, whereas Pauli tomography reduces this to $4^N - 1$. Whether fewer parameters suffice for entanglement classification is an open problem~\cite{Otfried_2009}. Recent techniques, such as shadow tomography~\cite{Aaronson_2018}, offer exponential compression and may further reduce measurement requirements.

\subsection{SLOCC entanglement classes} \label{sect.:SLOCC}

\subsubsection{Pure states} \label{sect.:SLOCC_pure}

Entanglement is a nonlocal resource arising from correlations among subsystems $\mathcal H = \bigotimes_{n=1}^N \mathcal H_n$. A pure state $\ket{\psi} \in \mathcal H$ is fully separable if it can be written as $\ket{\psi} = \bigotimes_{n=1}^{N} \ket{\psi_n}$, where $\ket{\psi_n} \in \mathcal H_n$; otherwise, it is entangled.

Two states are considered to exhibit the same type of entanglement if they can be converted into one another by local operations. In particular, SLOCC operations are commonly used for the purpose~\cite{Otfried_2009}. Two pure states $\ket{\psi}$ and $\ket{\phi}$ are SLOCC equivalent if and only if local invertible operators (LIOs) $A_n \colon \mathcal H_n \to \mathcal H_n$ exist such that~\cite{Dür_2000}
\begin{equation} \label{eq.:SLOCC}
    \ket{\phi} = \bigotimes_{n=1}^{N} A_n \ket{\psi} \, .
\end{equation}

At least in the $N = 2,3$ cases, the number of SLOCC classes is finite~\cite{Verstraete_2002}. To characterize each class, it is required to find a representative state for each of them. All other states in that class can then be obtained by applying LIOs.

For two qubits, only two entanglement classes exist: separable states, with representative $\ket{00}$; and entangled states, represented by the Bell state $\frac{1}{\sqrt{2}}(\ket{00}+\ket{11})$. The three-qubit case admits a richer structure, comprising six inequivalent classes. These include fully separable states, such as $\ket{000}$, and three biseparable classes, characterized by entanglement confined to a single qubit pair $(1,2)$, $(1,3)$, or $(2,3)$; for instance, entanglement between qubits $(2,3)$ is captured by $\ket{0} \otimes \frac{1}{\sqrt{2}}(\ket{00}+\ket{11})$. The remaining two classes exhibit genuine multipartite entanglement, namely the W and GHZ classes~\cite{Dür_2000}, with canonical states $\ket{\rm W} = \frac{1}{\sqrt{3}}(\ket{001}+\ket{010}+\ket{100})$ and $\ket{\rm GHZ} = \frac{1}{\sqrt{2}}(\ket{000}+\ket{111})$, respectively.

\subsubsection{Mixed states}\label{sect.:SLOCC_mixed}

A mixed state $\varrho$ is separable if it admits an ensemble decomposition $\varrho = \sum_i p_i \bigotimes_{n=1}^{N} \ket{\psi_n^i}\bra{\psi_n^i}$, with $\{p_i\}_i$ a probability distribution. Otherwise, the state is called entangled. Determining whether no separable decomposition exists is generally NP-hard~\cite{Gurvits_2003}. Classification into finer entanglement classes is even more challenging.

For bipartite systems, the Schmidt number provides a natural generalization of the Schmidt rank to mixed states~\cite{Terhal_2000}. For two qubits, this classification is a rephrasing of the definition of entanglement, reducing the problem to entanglement detection. This problem is solved, for instance by the positive partial transpose (PPT) criterion~\cite{Peres_1996,Horodecki_1996}. $\varrho$ is separable if and only if $(\mathrm{id} \otimes \top)(\varrho) \ge 0$, where $\top$ is the transposition map.

For three qubits, mixed-state SLOCC classes generalize the pure-state hierarchy~\cite{Acin_2001}, with a total of four classes. Mixed separable states admit at least one ensemble decomposition where all terms are separable pure states. Mixed biseparable states are not separable, and admit at least one ensemble decomposition where all terms are in either the separable or biseparable pure classes. Mixed W states are not biseparable, and admit at least one ensemble decomposition where all terms are in the separable, biseparable, or W pure state classes. Finally, mixed GHZ states are not in the W class, and admit at least one ensemble decomposition where all terms are in the separable, biseparable, W, or GHZ pure state classes.

Identifying the class of a given mixed state remains difficult. Entanglement witnesses provide experimentally accessible, although only sufficient conditions~\cite{Guhne_2009}. An entanglement witness $\mathcal W_C$ is a Hermitian operator such that if $\tr(\mathcal W_C \varrho) < 0$, then $\varrho$ is in class $C$. We use witnesses of the form
\begin{equation} \label{eq.:witness}
    \mathcal W_{\phi} = \alpha \mathbb I - P_{\phi} \, ,
\end{equation}

where $\left(\alpha, \ket \phi\right) = (3/4, \ket{\rm GHZ}), (2/3, \ket{\rm W}), (1/2, \ket{\rm GHZ})$, and $P_\phi := \ket{\phi}\bra{\phi}$.

\subsection{Shapley values} \label{sect.:pre._shapley}

Machine learning models are becoming increasingly prevalent in many domains. Nevertheless, the ``black box'' nature of these models poses significant risks.  Interpretability methods are desired for trust, fairness, debugging, and regulation. Among the most prominent approaches are Shapley values~\cite{Shapley_1953}, originally developed in cooperative game theory. They solve how to fairly distribute the total gain of a coalition game among players that contributes differently to the final outcome.

In the context of machine learning, the ``game'' is a single instance prediction task, the ``players'' are the features of that instance, and the ``payoff'' is model's output deviation from its mean prediction. The Shapley value for a feature $k$ is calculated as a weighted average of the difference between the payoff with and without feature $k$ across all possible subsets of features $S \subseteq F$, where $F$ is the complete set of features:
\begin{equation}\label{eq.:shap}
\chi_j = \sum_{S \subseteq F \setminus \{j\}} \frac{|S|! (|F| - |S| - 1)!}{|F|!} [ f(S \cup \{k\}) - f(S) ] \, .
\end{equation}

\begin{figure}
    \centering
    \includegraphics[width=\linewidth]{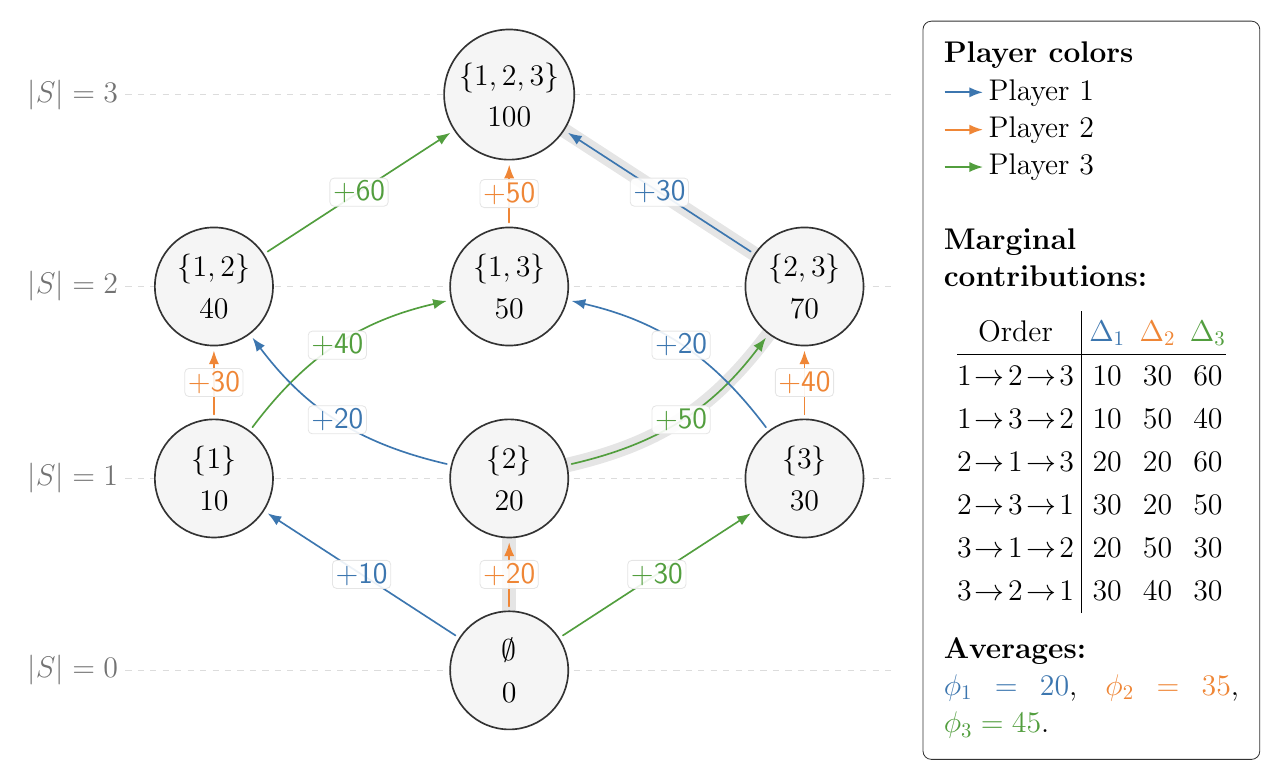}
    \caption{In this graph, nodes represent the different coalitions $S$ (subsets of players that may contribute to the game) with associated values $f(S)$. Arrows represent the process of adding one player $i$ to $S$. Labels on the arrows show the (marginal) contribution of each addition. That is the difference between the value of the coalition with and without such player $f(S\cup\{i\})-f(S)$. Any full path $\emptyset \to \{1,2,3\}$ corresponds to one ordering of players. To compute the Shapley value of a given player, list all the orders that can be constructed $\emptyset \to \{1,2,3\}$. For each path, extract the marginal contribution to each player to the game (this is shown in the table under marginal contributions on the figure). Then, compute the average of all the $3! = 6$ marginal contributions. This corresponds to the coefficient $\frac{|S|! (|F| - |S| - 1)!}{|F|!}$ in Eq.~\eqref{eq.:shap}.}
    \label{fig:shapley_schematic}
\end{figure}

Here $\chi_j$ is the Shapley value for feature $j$, $f(S)$ denotes the expected model prediction when only the features in $S$ are known, with the remaining features marginalized over their distribution~\cite{Strumbelj_2014}. In Fig. \ref{fig:shapley_schematic} we show a schematic of the computation of the Shapley values $\chi_j$ for a general simple game of three players. Shapley values satisfy desirable axioms such as efficiency, symmetry, the null-player property, additivity and being model agnostic.

However, while Shapley values provide an attractive theoretical framework, their direct application to machine learning is computationally prohibitive. Their exact calculation requires evaluating the model on all $2^{|F|}$ subsets, which is infeasible for high-dimensional inputs. Therefore, we employ the SHAP library~\cite{Lundberg_2017}, which provides tractable, model-specific approximations by linking Shapley values to local approximation models (e.g., LIME~\cite{Ribeiro_2016}) through carefully chosen loss functions, weighting kernels, and approximating the conditional expectations $f(S)$ of the model output.

For a single instance, SHAP returns a Shapley value $\chi_j$ for each feature. A positive (negative) sign indicates if the feature contributes to classifying (not classifying) the instance into the corresponding class, while the magnitude reflects the net strength of this feature contribution.

\section{Method} \label{sect.:method}

\begin{figure*}[t]
    \centering
    \includegraphics[width=\textwidth]{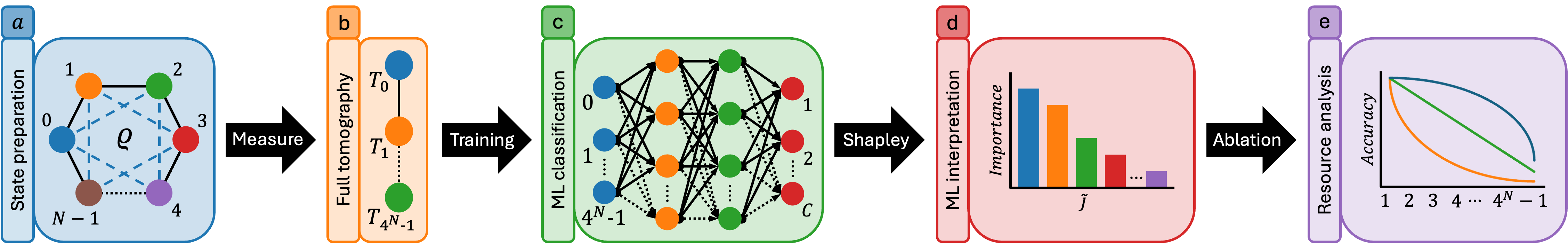}
    \caption{Schematic representation of the method presented. (a) We first prepare the dataset, which is divided into subsets corresponding to all SLOCC entanglement class. (b) For each state of the dataset, we compute its complete Pauli basis decomposition by simulating a full tomography measurement scenario. (c) The measuremets are used as data for different machine learning models, whose task is to perform SLOCC entanglement classification on the states of the dataset. (d) After the machine learning models are trained, we apply an interpretability algorithm (approximate Shapley values), to obtain the importance of each measurement setting for the entanglement classification. (e) Finally, we bechmark the importance distributions on number of measurements vs. accuracy numerical experiments. As a result, we find minimum sets of measurement required for entanglement classification as a function of the accuracy requirements.}
    \label{fig.:schematic_method}
\end{figure*}

This section details our methodology for training and interpreting neural network classifiers for entanglement. Our approach consists of three sequential components. First, we describe the generation of labeled datasets of quantum states (Sect.~\ref{sect.:dataset}), for both pure and mixed systems of two and three qubits. Second, we explain the selection and benchmarking of neural network architectures, justifying our final model choices based on the data structure and task requirements (Sect.~\ref{sect.:model}). Finally, we outline our implementation of the SHAP interpretability framework, including the specific algorithms and statistical procedures used to compute robust, model-agnostic measurement-importance scores from the trained networks (Sect.~\ref{sect.:shap}). Figure \ref{fig.:schematic_method} shows a schematic representation of the different steps of the method.

\subsection{Dataset generation} \label{sect.:dataset}

Our method requires a dataset of quantum states for training and evaluating the neural network models. Each state must be labeled according to its entanglement class. The networks do not receive the states directly, but rather they take as input the ideal expectation values of the corresponding Pauli operators, i.e. the correlation tensor $T$ in Eq.~\eqref{eq.:Pauli}.

\subsubsection{Pure states} \label{sect.:dataset_pure}

Canonical representatives of each pure-state entanglement class are known a priori (Sect.~\ref{sect.:SLOCC_pure}). We generate states within a class by applying random SLOCC transformations to these representatives (Eq.~\eqref{eq.:SLOCC}), followed by renormalization. Specifically, we sample $N$ independent operators from the set of invertible operators on $\mathbb{C}^2$. To ensure that the resulting ensemble provides uniform coverage of the target class, the sampling is performed with respect to the uniform measure on the local invertible operators. We refer to App.~\ref{app.:rand._LIO} for the details of the sampling procedure, which also provides an analysis of the induced distribution of entangled states, confirming that it covers the intended class.

Class labels follow directly from the dataset construction, since each generated state remains in the entanglement class of its representative. We then simulate idealized measurement statistics (Eq.~\eqref{eq.:meas.}) using a complete tomographic set of Pauli observables (Eq.~\eqref{eq.:Pauli}). To approximate realistic experimental conditions, we also generate non-ideal measurements by estimating the expectation values using a finite number of shots, benchmarking the method performance under non-ideal conditions (see App.~\ref{app.:non-ideal_meas} for details).

\subsubsection{Mixed states} \label{sect.:dataset_mixed}

Deterministically generating mixed states belonging to a predetermined entanglement class by explicitly constructing all its ensemble decomposition is computationally prohibitive.

For $N = 2$, we instead sample random density matrices from the Hilbert-Schmidt distribution (cf. App.~\ref{app.:rand._2dm}) and classify them using the PPT criterion. Although this approach is not asymptotically efficient, the two-qubit Hilbert space is small and contains only two entanglement classes, making the procedure sufficiently fast for our purposes. Note that in this simple setting, one can interpret the network task as learning to replicate the PPT criterion.

For $N = 3$, direct random sampling followed by classification is infeasible. Within the Hilbert-Schmidt distribution, the relative volume of non-GHZ classes (such as W or biseparable) is negligible, with respect to the entire set of mixed states. Generating a random state will place it in the GHZ class with near-unit probability. To circumvent this issue, we generate states of the form
\begin{equation}\label{eq.3q_states}
\varrho_\psi = \frac{\beta}{8} \mathbb I + (1 - \beta) P_\psi \, ,
\end{equation}

where $\ket{\psi}$ is a pure state and $\beta \in [0,1]$ is a mixing parameter. Both $\beta$ and $\ket{\psi}$ are sampled such that the resulting state $\varrho_\psi$ is detected by an entanglement witness given in the form of Eq.~\eqref{eq.:witness}. Although this method does not uniformly cover the full state space, it does serve as a proof-of-concept dataset for our method. We refer the interested reader to the App.~\ref{app.:witness_inequality} for the relationship between $\alpha$, $\beta$, $\ket{\psi}$, and $\ket{\phi}$ with respect to the condition $\mathrm{tr}(\mathcal{W}_\phi\,\varrho_\psi) < 0$. Furthermore, the complete algorithm for state generation is detailed in App.~\ref{app.:rand._3dm}.

For states of the form in Eq.~\eqref{eq.3q_states}, not all initial witnesses contribute equally. The witness $(1/2,\mathrm{GHZ})$ detects a strict superset of those detected by $(3/4,\mathrm{GHZ})$ (proof in App.~\ref{app.:W_redundancy}). Moreover, the sets of states detected by $(2/3, \mathrm{W})$ and $(1/2, \mathrm{GHZ})$ are disjoint (proof in App.~\ref{app.:W_disjointness}). Consequently, only $(2/3, \mathrm{W})$ and $(1/2, \mathrm{GHZ})$ are required to distinguish the relevant classes among the generated states. In this context, the task of the network consists in combining these two criteria into a single non-linear witness, corresponding to the union of two hyperplanes in the state space.

\subsection{Model selection} \label{sect.:model}

The choice of architecture is guided by three factors: the structure and the dimensionality of the input data, and the nature of the task. Tomographic data inherits a tensor-product structure from the Pauli operator expansion (Eq.~\eqref{eq.:Pauli}). This induces a natural multidimensional ordering in the correlation tensor. At the same time, entanglement is inherently a non-local property. It manifests as correlations across measurements that need not be adjacent in the chosen ordering of the observables.

To resolve this dilemma, we compare four distinct architectures: dense neural networks (DNNs), 1D convolutional neural networks (1D CNNs), 2D CNNs, and 3D CNNs. Because entanglement can depend on long-range correlations, a DNN—whose fully connected layers access all measurements simultaneously, might be better suited. CNNs, by contrast, impose locality through finite receptive fields determined by kernel size.

A further design constraint arises from the need to compute Shapley values, which require large numbers of forward evaluations. Model efficiency is therefore critical.

Overall, DNNs perform robustly across all scenarios, including mixed states. CNNs become competitive only for pure states, where the input dimension is smaller and the redundancies in tomographic data allow convolutional architectures to operate effectively. For an overview of model designs for both pure and mixed states, see App.~\ref{app.:net_design}.

\subsubsection{Pure states}

Among the CNN architectures, the 1D CNN provided the most effective and versatile solution. First, its architecture is the most amenable with the latter measurement reduction. Second, any $N$-qubit correlation tensor can be flattened to a 1D vector, eliminating the need to redesign higher-dimensional convolutional architectures for different system sizes. Third, achieving a given target accuracy requires far fewer parameters than an equivalent DNN. This compactness yields shorter training times and, critically, much faster evaluation times. Fourth, the 1D CNN exhibited substantially less overfitting than comparably accurate DNNs, leading to better generalization.

Attempts to reduce DNN overfitting did not succeed within reasonable efficiency constraints (see App.~\ref{app.:over-fitting_reduction}). Details of the CNN design and architecture are provided in App.~\ref{app.:CNN_design}, while the final architectures and their performance are explained in App.~\ref{app.:net._arch.}. In App.~\ref{app.:non-ideal_meas} we benchmark the method against non-ideal measurements. That is, instead of computing the expected value by Eq.~\eqref{eq.:meas.}, we assume a finite number of measurement shots that are then averaged to get the approximated expected value. The results indicate that the method offers robustness against approximated tomography, and converges to the ideal precision as more shots are taken as expected. 

\subsubsection{Mixed states}

The architectural performance changes significantly for mixed states. Pure states (rank-1 projectors) form a low-dimensional manifold within the full state space. Tomographic data for pure states exhibit redundancies on the information available across different measurements. Yet, in a generic mixed state, all measurements may contain unique or non-redundant information necessary for a complete characterization. As a result, convolutional architectures face three major difficulties.

\begin{enumerate}
    \item Kernel Size Requirement: Mixed-state classification requires integrating correlations across many measurements simultaneously, necessitating large kernels to effectively make the analysis less local.
    \item Input Order Dependence: Convolutions are local operations. Even for bigger kernel sizes, the ordering of the measurement as input strongly affects the optimization path of the network and the resulting Shapley rankings. Enlarged kernels mitigate but do not eliminate this bias (see App.~\ref{app.:input_order} for details).
    \item Data and Model Complexity: The increased complexity of the input space demands substantially more training data and larger models to achieve comparable accuracies to the pure-state case.
\end{enumerate}

Given these challenges, DNNs become the preferred and effectively necessary choice for mixed-state classification. In addition, it is worth mentioning that they are permutation-invariant with respect to the input features. They achieve comparable accuracy without large architectures and show similar levels of overfitting to CNNs. After an optimization process analogous to the pure-state case, data augmentation was selected as the primary method for overfitting reduction. Final model architectures and performance for two qubits are provided in App.~\ref{app.:two_qubits}. 

As described in Sect.~\ref{sect.:dataset_mixed}, the classification task for three qubits consists on simple entanglement witnesses, which is considerably simpler than the two-qubit problem. The dimension of the Hilbert space increases, but the classification rule simplifies substantially. Effectively, the decision boundary can be well approximated by a pair of hyperplanes in the feature space. Consequently, network size, dataset volume, and number of training iterations all decrease. Here, a simple DNN suffices, avoids the kernel-size biases inherent to CNNs, and provides comparable computational demands. The resulting architecture and its performance are reported in App.~\ref{app.:three_qubits}.

\subsection{SHAP} \label{sect.:shap}

As explained in Sect.~\ref{sect.:pre._shapley}, the exact calculation of the Shapley values using Eq.~\eqref{eq.:shap} is computationally intractable for high-dimensional data, as it requires evaluating the model $2^{|F|}$ times. To overcome this limitation for deep neural networks, we employ the \texttt{DeepExplainer} algorithm from the SHAP library~\cite{Lundberg_2017}. \texttt{DeepExplainer} is an enhanced variant of the DeepLIFT algorithm~\cite{Shrikumar_2017} specifically designed to approximate Shapley values. It leverages a connection between DeepLIFT's multiplicative rule and the Shapley value axioms. Essentially, \texttt{DeepExplainer} approximates the Shapley values for a subset of the dataset, called the sample dataset, by using a separate background (or reference) dataset. The background dataset provides a baseline for evaluating the marginal contribution of each feature. For further information on the underlying approximation, we refer to App.~\ref{app.:deepexplainer}.

For large models and complex datasets, \texttt{DeepExplainer} becomes computationally prohibitive due to a growing size of the sample and background datasets to be good representatives of the entire dataset distribution. This required increase in dataset sizes increases the computational effort. Keeping the datasets in tractable sizes introduces a source of approximation error and variability, as a small sample may not capture the full diversity of patterns. To mitigate this sampling variability, we employ a repeated sampling strategy. For a single trained model, we run \texttt{DeepExplainer} multiple times, each time with a newly and independently drawn pair of datasets of fixed sizes. The resulting Shapley values are then averaged across these iterations to produce a more stable estimate for that specific model.

\begin{figure}
    \centering
    \includegraphics[width=\linewidth]{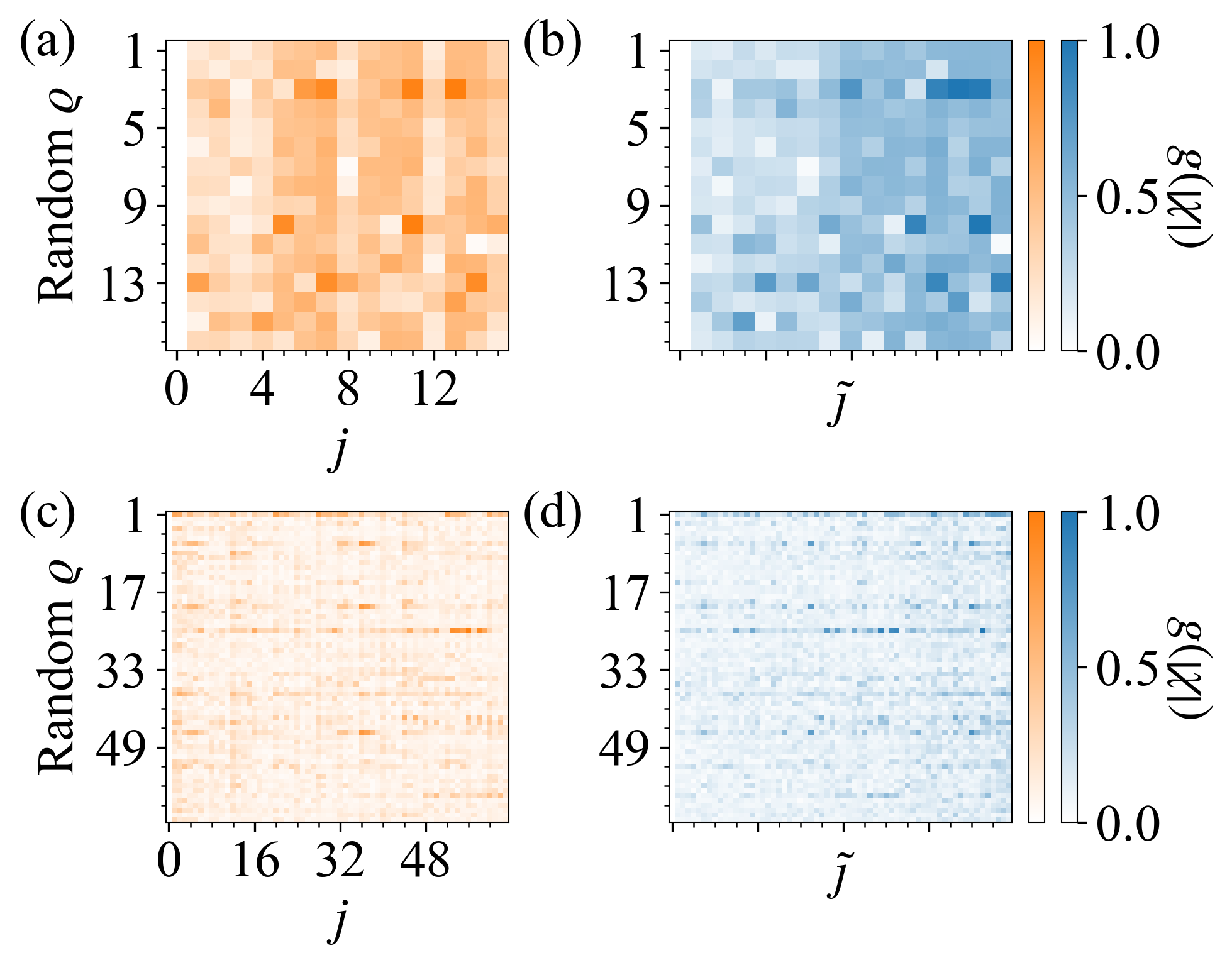}
    \caption{Matrix plots showing the re-scaled Shapley values for individual states. The re-scaling is done through the function $g = \sqrt{1 - (|\chi| - 1)^2}$, where $|\chi|$ denotes Shapley values normalized to the interval $[0, 1]$. Orange plots show the measurement settings ordered by lexicographic order $j$, while blue plots show measurement settings ordered by average Shapley value importance $\tilde \jmath$. Panels (a) and (b) are for mixed two-qubit states, while panels (c) and (d) are for pure three-qubit states.}
    \label{fig.:shap_box_plot}
\end{figure}

This global importance metric is not a perfect predictor for individual cases. By eliminating state-dependent information, we capture global patterns that may not align with the measurement importance for a particular state. For instance, Fig.~\ref{fig.:shap_box_plot} shows the distribution of state-specific re-scaled Shapley values $\chi_j$ for each measurement setting ordered both by lexicographic ($j$) and global importance ($\tilde \jmath$) order. Although ordering by global importance provides an overall trend, it may happen that the most important measurement settings for a particular state are ranked low by it. The global order is a good metric for identifying global patterns but should not be over-interpreted for specific instances. Interestingly, for the majority of states only a few measurements contribute significantly to the classification.

An additional, independent source of variability arises from the random initialization of the neural network weights and the stochastic nature of the optimization algorithm. Different initializations can lead the models to converge to distinct local minima, each potentially utilizing the input features in slightly different ways to achieve similar predictive performance. This results in different Shapley value profiles across model initializations. To ensure our findings about measurement importance are not specific to a single model, we repeat the entire procedure (from training to the repeated Shapley approximation) for multiple independently trained models. The final importance metric is derived from an aggregation across all models. For all trained models, we assign a score to each measurement setting according to the global Shapley order. Then, the score of each measurement is averaged over all models to optain the final aggregated order.

This two-tiered approach (averaging over samplings and over models) allows us to account for variation from both random sampling and random model initialization, yielding a more reliable and model-agnostic interpretation of measurement importance. We refer to App.~\ref{app.:shap_approx.} for a comprehensive, step-by-step explanation of this procedure, where we also show the results of training independent models for both pure and mixed states of two and three qubits.

\section{Results} \label{sect.:results}

Applying the methodology outlined in Section~\ref{sect.:method}, we obtain two primary results: the performance of our neural network classifiers and, more importantly, an interpretable ranking of the physical measurements that drive their decisions. We structure our findings by first presenting the Shapley value analysis, which quantifies the importance of each Pauli measurement setting for classification. We then validate these rankings through measurement-reduction studies, where we systematically remove measurements to test their impact in entanglement classification. This analysis is performed separately for pure states (Sect.~\ref{sect.:results_pure}) and mixed states (Sect.~\ref{sect.:results_mixed}) of two and three qubits.

\subsection{Pure states} \label{sect.:results_pure}

\subsubsection{Variability of Shapley values across models}

We have compensated the variability of the Shapley values due to (i) approximating with a background and sample datasets, and (ii) random processes of the network. After that, the distribution of Shapley values across measurement settings still exhibits significant variability between different model instances, both in the relative ordering of settings and the magnitude of their assigned importance. Even if we do not accomplish complete compensation of the variability effects (some degree of variation is still expected), one might anticipate that fundamentally important measurement settings would consistently receive high global importance. However, our results indicate a more complex picture.

\begin{figure}
    \centering
    \includegraphics[width=\linewidth]{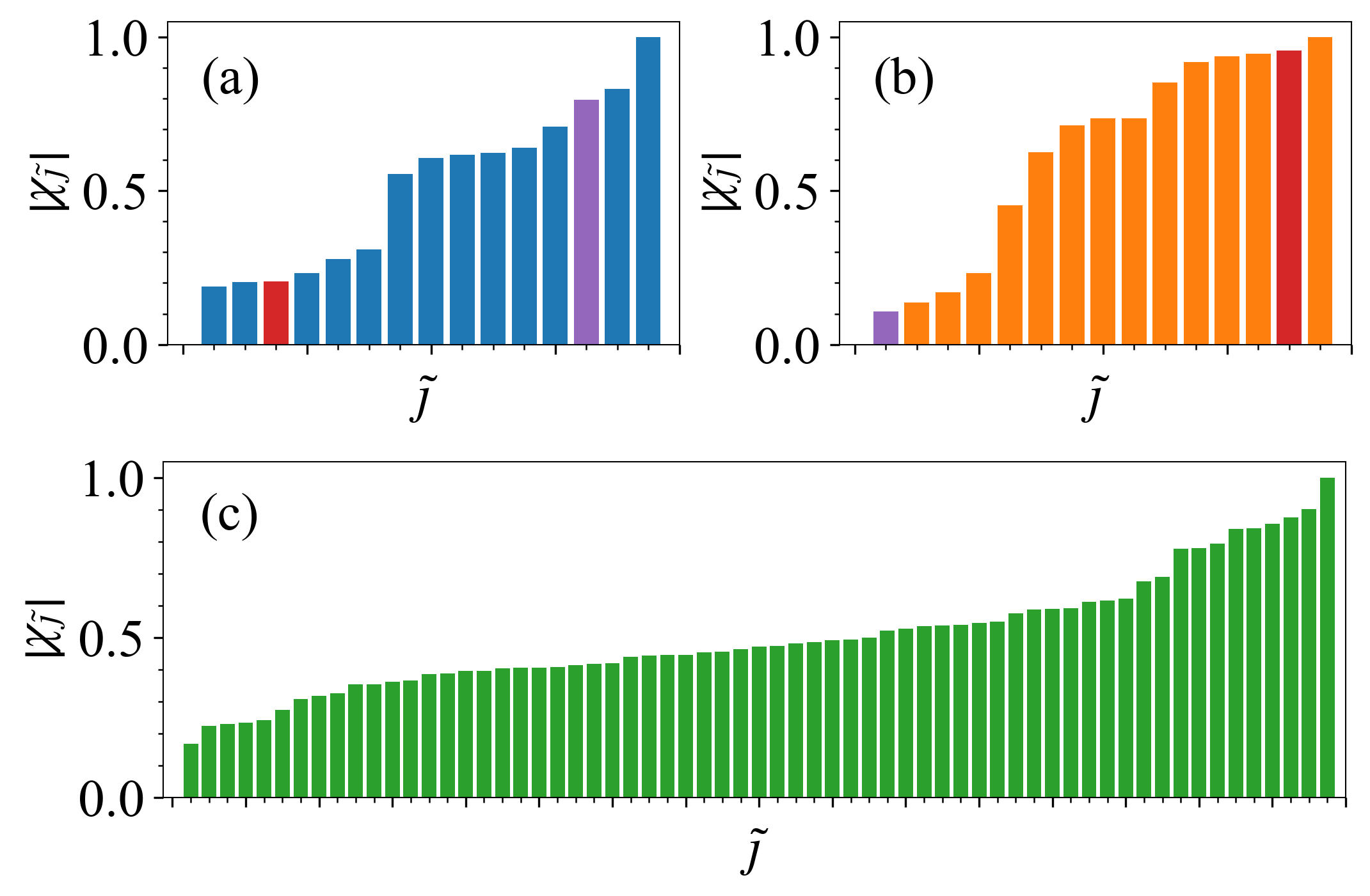}
    \caption{Normalized distribution of Shapley values for different models sorted in increasing order of importance, $\tilde{\jmath}$. Panels (a) and (b) show the distributions for two different two-qubit models. We have highlighted with different colors the Shapley values corresponding to measurement settings $j = 8$ (red), and $j=15$ (purple). Due to variability between models orders, they appear in different positions when ordered by importance ($\tilde \jmath$) deppending on the model. Measurement setting $j = 8$ and $j = 15$ are assigned importance orders $\tilde\jmath = 3, 14$ and $\tilde\jmath = 13, 1$ on models models (a) and (b), respectively. Panel (c) corresponds to a three-qubit model.}
    \label{fig.:pure_shap}
\end{figure}

Figures~\ref{fig.:pure_shap}(a) and~\ref{fig.:pure_shap}(b) illustrate the normalized Shapley value distributions for two distinct models in the two-qubit case. They present similar but different profiles, but they have been selected specifically for their divergent importance orders $\tilde \jmath$, which are are $[0, 1, 4, 8, 2, 3, 12, 11, 7, 5, 10, 9, 6, 15, 13, 14]$ and $[0, 15, 1, 2, 3, 12, 11, 14, 13, 7, 6, 4, 5, 9, 8, 10]$ respectively. Figure~\ref{fig.:pure_shap}(c) shows a representative distribution for a three-qubit model. Note that the trivial $j=0$ measurement setting, corresponding to the identity operator $T_{0, \ldots, 0} = \langle \sigma_0^{\otimes N} \rangle$, is included solely for benchmarking purposes, since $\chi_0=0$ as it carries no information and thus should always output zero importance.

The local receptive fields of CNNs are not the cause of the substantial variability in Shapley value profiles (see App.~\ref{app.:input_order}). Indeed, the results suggests that the network can successfully extract the requisite information for classification from many different combinations of measurement settings. The optimization process does not converge to a single, predominant parameter configuration that relies on a fixed set of measurements. This observation points towards the redundancy of information in tomographic data for pure states. This redundancy can be understood from a dimensionality perspective. A general pure state is fully parameterized by a complex vector of dimension $2^N - 1$ (after the normalization condition), equivalent to $2^{N+1} - 1$ real parameters. In contrast, the complete tomography protocol outputs $4^N - 1$ real values.

Despite these factors, the aggregated Shapley value analysis reveals a consistent preference for certain measurement directions on the Bloch sphere over others. To further test this consistency, we conduct a controlled experiment. We hold constant: the dataset, the input feature order, and the network initialization. We then applied a rotation to all states in the dataset. Specifically, a permutation of the measurement indices, $\pi$, which relabels the non-identity Pauli operators. After carrying out the full Shapley analysis with this new transformed representation, the measurement settings retained their relative importance, but their labels were transformed according to the $\pi$ rotation. Thus, while subject to methodological influences, the Shapley value extraction process is robust, self-consistent, and correctly tracks changes in the physical representation of the system.

\subsubsection{Measurement-reduction studies for measurement importance validation}

To validate the utility of the aggregated measurement-importance rankings, we perform measurement reduction on the input measurement settings. These are based on removing measurement settings according to the different importance rankings and evaluating the impact on classification performance.

\begin{figure}[t!]
    \centering
    \includegraphics[width=\linewidth]{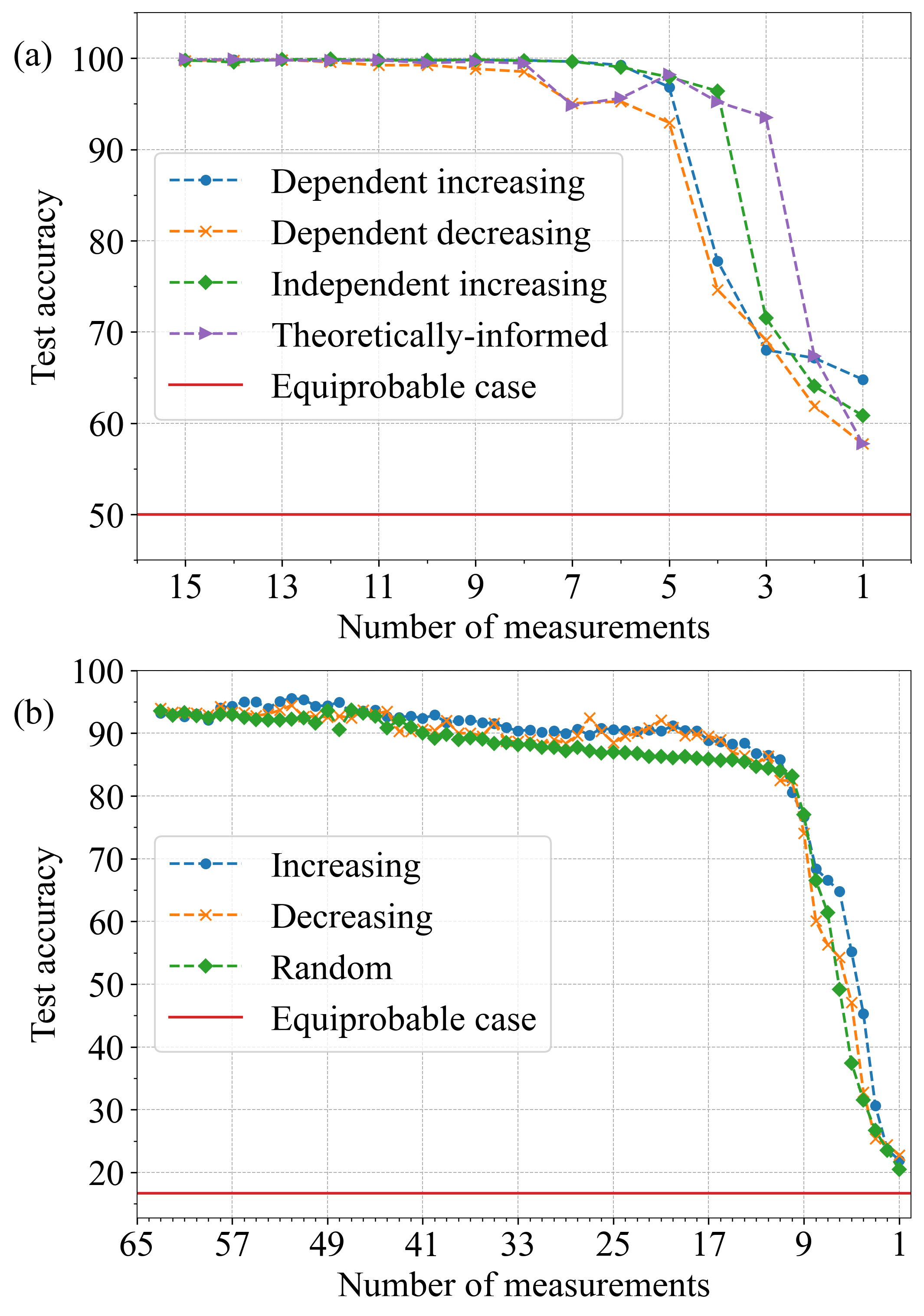}
    \caption{Test accuracies of neural networks trained with a progressively reduced number of measurements. (a) For two-qubit states, results are shown for measurement removal in both increasing and decreasing order of importance for a single model (model-dependent), and for the aggregated order across all models (model-independent). (b) For three-qubit states, results are shown for the aggregated increasing/decreasing orders and a random order for comparison.}
    \label{fig.:pure_ablation}
\end{figure}

Note, however, that in order to perform measurement reduction one cannot simply mask the inputs of the previously trained NN (see App.~\ref{app.:ablation_mask}). Indeed, one needs to train a new NN each time a change is done to the set of inputs. Thus, we train neural networks using only designated subsets of the original measurement settings. In this manner, measurement reduction is simply implemented by excluding the corresponding data from the input dataset and architecturally reducing the size of the input layer. For consistency across all experiments, the identity measurement (setting $j=0$) was always removed first, as it carries no information. The results for two- and three-qubit pure states are presented in Fig.~\ref{fig.:pure_ablation}. The model architecture and hyperparameters are consistent with App.~\ref{app.:pure_arch._opt.}, with the exception of the input layer size.

The classification accuracy remains high until a critical number of measurements is removed, after which a sharp decline, or ``elbow,'' occurs. In the two-qubit case, Fig.~\ref{fig.:pure_ablation}(a), a clear performance hierarchy emerges between the model-dependent increasing and decreasing orders. Both exhibit an accuracy elbow at approximately $4$ remaining measurement settings, although the decreasing order shows a preliminary performance drop at $7$ settings. This indicates that, while the network can achieve baseline classification without the most important individual inputs, these measurements are crucial for optimally defining the complex decision boundaries between classes. Notably, the model-independent increasing order (aggregated across all models) outperforms the model-specific orders, delaying the accuracy elbow until only $3$ measurement settings remain, indicating that the consensus ranking captures more robustly generalizable measurements.

An intriguing discrepancy arises when comparing these empirical results with theoretical expectations from quantum information theory. From the Schmidt decomposition~\cite{Nielsen_2010}, a two-qubit pure state is entangled if and only if $\langle\sigma_{01}\rangle^2 + \langle\sigma_{02}\rangle^2 + \langle\sigma_{03}\rangle^2 < 1$ (and similarly for the other subsystem). Therefore, a network should be able to perform perfect classification using only these three specific measurement settings. This is precisely what is used in the theoretically-informed curve in the Fig.~\ref{fig.:pure_ablation}(a), where we manually prioritized these three settings at the end of the removal sequence. In this scenario, high accuracy is maintained until the elbow at $2$ measurements is reached, which aligns perfectly with the theoretical minimum. With $2$ or fewer measurements, the performance of the network degrades significantly, although a single correlation measurement still provides non-trivial discriminatory power. The failure of the standard Shapley method to autonomously identify this theoretically optimal set suggests that the network, during its standard training process, learns to utilize the available measurements in a collective, distributed manner rather than isolating the minimal sufficient statistic. While classification with a minimal set is theoretically possible, the optimization process naturally converges to solutions that leverage the full set of measurements to construct more accurate robust decision boundaries.

The three-qubit case, shown in Fig.~\ref{fig.:pure_ablation}(b), serves as a demonstration of the information redundancy inherent in pure state tomography. We compared three removal orders: increasing importance, decreasing importance, and random. While one might hypothesize a clear performance hierarchy (increasing best, decreasing worst, random intermediate), the observed differences are notably subtle. A high importance ranking for a measurement setting indicates that it is frequently useful across the diverse optimization paths discovered during training. However, the high performance maintained even with random removal orders suggests that the essential information for entanglement classification is widely distributed across many measurement settings. Minor variations in accuracy are consequence of random model initialization. Partial proof of this is presented in App.~\ref{app.:shap_approx.}, where we plot the training and test accuracies for models initialized randomly.

\begin{figure}
    \centering
    \includegraphics[width=\linewidth]{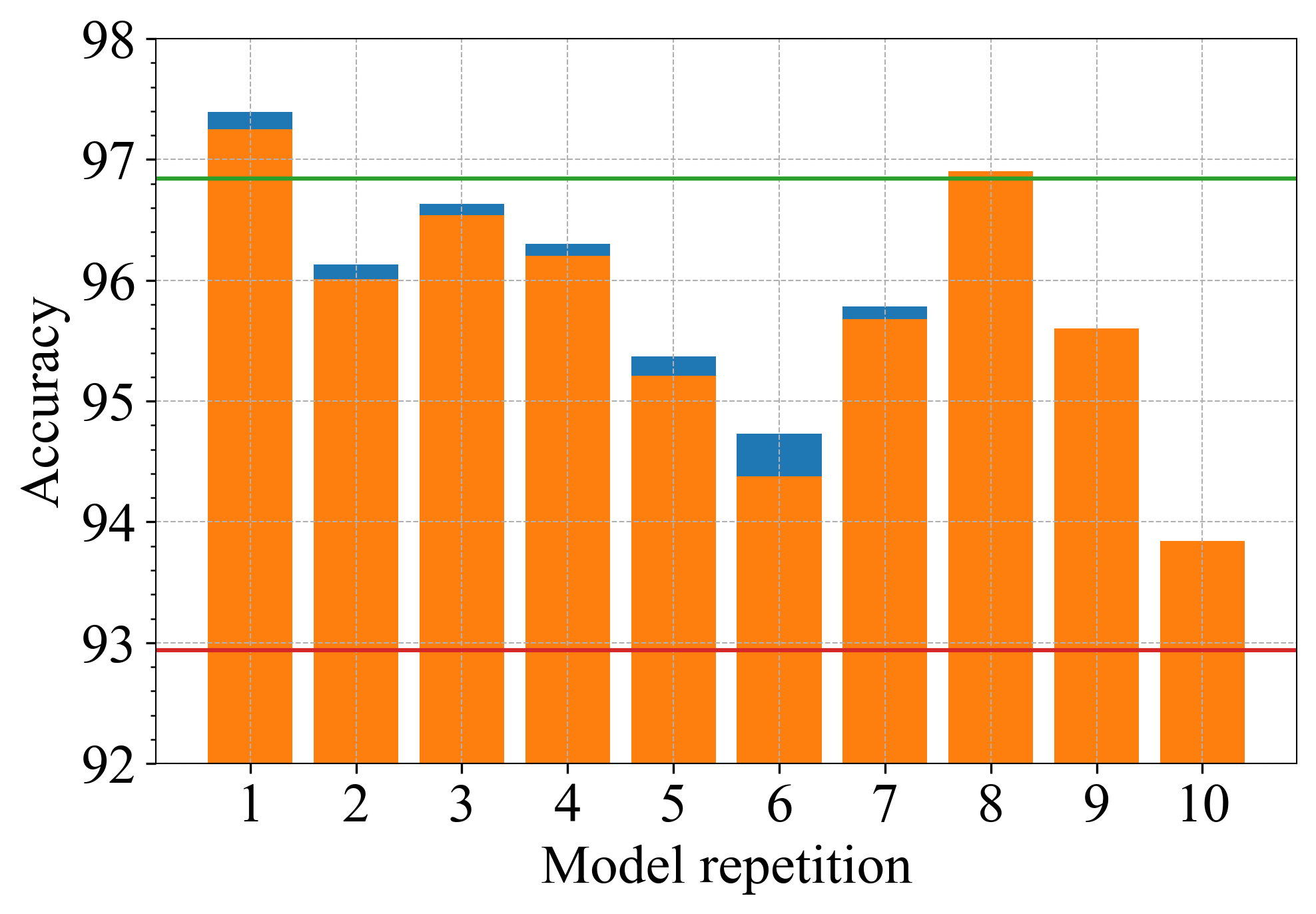}
    \caption{Test (blue) and training (orange) accuracy in entanglement classification for 10 independent pure-state two-qubit models, each trained with a unique random set of $5$ measurement settings. The green line sets the test accuracy for the model trained using the measurements from the independent increasing order. The red line sets an approximate lower bound, corresponding to the test accuracy of the model trained using the measurement from the dependent decreasing order.}
    \label{fig.:pure_2q _reduced}
\end{figure}

These measurement-reduction numerical experiments offer practical insights for scenarios where one aims to classify entanglement with limited control or knowledge over the available measurement settings. By examining the curves corresponding to random or decreasing importance orders, one can establish an approximate lower bound for the classification accuracy achievable when measuring a system with a given number of randomly selected measurement settings. For instance, measuring just $5$ random settings in the two-qubit case, or approximately $10$-$11$ settings in the three-qubit case, appears to guarantee a baseline level of classification confidence. Figure~\ref{fig.:pure_2q _reduced} shows the results of training $10$ independent models to classify entanglement in a two-qubit system based on only $5$ randomly chosen measurement settings. The resulting test accuracies are consistently lower-bounded by the performance of the model trained with the $5$ least important settings, cf. Fig.~\ref{fig.:pure_ablation}(a).

This implies that, for example, in the two-qubit case, by choosing any $5$ measurement settings at random, one can perform entanglement classification with approximately $93\%$ confidence. Alternatively, with control over the measurements one can perform, it is possible to maximize the accuracy while minimizing the number of measurements. We emphasize that these results are agnostic to the specific quantum system. For instance, in the three-qubit case, more accurate and efficient measurement strategies could be devised with prior knowledge of the states under study (as we will actually see later in the tree-qubit mixed-state scenario).

\subsection{Mixed states} \label{sect.:results_mixed}

\subsubsection{Two-qubits}

For mixed states, the distribution of Shapley values across measurement settings reveals a remarkably consistent and structured pattern. While minor variations exist between models, attributable to the stochastic elements of training, the analysis identifies a set of fundamentally important measurement settings that consistently receive high rankings across all model instances.

\begin{figure}
    \centering
    \includegraphics[width=\linewidth]{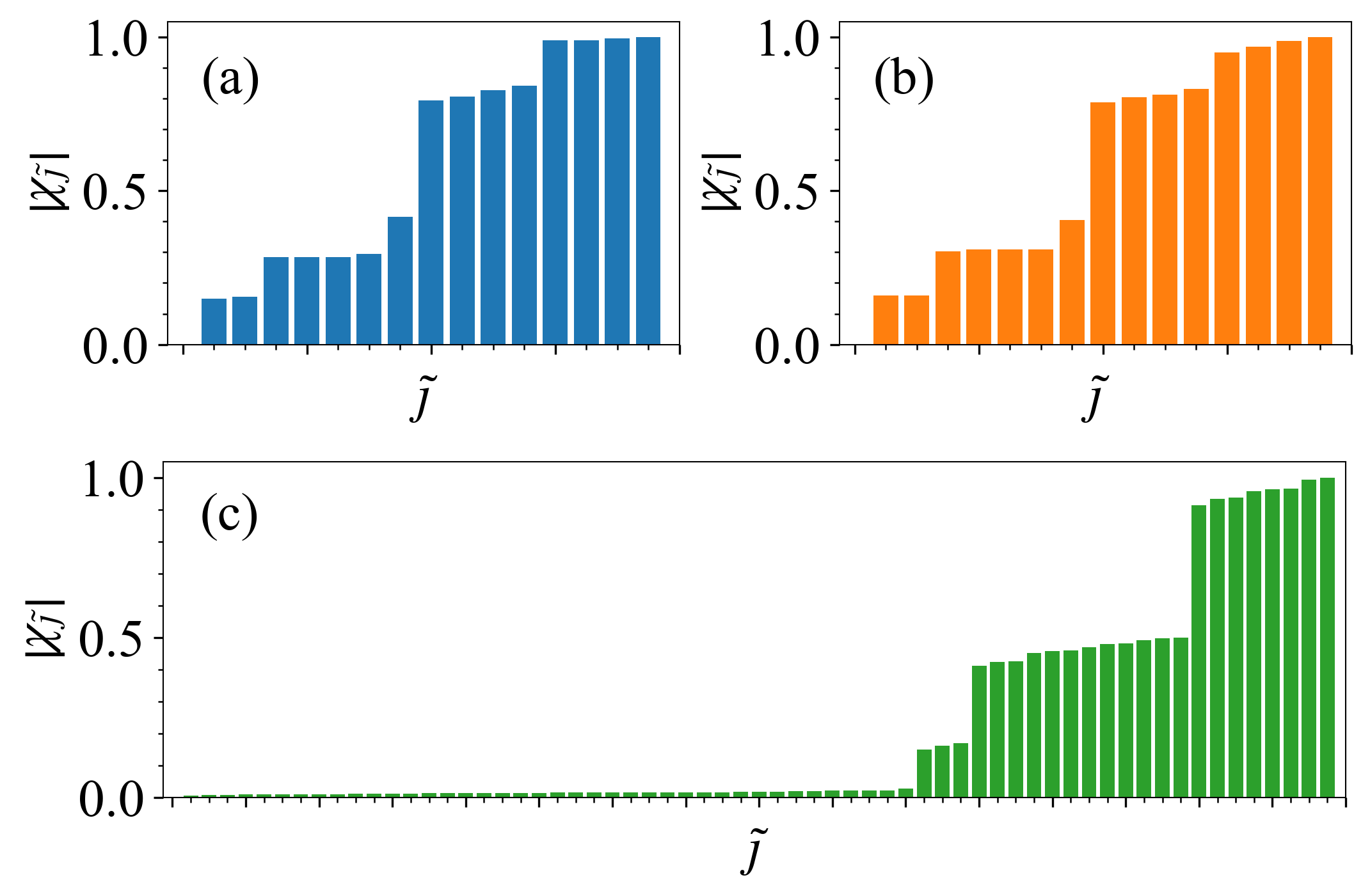}
    \caption{Normalized distributions of Shapley values for mixed state models sorted by increasing importance $\tilde\jmath$. Panels (a) and (b) show distributions for two different two-qubit models, while (c) corresponds to the distribution for a three-qubit model. Note that, although importance order may vary, the block structure is consistently preserved across independent repetitions.}
    \label{fig.:mixed_shap}
\end{figure}

Figures~\ref{fig.:mixed_shap}(a) and~\ref{fig.:mixed_shap}(b) illustrate the normalized Shapley value distributions for two distinct models, demonstrating this high degree of reproducibility. The $16$ measurement settings naturally partition into six distinct subsets characterized by comparable internal importance. These blocks, in ascending order of importance, are: (i) identity $\{\sigma_0 \otimes \sigma_0\}$; (ii) single-qubit Pauli-$Z$ measurements $\{ \sigma_3 \otimes \sigma_0, \sigma_0 \otimes \sigma_3\}$; (iii) single-qubit Pauli-$X$ and -$Y$ measurements $\{ \sigma_1 \otimes \sigma_0, \sigma_2 \otimes \sigma_0, \sigma_0 \otimes \sigma_1, \sigma_0 \otimes \sigma_2 \}$; (iv) $\{\sigma_3 \otimes \sigma_3\}$; (v) correlation measurements without $Z$ measurements: $\{\sigma_1 \otimes \sigma_1, \sigma_1 \otimes \sigma_2, \sigma_2 \otimes \sigma_1, \sigma_2 \otimes \sigma_2 \}$; (vi) correlation measurements containing $Z$ measurements: $\{ \sigma_1 \otimes \sigma_3, \sigma_2 \otimes \sigma_3, \sigma_3 \otimes \sigma_1, \sigma_3 \otimes \sigma_2 \}$. Similarly to what happened on pure states, applying a rotation to the states in the dataset does not change the structure of the importance patterns, rather it modifies the indices with respect to the proper permutation. This thus reflects that the results are influenced by the representation basis.

The relative ordering of settings within each block may vary slightly from model to model, but the block membership itself and the global ordering between blocks are preserved.

This structure, in contrast to the high variability observed for pure states, suggests a fundamental difference in the information content of the measurements. For mixed states sampled uniformly from the space of full-rank density matrices, the dataset possesses no inherent redundancy or low-dimensional structure. Consequently, the network must rely on a consistent set of measurements to perform the classification. The emergence of a stable block structure indicates that, on average, certain measurements provide more discriminative power for distinguishing separable from entangled states across the entire Hilbert space.

\begin{figure}
    \centering
    \includegraphics[width=\linewidth]{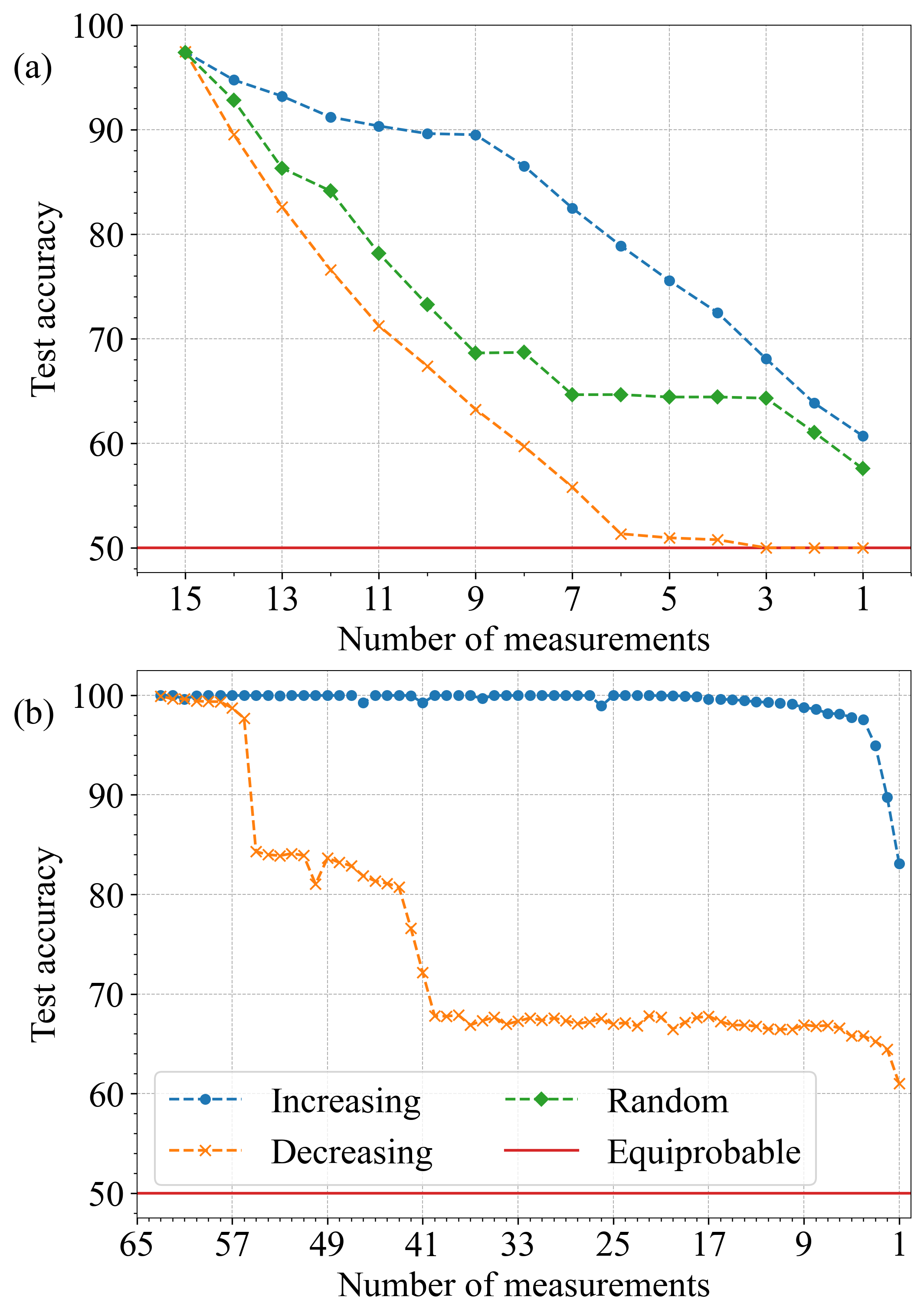}
    \caption{Test accuracies of neural networks trained with a decreasing number of measurement settings for mixed states of (a) two- and (b) three-qubits. Different points correspond to the classification accuracy when removing measurement settings in an increasing or decreasing order of importance, as well as for a randomly chosen order. Notice that the same legend is used for both panels, and panel (b) doesn't show the random order.}
    \label{fig.: mixed ablation}
\end{figure}

To quantitatively validate these measurement-importance rankings, we conducted a retraining measurement-reduction study, similar to the pure state scenario. Figure~\ref{fig.: mixed ablation}(a) shows the results of this numerical experiment, which aligns well with theoretical expectations and differ significantly from the pure state case. Model architectures and hyperparameters are consistent with App.~\ref{app.:mixed_arch._opt.}, with the input layer size adjusted accordingly.

Three important insights can be extracted from the results of the measurement-reduction study in Fig.~\ref{fig.: mixed ablation}. First, all tested orders achieve significantly lower accuracy for a given number of measurements compared to the pure state scenario. This expected behavior reflects the intrinsic complexity of mixed state entanglement detection. Even the reverse importance order for pure states outperforms the increasing importance order for mixed states. Second, a pronounced performance hierarchy emerges between the different removal orders. The increasing importance order maintains the highest accuracy, with a sharp decline occurring after the removal of the fourth most important block. This indicates that the highest-ranked measurements alone contain substantial, but not complete, information for general mixed state characterization. The decreasing importance order shows near-random performance when only the least important blocks remain, confirming that these measurements provide negligible standalone discriminative power. Their non-zero Shapley values likely reflect their utility in fine-tuning decision boundaries when used in conjunction with more informative settings. Finally, as expected, a random removal order exhibits intermediate performance, falling between the optimal (increasing) and worst-case (decreasing) scenarios. In view of these results, we can conclude that having control over the measurement settings one can perform, it is possible to maximize the accuracy while minimizing the number of measurements. As an example, one can reach $90 \%$ accuracy when measuring the top $9$ most important measurements.

\subsubsection{Three-qubits}

In the following we tackle the three-qubit mixed state scenario. We recall that the three-qubit mixed state dataset considered here is generated with a specific structure, as detailed in Sect.~\ref{sect.:dataset_mixed}. The distribution of Shapley values reflects this structure, exhibiting a fixed and interpretable pattern across all model instances, as illustrated in Fig.~\ref{fig.:mixed_shap}(c). The measurements again partition into four distinct importance subsets. Similarly, while the internal ordering within blocks may vary, the block structure and their global hierarchy remain consistent across models, which in turn is reflected in the measurement-reduction study (cf. Fig.~\ref{fig.: mixed ablation}(b)).

\begin{figure}
    \centering
    \includegraphics[width=\linewidth]{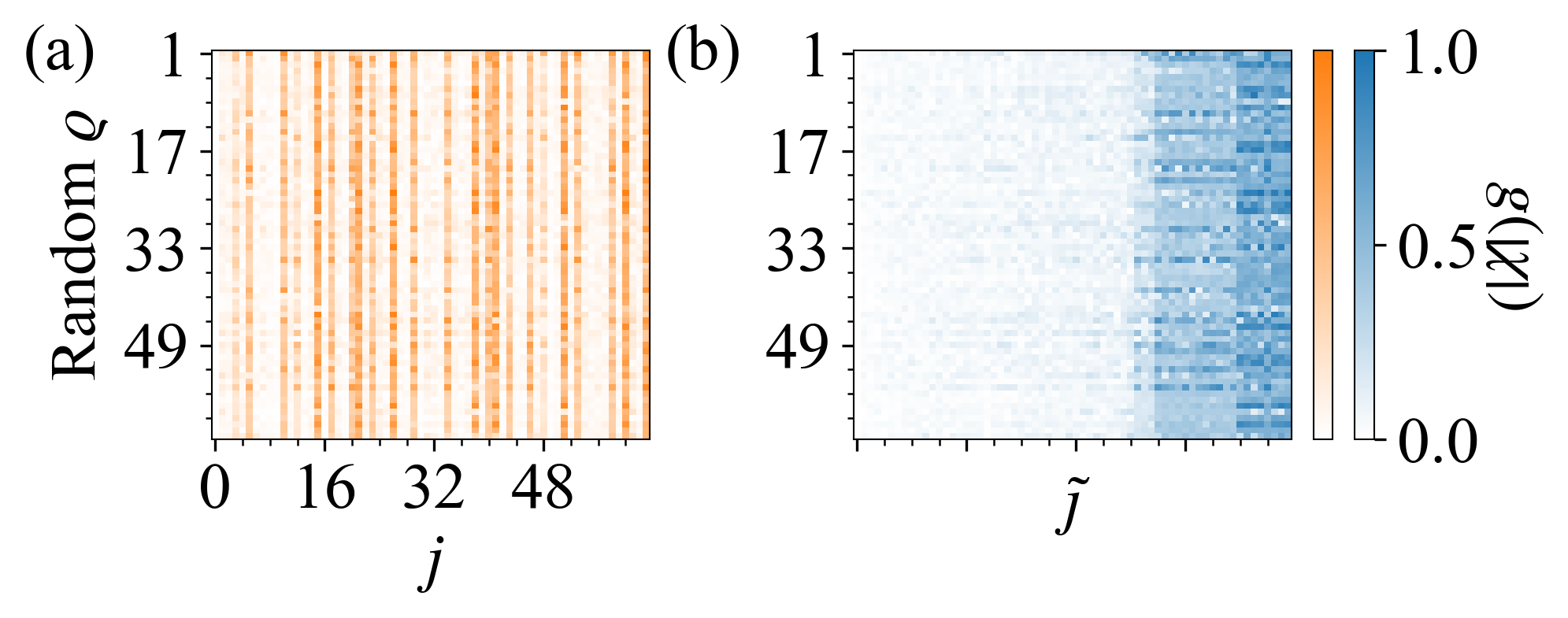}
    \caption{Matrix plots showing the re-scaled Shapley values for individual states in the three-qubit mixed state system. The re-scaling is done through the function $g = \sqrt{1 - (|\chi| - 1)^2}$, where $|\chi|$ denotes Shapley values normalized to the interval $[0, 1]$. Panel (a) shows measurement settings in lexicographic order $j$. Panel (b) shows the same data with settings ordered by their average Shapley value $\tilde \jmath$, revealing the underlying block structure.}
    \label{fig.:mixed_3q_shap_box_plot}
\end{figure}

This block structure is directly explainable by the dataset construction. The entanglement classification depends critically on a specific subset of measurement settings required to decompose the witnesses (see App.~\ref{app.:wit._decomposition} for further details). The top three highest-importance blocks correspond precisely to these essential settings. These results demonstrate the power of the Shapley value analysis for systems where some prior physical knowledge exists. This interpretation is further supported by Fig.~\ref{fig.:mixed_3q_shap_box_plot}, which illustrates the Shapley values at the individual state level, both for lexicographic ($j$) and global importance ($\tilde \jmath$) order, confirming that the block structure is not merely an aggregate model property but is consistently manifested across individual states. Surprisingly, measurement settings not present in the witness decompositions still attain non-zero (although small) Shapley values. This may result from the network approximate nature or from these settings providing auxiliary information that slightly refines the classification boundary. The measurement-reduction study will clarify their actual utility.

We validate these findings with the retraining measurement-reduction method. Figure~\ref{fig.: mixed ablation}(b) shows the results. Given the relative simplicity of the classification rule (based on linear witnesses), all orders perform well initially. However, a clear hierarchy emerges as more measurements are removed. The increasing importance order maintains near-perfect accuracy until the removal of settings from the highest-rated block, after which performance drops significantly but remains well above equiprobable choice. This type of entanglement can be witnessed with almost perfect classification using only $4$ measurement settings. This confirms that the most important block contains the core information for classification, and that all measurements within them contain significant information for classification. The decreasing importance order exhibits a distinctive ``ladder'' structure, where plateaus in accuracy are interrupted by sharp drops corresponding to the complete removal of all settings within a given importance block. The flatness of these plateaus indicates that measurements within the same block contribute roughly equivalent information, as their sequential removal does not gradually degrade performance. Again, we conclude that, with control over the measurements one can perform, it is possible to maintain nearly perfect accuracy with only $4$ measurements given the considered witness.

We further investigated the behavior of the network when classifying states with relative phases with respect to the canonical $\ket{\rm GHZ}$ and $\ket{\rm W}$ states. The network classification boundaries perfectly reproduce the theoretical predictions derived from the witness inequalities (cf. App.~\ref{app.:rel._phases} for further details). 

\begin{figure}
    \centering
    \includegraphics[width=\linewidth]{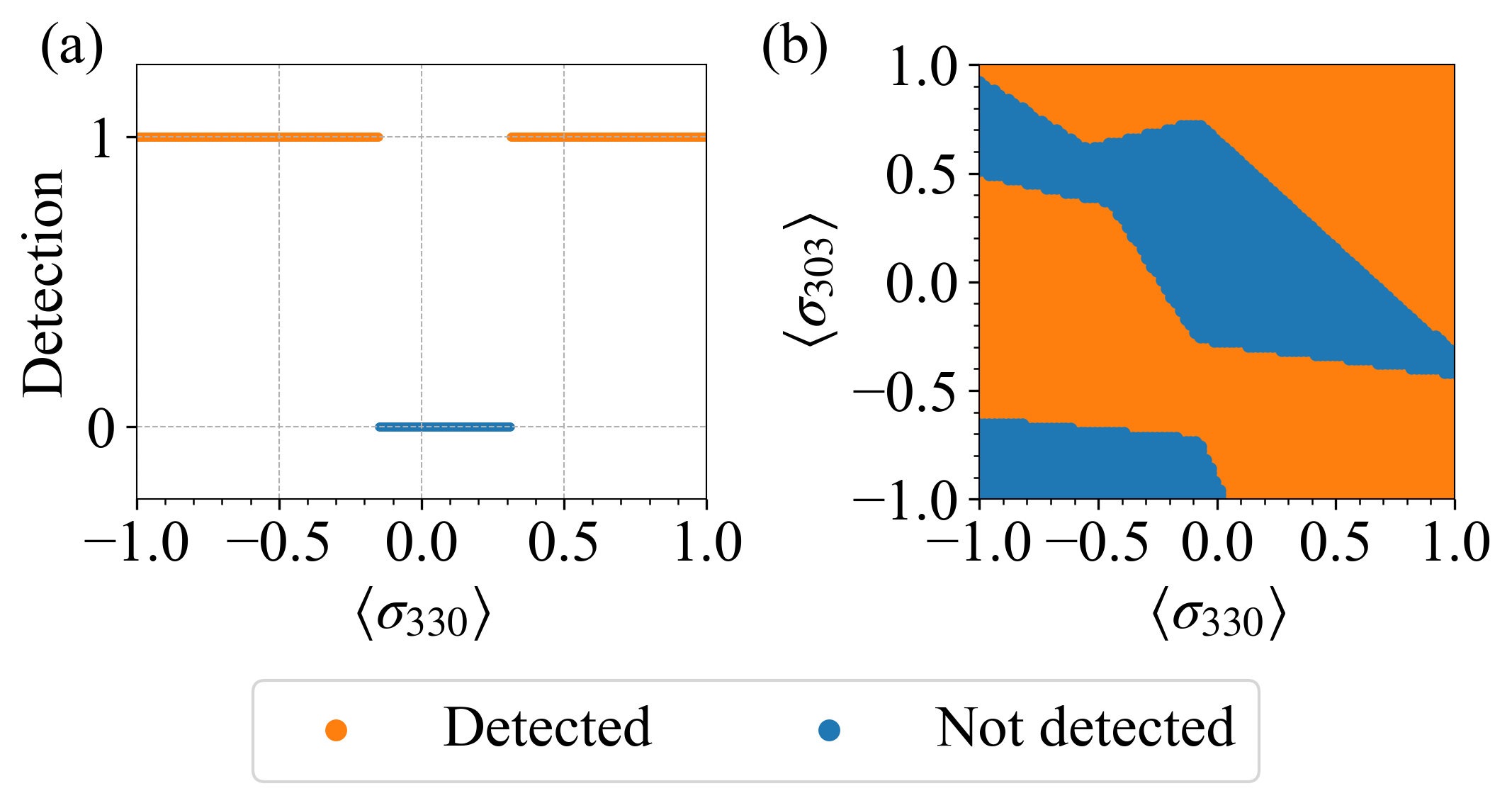}
    \caption{Visualization of neural network decision boundaries for three-qubit mixed state entanglement classification. Panel (a) shows the boundary when using the most important measurement setting. Panel (b) shows the boundary when using the $2$ most important measurement settings.}
    \label{fig.:mixed_3q_reduced}
\end{figure}

Finally, given the high-accuracy classification even for a reduced number of measurements, we analyze its behavior in these low-measurement regimes to visualize the decision boundaries it approximates. Figure~\ref{fig.:mixed_3q_reduced} illustrates the regions of the parameter space for which the network detects entanglement in three-qubit mixed states in the $1$ and $2$ measurements case. While similar analysis could be performed for three measurement settings, the resulting decision boundaries become difficult to visualize due to their higher-dimensional nature.

\section{Conclusions} \label{sect.:conclusion}

We have introduced a machine-learning framework for SLOCC entanglement classification using dense and convolutional neural networks trained on Pauli measurements. Our results show that CNNs perform competitively for pure states, benefiting from the inherent redundancy of tomographic data and exhibiting reduced overfitting. DNNs maintain stable performance across all scenarios, particularly for mixed states where measurement information is less redundant. We additionally find that the ordering of the input measurements substantially influences the resulting Shapley-value hierarchies when using CNNs. We employ data augmentation as the primary tool to mitigate overfitting across both architectures.

To interpret the learned models, we compute Shapley values and introduce an importance metric that reduces variance arising from neural-network initialization and the approximate computation of Shapley values. Although this metric reliably captures global patterns of measurement relevance, it is not a perfect predictor at the level of individual input instances. For pure states, the global importance patterns exhibit considerable model-to-model variability, reflecting the high redundancy of entanglement information in the measurement data. Nevertheless, aggregated Shapley distributions consistently favor specific measurement directions on the Bloch sphere. For mixed states, the Shapley values display a clear block structure, indicating the absence of redundancy and implying that the network must consistently leverage the same subsets of measurements to classify entanglement.

We validate the measurement rankings through a systematic measurement-reduction study in which networks are trained on restricted subsets of measurement settings. For pure states, classification accuracy remains high until a critical fraction of measurements is removed, after which it drops sharply. In the two-qubit case, networks can still reach baseline accuracy even without the most highly ranked inputs. Comparisons with analytical entanglement criteria suggest that the networks make collective use of the available measurements to form decision boundaries tailored to the numerical setting. For three-qubit pure states, accuracy remains largely unaffected by the order of removal, consistent with entanglement information being distributed broadly across the measurement set. In mixed states, however, a clear hierarchy emerges. Moreover, the highest-ranked measurements contain significant—but not sufficient—information for full-state classification. The removal order determines the achievable accuracy for reduced measurement budgets. These results allow us to estimate lower bounds on the classification accuracy attainable when only a given number of randomly chosen Pauli settings are available.

 Overall, our study provides a unified analysis of entanglement classification across multiple state families, clarifies when and how CNNs and DNNs extract entanglement-relevant measurements from incomplete data, and demonstrates the strengths and limitations of Shapley-based interpretability tools. Beyond offering practical measurement-reduction guidelines, our findings highlight the conditions under which neural networks recover physically meaningful measurement structures—and when redundancy or model variability limits the interpretability of learned patterns.

\begin{acknowledgments}
We acknowledge financial support form the Spanish Government via the project PID2024-161371NB-C21 (MCIU/AEI/FEDER, EU) and project TSI-069100-2023-8 (Perte Chip-NextGenerationEU). E.T.,
R.P. and Y.B. acknowledge the Ram{\'o}n y Cajal (RYC2020-030060-I), (RYC2023-044095-I) and (RYC2023-042699-I) research fellowships. 
\end{acknowledgments}


\appendix

\section{Dataset generation}

\subsection{Generating random LIO}\label{app.:rand._LIO}

\begin{figure}
    \centering
    \includegraphics[width=\linewidth]{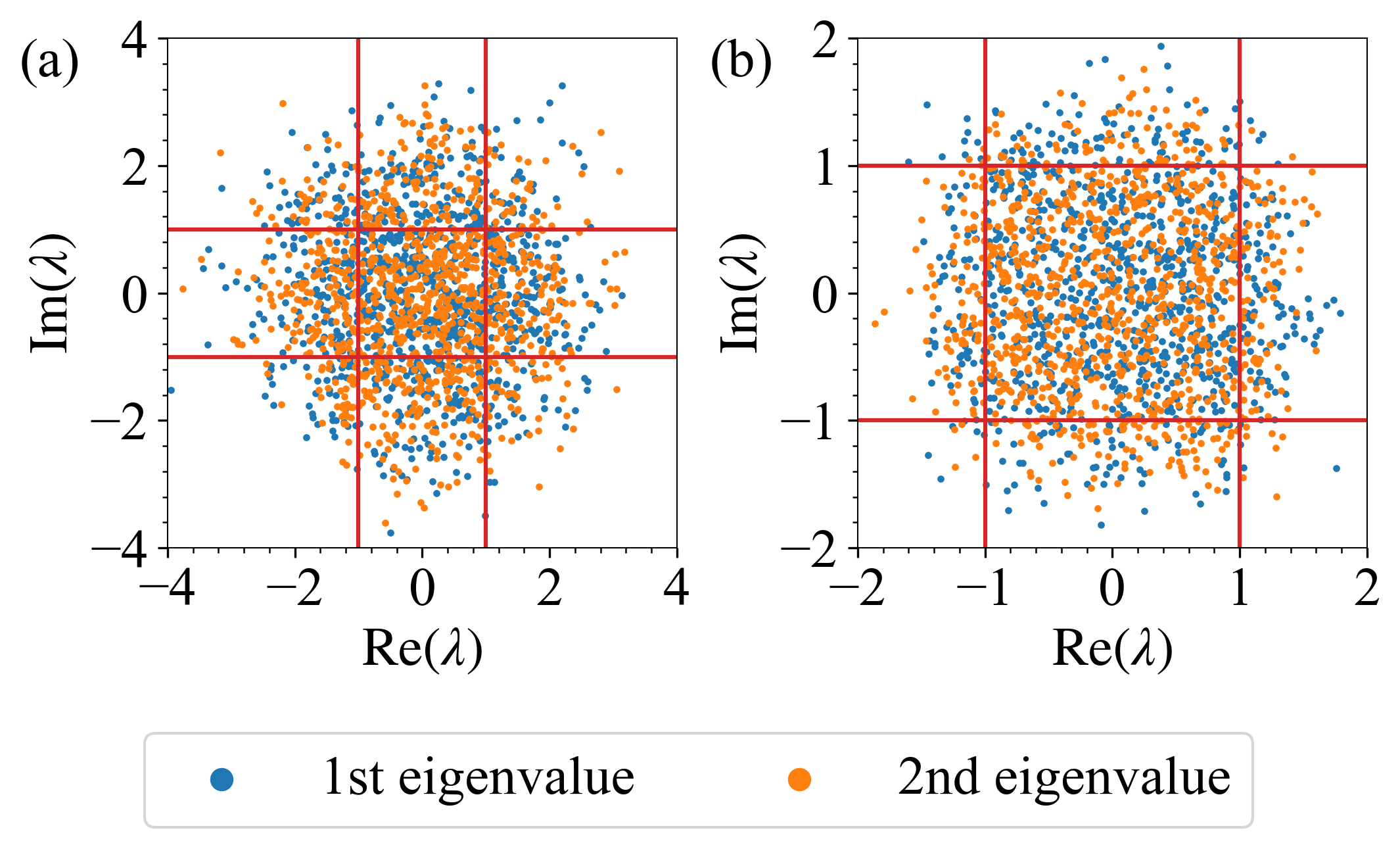}
    \caption{Distribution of eigenvalues for randomly generated invertible matrices of dimension two. Panel (a) for uniformly distributed matrix elements, and panel (b) for normally distributed matrix elements.}
    \label{fig.:eig._dist.}
\end{figure}

\begin{figure}
    \centering
    \includegraphics[width=\linewidth]{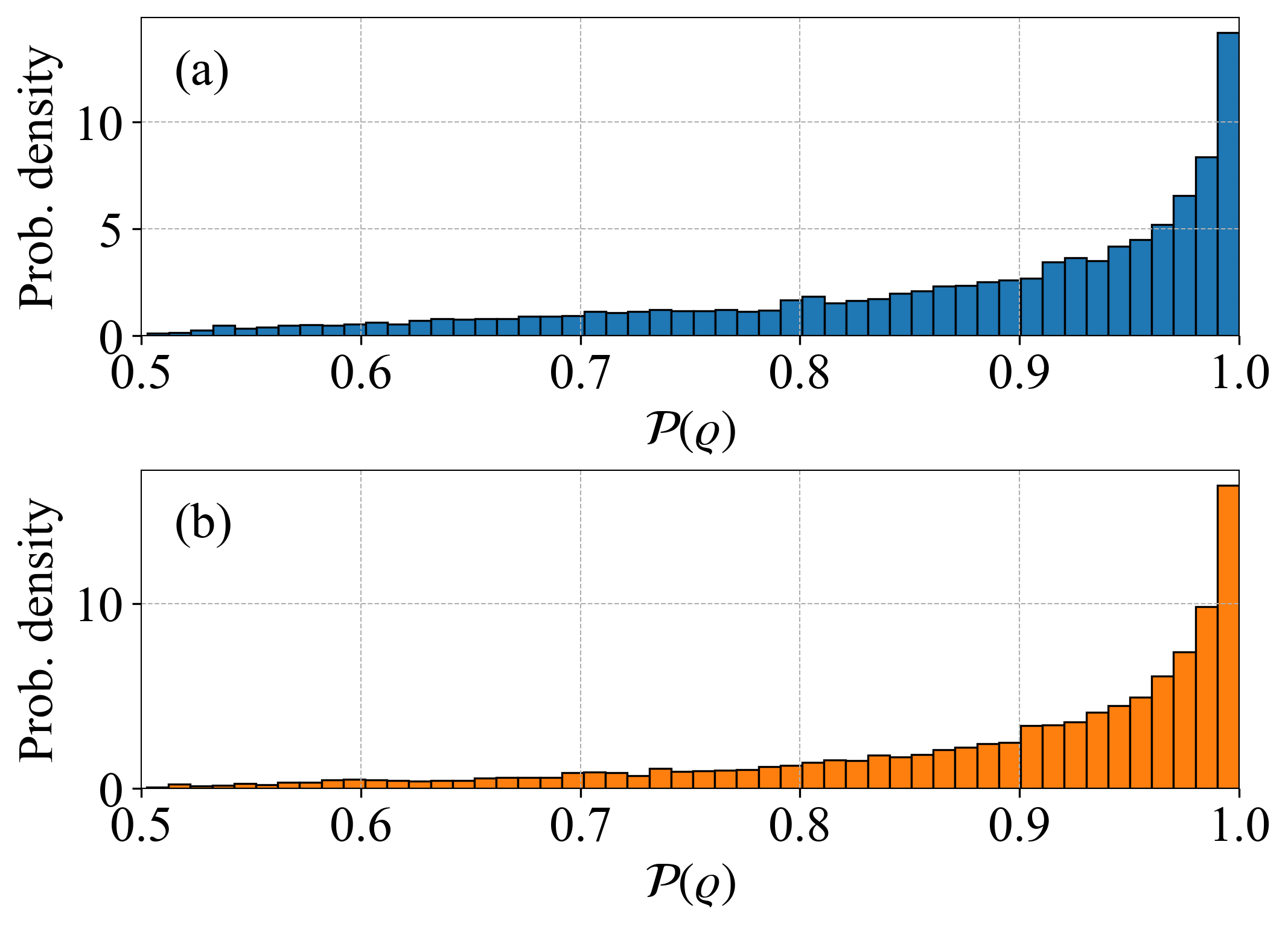}
    \caption{Distribution of the reduced purity for states generated by randomly sampled invertible matrices of dimension $2$ for both normally (a) and uniformly (b) distributed matrix elements.}
    \label{fig.:purity_dist.}
\end{figure}

Given the Hilbert space $\mathbb C^2$, we require an efficient algorithm to sample uniformly from the set of invertible transformations $A \colon \mathbb C^2 \to \mathbb C^2$. This task reduces to sampling uniformly from the general linear group $GL(2)$ with respect to the Haar measure.
The number of non-invertible matrices in $\mathbb C^{2 \times 2}$ is negligible with respect to the number of invertible ones~\cite{Ginibre_1965, Edelman_1988}. 
Thus, an efficient algorithm to generate random invertible matrices uniformly consists of randomly sampling from $\mathbb C^{2 \times 2}$ ensuring $\det A\neq 0$.
Sampling from the standard complex Ginibre ensemble in $\mathbb C^{2 \times 2}$ is equivalent to sampling both the real and imaginary parts of each matrix element from a normal distribution with mean zero and variance 1/2~\cite{Ginibre_1965, Mehta_2004}. Additionally, we have tested sampling the real and imaginary part of each element of the matrix from the uniform distribution. Figure~\ref{fig.:eig._dist.} shows the distribution of eigenvalues that arises from both methods. Also, Fig.~\ref{fig.:purity_dist.} shows distribution of entanglement in two-qubit pure states, that arises from both distributions.

Both methods generate similar distributions of entangled states, with higher probability the more separable the state. This is in fact beneficial, since for classification purposes, the error of the network will lie in the boundary between separable and entangled states (in this case, in the neighborhood of $\mathcal P (\varrho_n) = 1$). Therefore, by generating more states in this region, we aid the training of the network.

\subsection{Generating random density matrices}\label{app.:rand._2dm}

Given the Hilbert space $\mathbb C^4$, we require an efficient algorithm to sample uniformly from the set of quantum states in $\mathbb C^4$. Mathematically, it translates to generating linear operators $\varrho\colon \mathbb C^4 \to \mathbb C^4$ such that $\tr \varrho = 1$, $\varrho^\dagger = \varrho$, and $\varrho \ge 0$. For this, we use that given any matrix $X \in \mathbb C^{4 \times m}$, the operator
\begin{equation}
    \varrho = \frac{XX^{\dagger}}{\tr(XX^{\dagger})},
\end{equation}
is a quantum state. In order to generate full rank operators, a simple choice is to set $m = 4$. A standard method to generate $X$ is to sample according to the Ginibre ensemble. In this case, the matrix representing the operator $X$ has complex entries, which real and imaginary parts are sampled from a normal distribution of zero mean and standard deviation one. Using $m = 4$, the measure induced by the Ginibre ensemble coincides with the Hilbert–Schmidt measure and the average purity scales as $1/N$~\cite{Bruzda_2009}.

\begin{figure}
    \centering
    \includegraphics[width=\linewidth]{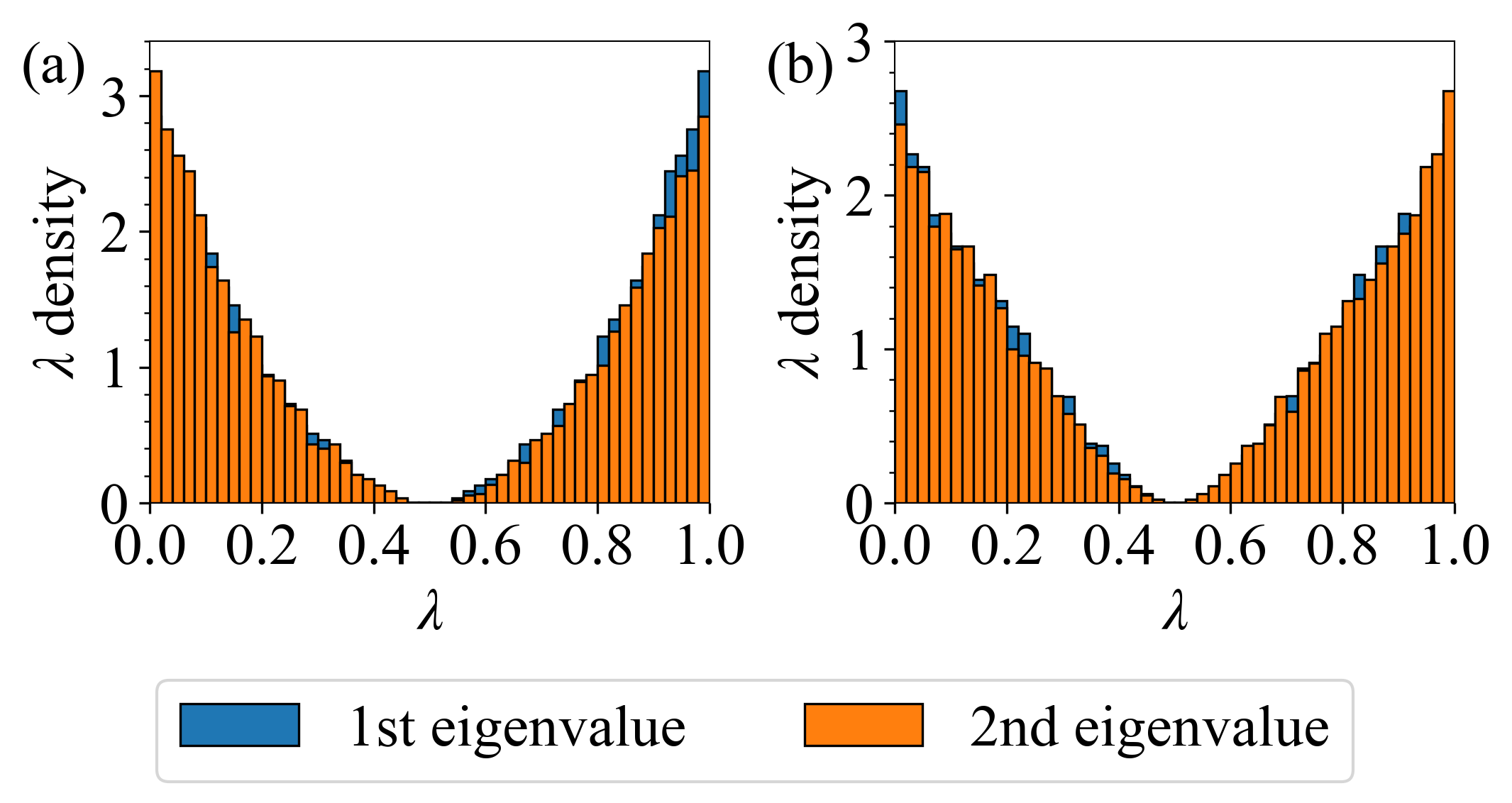}
    \caption{Distribution of eigenvalues for randomly generated invertible matrices of dimension two. Panel (a) for normally distributed matrix elements, and panel (b) for uniformly distributed matrix elements..}
    \label{fig.:dist._eig._states}
\end{figure}

Another possibility consists in sampling real and imaginary parts of the complex entries of $X$ from a uniform distribution, fixing maximum and minimum values. Figure~\ref{fig.:dist._eig._states} shows the distribution of eigenvalues of $2 \times 2$ quantum states for both normal and uniform distributions. Both samplings of the elements of $X$ give similar distributions for the eigenvalues of $\varrho$. Note that the majority of the states generated are pure. An evaluation of the performance of each distribution for training is given in App.~\ref{app.:two_qubits}.

\subsection{Three qubit states generation algorithm} \label{app.:rand._3dm}

Here we explain the method to generate the dataset for the three qubit case. We employ the witness inequality $\tr W_\phi \, \varrho_\psi < 0$ to label the states. Recall that  $W_\phi=\alpha\mathbb{I}-P_\phi$, while the state reads $\varrho_\psi= \beta \, \mathbb I/8 + (1 - \beta) P_\psi$ being $P_x=\ket{x}\bra{x}$. As explained in App.~\ref{app.:witness_inequality}, detected states correspond to $\alpha < \gamma$ and $\beta \leq \frac{\alpha - \gamma}{1/8 - \gamma}$.

The size of the detected class is negligible with respect to the non-detected one. Thus to generate the non-detected class, one could sample random states (and applying the witness as a check). Although this is  efficient, we consider instead a different procedure to  keep the same structure in both sides of the dataset. Thus, for the non-detected class we sample states with the same structure, with values $\alpha \geq \gamma$ or $\alpha < \gamma$ and $\beta > \frac{\alpha - \gamma}{1/8 - \gamma}$.

The second subdivision of the dataset comes from mixing these states among them, within the same class. Given two states $\varrho_1$ and $\varrho_2$ such that $\tr( \mathcal W \varrho_1 ) < 0$ and $\tr(\mathcal W \varrho_2) < 0$, and a probability $p \in [0, 1]$, it follows that
\begin{equation*}
    \tr(\mathcal W [p \varrho_1 + (1 - p) \varrho_2]) = p \tr(\mathcal W \varrho_1) + (1 - p) \tr( \mathcal W \varrho_2) < 0 \, .
\end{equation*}
Thus, the algorithm requires two steps. First, it generates $\varrho_\psi$ states, and then it mixes them.

In order to generate states $\varrho_\psi = \beta \, \mathbb I/8 + (1 - \beta) P_\psi$, we first sample pure states $\ket{\psi}$ uniformly from the set of all pure states with $\alpha < \gamma$, and then sample $\beta$ from the uniform distribution in $\left[\frac{\alpha - \gamma}{1/8 - \gamma}, 1\right]$. This requires  sampling a random unit vector $\ket{\psi}$ such that the squared overlap with a given unit vector $\gamma := |\langle \phi | \psi \rangle|^2$ has a prescribed value. The algorithm constructs the state
\begin{equation*}
\ket{\psi} = a |\phi\rangle + \sqrt{1 - |a|^2} |\phi^\perp\rangle, 
\end{equation*}
where $|\phi^\perp\rangle$ denotes the orthogonal complement of $|\phi\rangle$, and $|a| \in [\sqrt{\alpha}, 1]$, with a phase $\theta$ uniformly sampled in $\theta\in[0,2\pi)$ such that $a = |a| e^{i \theta}$. The practical implementation builds a full orthonormal basis $\{\ket{\phi}, \ket{\phi_2}\, \ldots, \ket{\phi_d}\}$ of $\mathbb C^d$ using Gram-Schmidt. Then it creates a unitary matrix $U$ from these basis states, and a vector
\begin{equation}
    |\psi'\rangle = \begin{pmatrix} |a| e^{i\theta} & v_1 & \cdots & v_{d - 1} \end{pmatrix}^T \, ,
\end{equation}
where $\ket{v} = \begin{pmatrix} v_1 & \cdots & v_{d - 1} \end{pmatrix}^T$ is sampled from the Haar measure in $\mathbb C^{d - 1}$ and satisfies $|\langle v | v \rangle|^2 = 1$. The state $\ket{\psi'}$ fulfills $|\langle 0| \psi'\rangle|^2=|a|^2$ with $\ket{0} = \begin{pmatrix} 1 & 0 & \cdots & 0 \end{pmatrix}^\top$. Thus, the final step is to apply the unitary $\ket{\psi} = U^\dagger \ket{\psi'}$ to move from the canonical basis to $\{\ket{\phi}, \ket{\phi_2}\, \ldots, \ket{\phi_d}\}$.

\begin{figure}
    \centering
    \includegraphics[width=\linewidth]{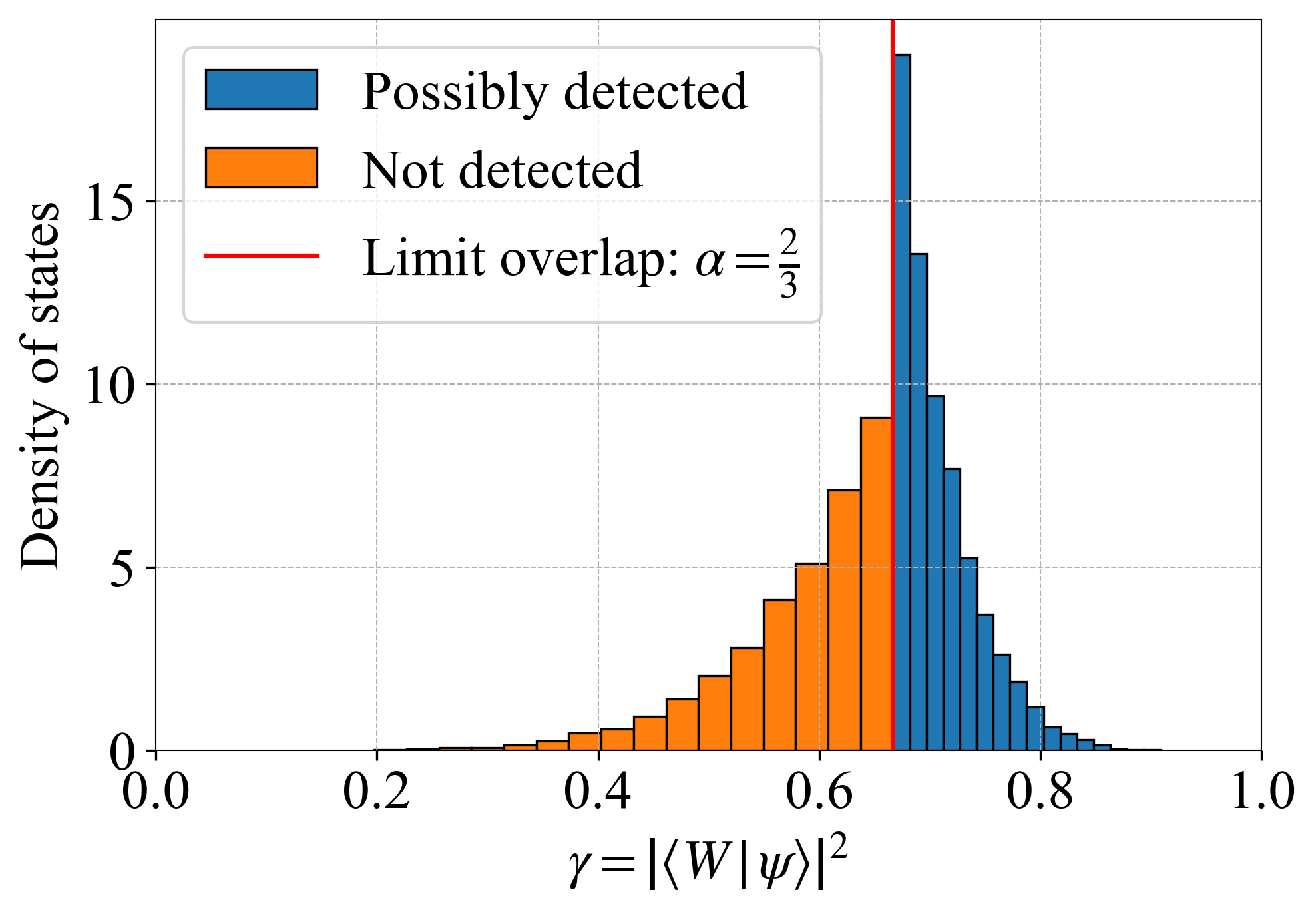}
    \caption{Histogram of sampled values of $\gamma$ for both conditional density functions of cases $\gamma \leq \alpha$ and $\alpha < \gamma$. The distribution ensures uniform sampling of the states $\ket{\psi}$. We have chosen $\ket{\phi} = \ket{\rm W}$ for illustration purposes.}
    \label{fig.:beta_dist.}
\end{figure}

To ensure the resulting states cover uniformly the cap $\{\ket{\psi} \in \mathbb C^d \, | \, |\langle \phi | \psi \rangle|^2 > \alpha\}$, we must sample $|a|$ from the beta distribution $\mathrm{Beta}(1, d-1)$. Figure~\ref{fig.:beta_dist.} shows the histogram of sampled values of $\gamma$.

As mentioned before, we proceed now to mix the states within a class between them. To do so, we just sample $m \in [2, 20]$ elements from this set at random. Then, generate a random probability distribution with $M$ elements. That is, a set $\{p_m \, | \, m = 1, \ldots, M\}$ satisfying $p_m \in [0, 1]$ and $\sum_m p_m = 1$. Finally compute the new state $\varrho = \sum_m p_m \, \varrho_m$.

\section{Model optimization}

\subsection{Network design}\label{app.:net_design}

This appendix describes the dataset organization, labeling scheme, and optimization components used for all neural network architectures. Let $C$ denote the number of entanglement classes for an $N$-qubit system, and let $E_c$ be the number of training examples generated per class. The total train set size is therefore $E = C E_c$. The test set contains $E_c/10$ states per class, yielding a total size of $E/10$. The corresponding data arrays, $X_{\mathrm{tr}}$ and $X_{\mathrm{tst}}$, have shapes $(E, 4^N)$ and $(E/10, 4^N)$, with each row containing the $4^N$ components of a single state's correlation vector.

To ensure unbiased model selection and prevent leakage between development and evaluation, we employ two additional datasets, both independent of the training and test sets. A development set is used for iterative tuning and may vary in size depending on computational constraints. A separate validation set, with size $1/10$ times the development set, is used exclusively to monitor performance during optimization. The use of distinct data for development and testing phases ensures independent and statistically reliable results. The specific sizes for these auxiliary sets are reported in the corresponding experimental sections.

Class labels follow a fixed, arbitrary ordering. The fully separable class is assigned index $c = 0$, and the remaining classes are indexed consecutively up to $c = C-1$. Labels for the training and test sets are stored in vectors $y_{\mathrm{tr}}$ and $y_{\mathrm{tst}}$, with shapes $(E,1)$ and $(E/10,1)$. The network outputs are encoded using one-hot encoding scheme. The target output for a state in class $c$ is a vector of length $C$ with a ``1'' in the $c$-th position and ``0''s elsewhere. In the two-qubit case ($C=2$), the separable and entangled classes correspond to $(1,0)^T$ and $(0,1)^T$, respectively. For three-qubit pure states, the indices are: fully separable ($c=0$), the three biseparable classes ($c=1,2,3$), and the W and GHZ classes ($c=4,5$).

The input layer always contains $4^N$ neurons to match the dimension of the correlation vector. The output layer contains $C$ neurons, one for each class. Ideally, the $c$-th output neuron activates strongly for states of class $c$, while all others remain near zero.

All models are trained using the multi-class Cross-Entropy Loss~\cite{Cover_2005}. It provides a unified objective for systems with different numbers of classes and avoids architectural adjustments when moving between binary and multi-class regimes. Optimization is performed using the Adam algorithm~\cite{Zhang_2018}. Empirical comparison with SGD and RMSProp showed consistently faster convergence and higher final accuracies.

We monitor classification accuracy on all datasets as
\begin{equation*}
    \mathrm{Accuracy} = \frac{\text{Number of correctly classified states}}{\text{Total number of states}} \, .
\end{equation*}

We adopt the Rectified Linear Unit (ReLU) activation function for all hidden layers. ReLU provides computational efficiency and mitigates vanishing gradients. A comparative experiment confirmed equal or superior performance compared to tanh and sigmoid~\cite{LeCun_2015}.

\subsection{Pure states: architecture optimization}\label{app.:pure_arch._opt.}

This section serves two primary purposes. First, to justify the choice of Convolutional Neural Networks (CNNs) over Dense Neural Networks (DNNs), given DNNs’ susceptibility to over-fitting. And second, to summarize the hyperparameter optimization process leading to the final CNN design.

\subsubsection{DNN over-fitting reduction}
\label{app.:over-fitting_reduction}

We initially performed hyperparameter optimization to mitigate over-fitting. Table~\ref{tab.:dnn_arch._test} reports development and validation accuracies for a range of architectures. Broadly, accuracy increases as the number of neurons per layer does. Interestingly, while development loss decreases equally, validation loss increases with validation accuracy; signature of over-fitting. Wide early layers improved performance more than increased depth. This suggests that early-layer width is critical for capturing the complex, non-local features characteristic of entanglement. Note that individual accuracy values are subject to variability due to randomness from network initialization and the stochastic optimization process.

\begin{table}
    \centering
    \begin{tabular}{ | c c c | }
    \hline
    Arch. & Val. acc. (\%) & Dev. acc. (\%) \\
    \hline
    $(64^2, 50, 15, 6)$ & $78$ & $75.7$ \\
    $(64^2, 50^2, 35^2, 15^2, 6)$ & $75$ & $71.5$ \\
    $(64^3, 50^3, 35^3, 15^3, 6^3)$ & $78.5$ & $76.2$ \\
    $(64^4, 50^4, 35^4, 15^4, 6)$ & $76.5$ & $74.2$ \\
    $(64^4, 50^4, 35^4, 15^4, 6^3)$ & $75.4$ & $72.5$ \\
    $(64, 48, 32, 16, 6)$ & $78.2$ & $76$ \\
    $(100^2, 85, 55, 25, 6)$ & $88.7$ & $79$ \\
    \hline
    $(512, 128, 64, 6)$ & $97.6$ & $78.6$ \\
    $(512, 128^2, 6)$ & $98$ & $79.1$ \\
    $(512, 256, 64, 6)$ & $98.7$ & $78.7$ \\
    $(512, 256, 128, 6)$ & $98.6$ & $79.2$ \\
    $(512^2, 32, 6)$ & $98.9$ & $78.5$ \\
    $(512^2, 64, 6)$ & $99.2$ & $78.9$ \\
    $(512^2, 128, 6)$ & $99.1$ & $79.4$ \\
    $(512^2, 128, 32, 6)$ & $99.4$ & $79.1$ \\
    $(512^2, 128, 64, 6)$ & $100$  & $80.4$ \\
    \hline
    \end{tabular}
    \caption{Validation and development accuracy for several architectures of DNN.}
    \label{tab.:dnn_arch._test}
    \end{table}

Using the best-performing architecture, we individually optimized key hyperparameters while keeping others fixed to a learning rate of $\mathrm{lr} = 10^{-3}$, batch size of $\mathrm{bs} = 64$, and $\mathrm{ep} = 100$ epochs. We have identified settings that generalize well across all pure and mixed quantum systems.

\begin{figure}
    \centering
    \includegraphics[width=\linewidth]{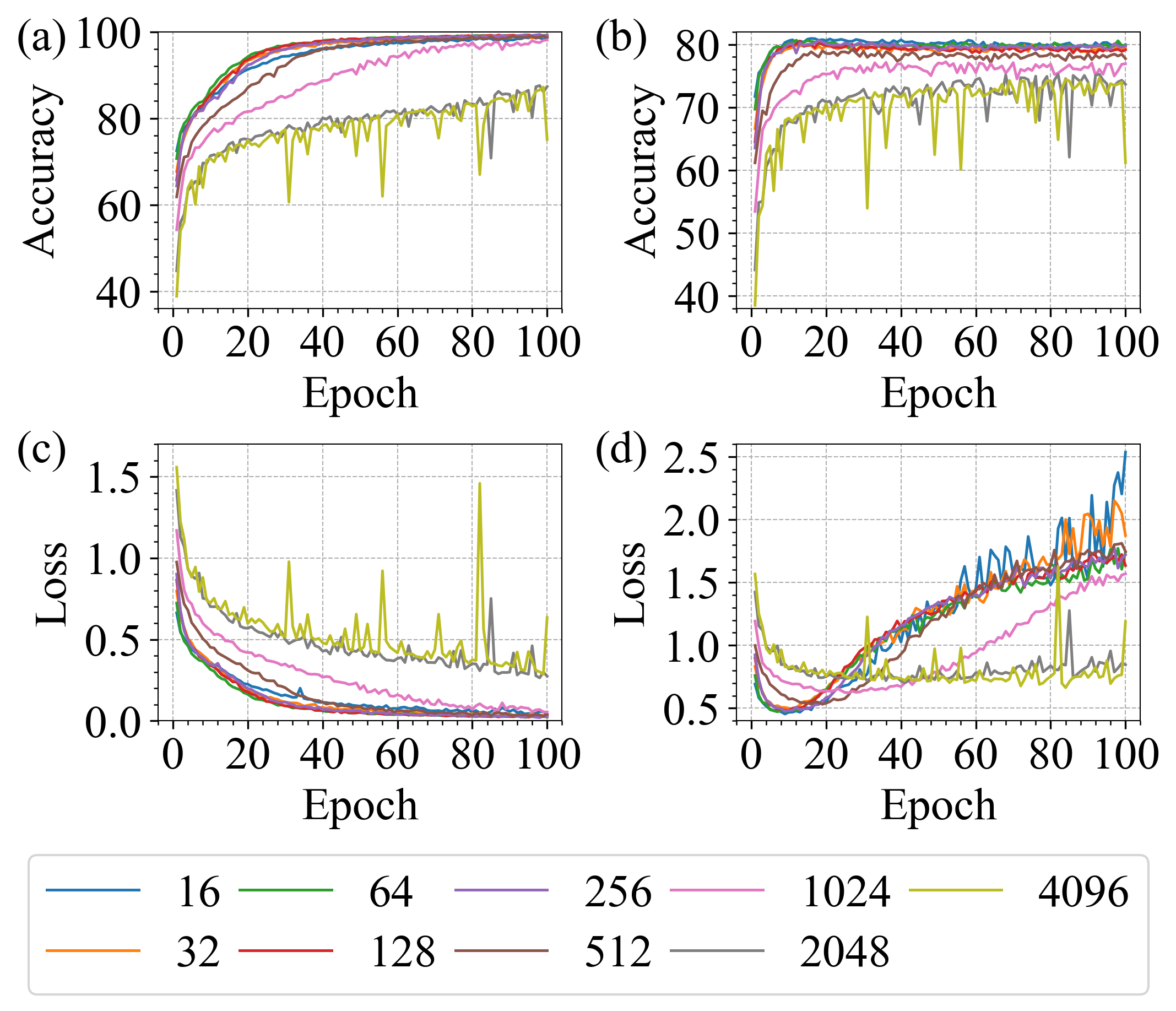}
    \caption{Results of the batch size optimization process. Panels (a) and (b) show the development and validation accuracy, while panels (c) and (d) show the development and validation losses respectively; all of them for different values of the batch size.}
    \label{fig.:dnn_tunning_batch}
\end{figure}

Figure~\ref{fig.:dnn_tunning_batch} shows the results batch size optimization, where the values explored range from $2^4 = 16$ to $2^{12} = 4096$ in powers of two, i.e., $\mathrm{bs} \in \{2^k \ | \ k = 4, 5, \ldots, 12\}$. Larger batch sizes reduce over-fitting. This is evidenced by: (i) a smaller gap between development and validation accuracy; and (ii) lower, more stable validation loss values, avoiding concave profiles indicating over-optimization. However, this reduction in over-fitting comes at the cost of lower overall development and validation accuracy and higher development loss. The trade-off arises since smaller batch sizes provide more steps of parameter update each epoch. Larger batch sizes lead to fewer updates and often converge to flatter minima. Additionally, larger batch sizes significantly reduce training time per epoch, an important factor for larger problems. For our purposes, batch sizes in the range $\mathrm{bs} \in [64, 512]$ perform adequately, with $\mathrm{bs} = 64$ yielding slightly better accuracy and $\mathrm{bs} = 512$ offering superior efficiency.

\begin{figure}
    \centering
    \includegraphics[width=\linewidth]{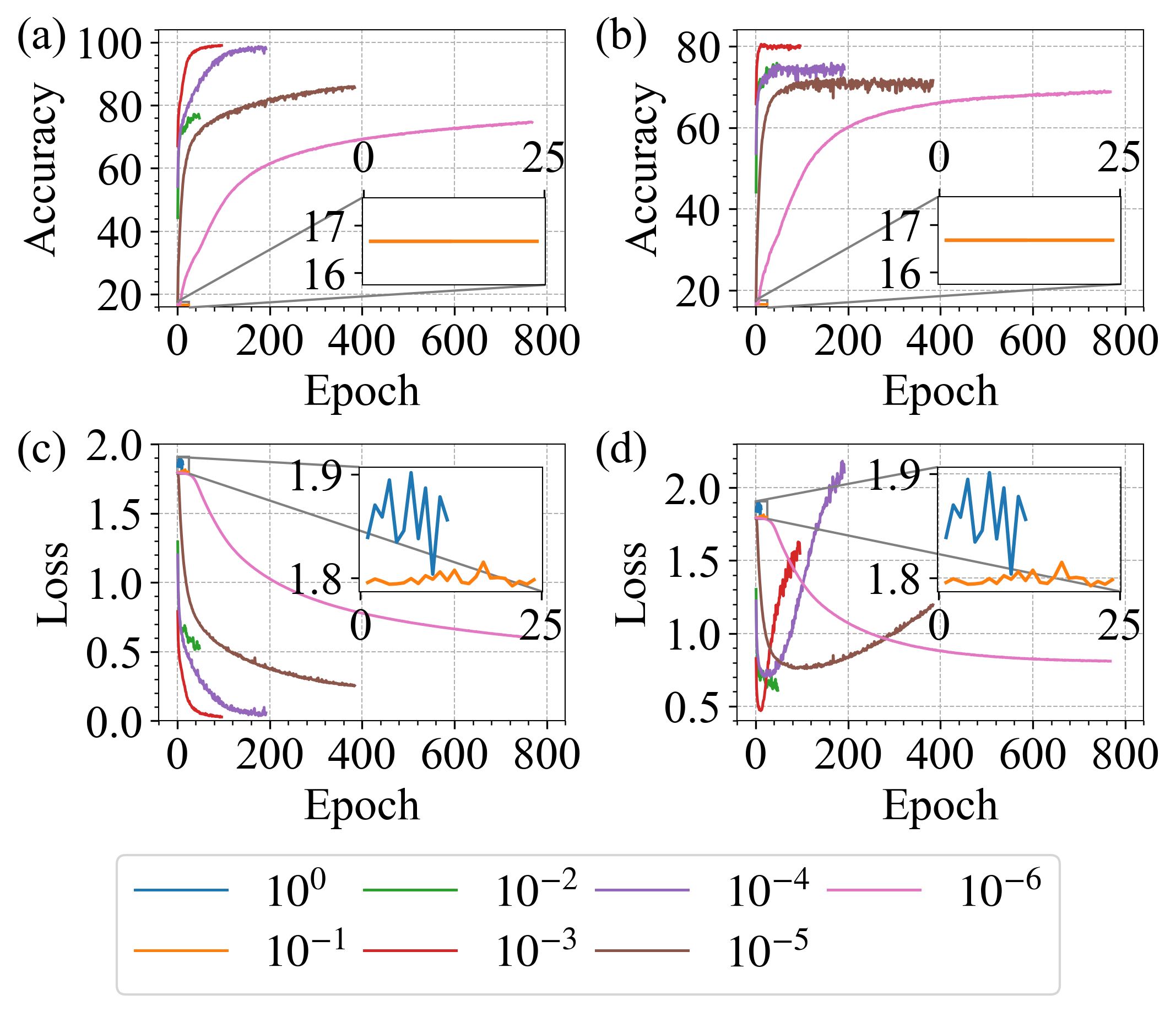}
    \caption{Results of the learning rate optimization process. Panels (a) and (b) show the development and validation accuracy respectively, while panels (c) and (d) show the development and validation losses respectively, all for different values of the learning rate. Depending on the value of the learning rate, different epochs have been considered. In panels (a) and (b), the plots $lr = 1, 0.1$ overlap, and that is why only one line is shown.}
    \label{fig.:dnn_tunning_lr}
\end{figure}

Figure~\ref{fig.:dnn_tunning_lr} shows the optimization results for the learning rate. Experiments were conducted for learning rates from $10^0 = 1$ down to $10^{-6}$ in multiplicative steps of ten, i.e., $\mathrm{lr} \in \{10^{-k} \ | \ k = 0, 1, \ldots, 6\}$. Smaller learning rates lead to slower convergence. To account for this, we scaled the number of training epochs proportionally: for a learning rate of $10^{-k}$, the network was trained for $12 \cdot 2^k$ epochs. This scaling is a rough estimate of the epochs needed for convergence and is sufficient for a comparative study. A direct consequence of this scaling is a significant increase in total training time for smaller learning rates, though evaluation time remains unaffected.

Extremely high learning rates ($\mathrm{lr} = 1, 0.1$) fail to converge. Low rates either converge too slowly or to suboptimal minima. The optimal balance was achieved at $\mathrm{lr} = 10^{-3}$, with the best accuracies and development. The value $\mathrm{lr} = 10^{-2}$ exhibits slightly less over-fitting, but achieved a lower final validation accuracy. Since the ultimate goal is to maximize generalizable performance $\mathrm{lr} = 10^{-3}$ remains the superior choice.

\begin{figure}
    \centering
    \includegraphics[width=\linewidth]{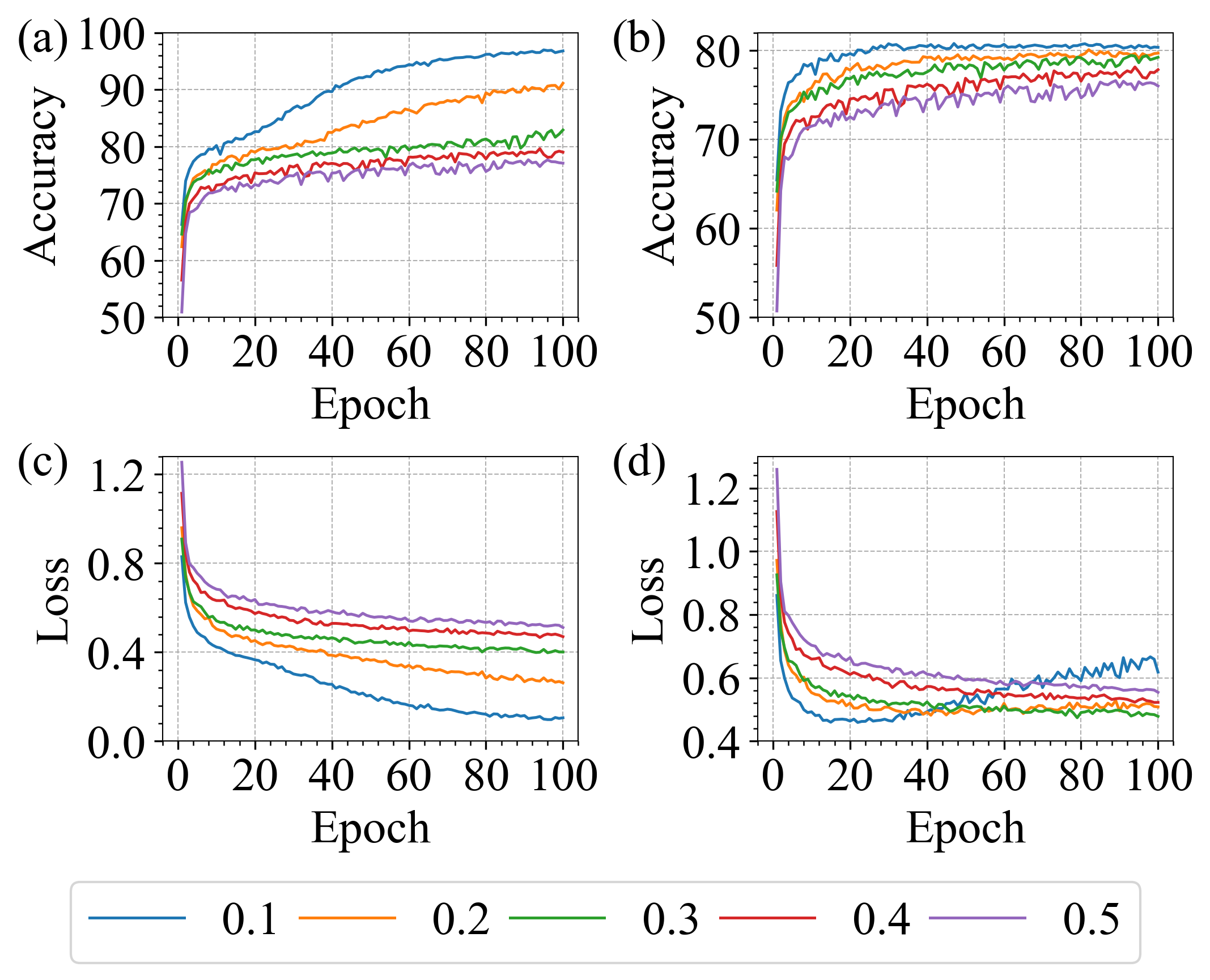}
    \caption{Results of the dropout optimization process. Panels (a) and (b) show the development and validation accuracy respectively, while panels (c) and (d) show the development and validation losses respectively; all of them for different values of the dropout probability.}
    \label{fig.:dnn_tunning_drop.}
\end{figure}

Figure~\ref{fig.:dnn_tunning_drop.} shows the results of incorporating dropout layers after each dense layer in the DNN. All dropout layers shared the same dropout probability $p$, tested for values $p \in \{0.1, 0.2, 0.3, 0.4, 0.5\}$. Dropout reduced over-fitting: the gap between development and validation accuracy narrowed as the dropout probability increased. However, this regularization came at the cost of reduced accuracy overall. The best validation accuracy achieved with dropout did not surpass that of the standard DNN without dropout, indicating that the benefits of reduced over-fitting were outweighed by a loss in expressive power.

None of the standard techniques for mitigating DNN over-fitting provided a validation performance higher than the unregularized DNN. Data augmentation remains feasible and effective. However, this approach carries a significant computational cost for training. Given that CNNs are architecturally more efficient and show less over-fitting, we pivoted our focus to optimizing a CNN architecture. Furthermore, data augmentation consistently improves accuracy and reduces over-fitting in every quantum systems and network tested, usable also in CNNs.

\subsubsection{CNN optimal design}
\label{app.:CNN_design}

CNNs offer inherent over-fitting resistance, reducing the need for additional regularization techniques. Optimal hyperparameters mirror those of the DNN: $\mathrm{lr} = 10^{-3}$ and $\mathrm{bs} = 64$. For CNN-specific design, we empirically found design rules for CNN-based entanglement classification across the studied problems.

Pooling was generally ineffective. Both max and average pooling degrade performance by a loss of relevant information. Dimensionality reduction is better handled by the final DNN layers.
Stride of $1$ provides maximal information extraction which preserves local correlations crucial for entanglement features. Larger strides risk skipping relevant information.
Increasing the number of feature channels beyond a threshold saturates accuracy gains while increasing computational cost and over-fitting. An optimal number exists that balances accuracy and efficiency.
A filter size of $3$ suffices; larger filters do not improve accuracy. Likely because input correlation vectors contain redundant information localizing the correlations. This choice implicitly sets the padding $1$ to preserve the spatial dimensions of the feature maps.
Increasing the number of convolutional layers increases accuracy by capturing hierarchical features. Similarly, increasing the number of DNN layers improves accuracy but are a big source of over-fitting.
Finally, we employed two phases training. An initial phase with $\mathrm{lr} = 10^{-3}$ for 100 epochs to achieve rapid initial convergence. Then, a second phase of $\mathrm{lr} = 10^{-4}$ for 50 epochs to fine-tune weights and approach sharper minima.

Table~\ref{tab.:pure_cnn_arch.} summarizes the final architectures and hyperparameters for two- and three-qubit systems. The performance results are shown in Fig.~\ref{fig.:pure_model}.

\begin{table}[hbtp]
    \centering
    \begin{tabular}{|c|c|c|c|}
    \hline
    \multirow{4}{*}{two qubits} & kernel & stride & padding \\
                             & $3$ & $1$ & $1$ \\
                             \cline{2-4}
                             & $\mathrm{lr}$ & $\mathrm{bs}$ & $\mathrm{ep}$ \\
                             & $(10^{-3}, 10^{-4})$ & $64$ & $(100, 50)$ \\
                             \cline{2-4}
                             & \multicolumn{3}{c|}{Architecture} \\
                             & \multicolumn{3}{c|}{$(4, 8^3) + (64, 2)$} \\
    \hline
    \hline
    \multirow{4}{*}{3 qubits} & kernel & stride & padding \\
                             & $3$ & $1$ & $1$ \\
                             \cline{2-4}
                             & $\mathrm{lr}$ & $\mathrm{bs}$ & $\mathrm{ep}$ \\
                             & $(10^{-3}, 10^{-4})$ & $64$ & $(100, 50)$ \\
                             \cline{2-4}
                             & \multicolumn{3}{c|}{Architecture} \\
                             & \multicolumn{3}{c|}{$(4, 8^7) + (128, 64, 2)$} \\
    \hline
    \end{tabular}
    \caption{Final values of the CNN for 2 and 3 qubits pure states SLOCC entanglement classification.}
    \label{tab.:pure_cnn_arch.}
\end{table}

\subsubsection{Final network architecture}\label{app.:net._arch.}

\begin{figure}
    \centering
    \includegraphics[width=\linewidth]{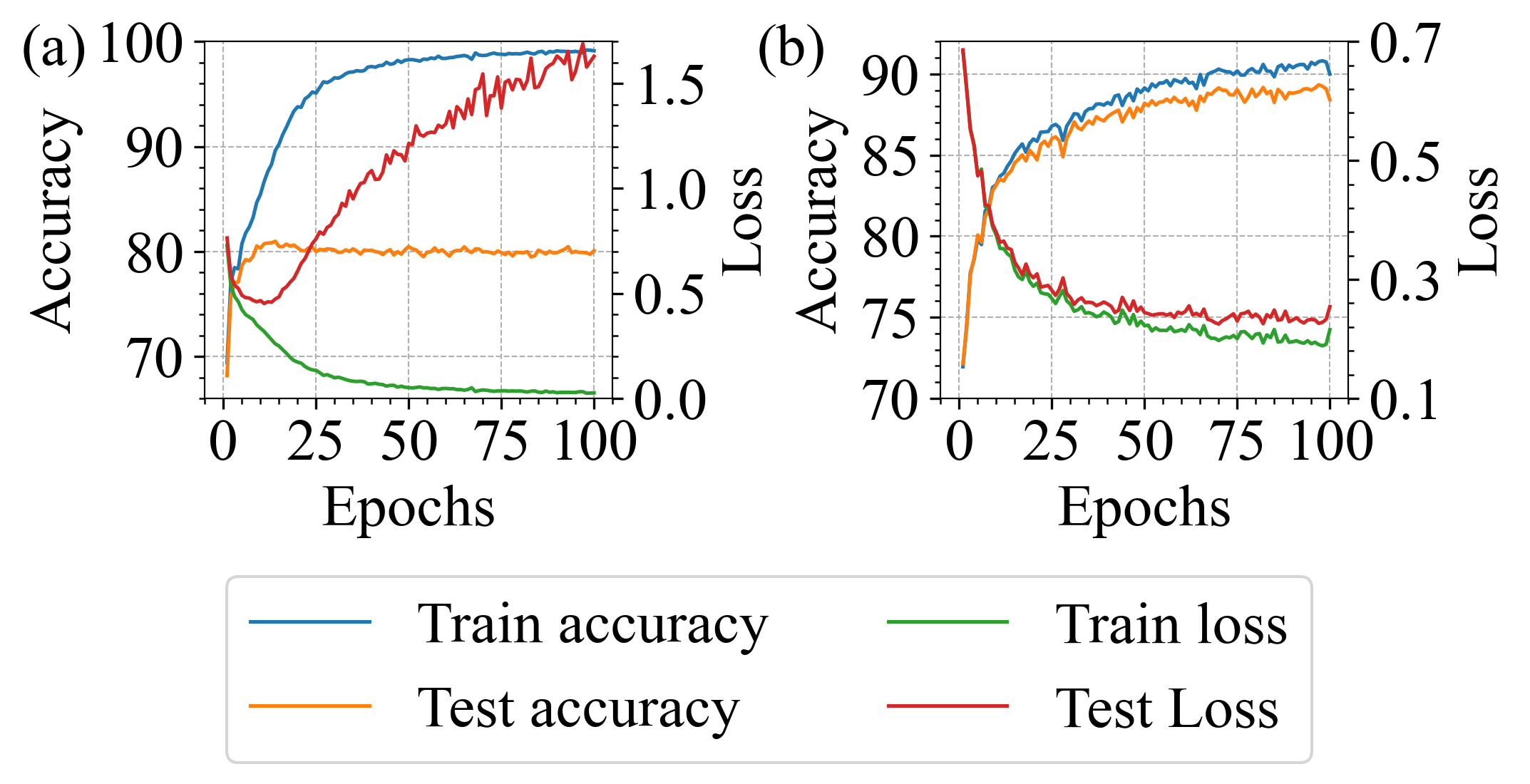}
    \caption{Comparative analysis of DNN and CNN performance on three-qubit pure state entanglement classification. (a) Test and train accuracies and losses during the training process of a DNN. The architecture is $(512, 512, 128, 64, 6)$. (b) Test and train accuracies and losses during the training process of a CNN. The architecture is $(32, 16, 8, 8, 8)$ filters for the CNN section, and $(128, 64, 6)$ for the DNN section. Both networks are using ReLU activation function. The values of the hyperparameters are: $\mathrm{bs} = 64$, $\mathrm{lr} = 10^{-3}$, $\mathrm{ep} = 100$. They were trained with $E_c = 3 \cdot 10^4$ states per class.}
    \label{fig.:pure_3q_dnn_vs._cnn}
\end{figure}

The final model consists of two stages: a 1D CNN for feature extraction, followed by a DNN for final classification. Figure~\ref{fig.:pure_3q_dnn_vs._cnn} compares training trajectories of representative DNN and CNN architectures for three qubits. The DNN achieves marginally higher training accuracy but suffers from over-fitting, as evidenced by rising test loss after initial epochs. The CNN exhibits stable training and test loss, indicating robust generalization. Differences in argmax-based classification prevent catastrophic drops in DNN test accuracy despite increasing loss.  

\begin{figure}
    \centering
    \includegraphics[width=\linewidth]{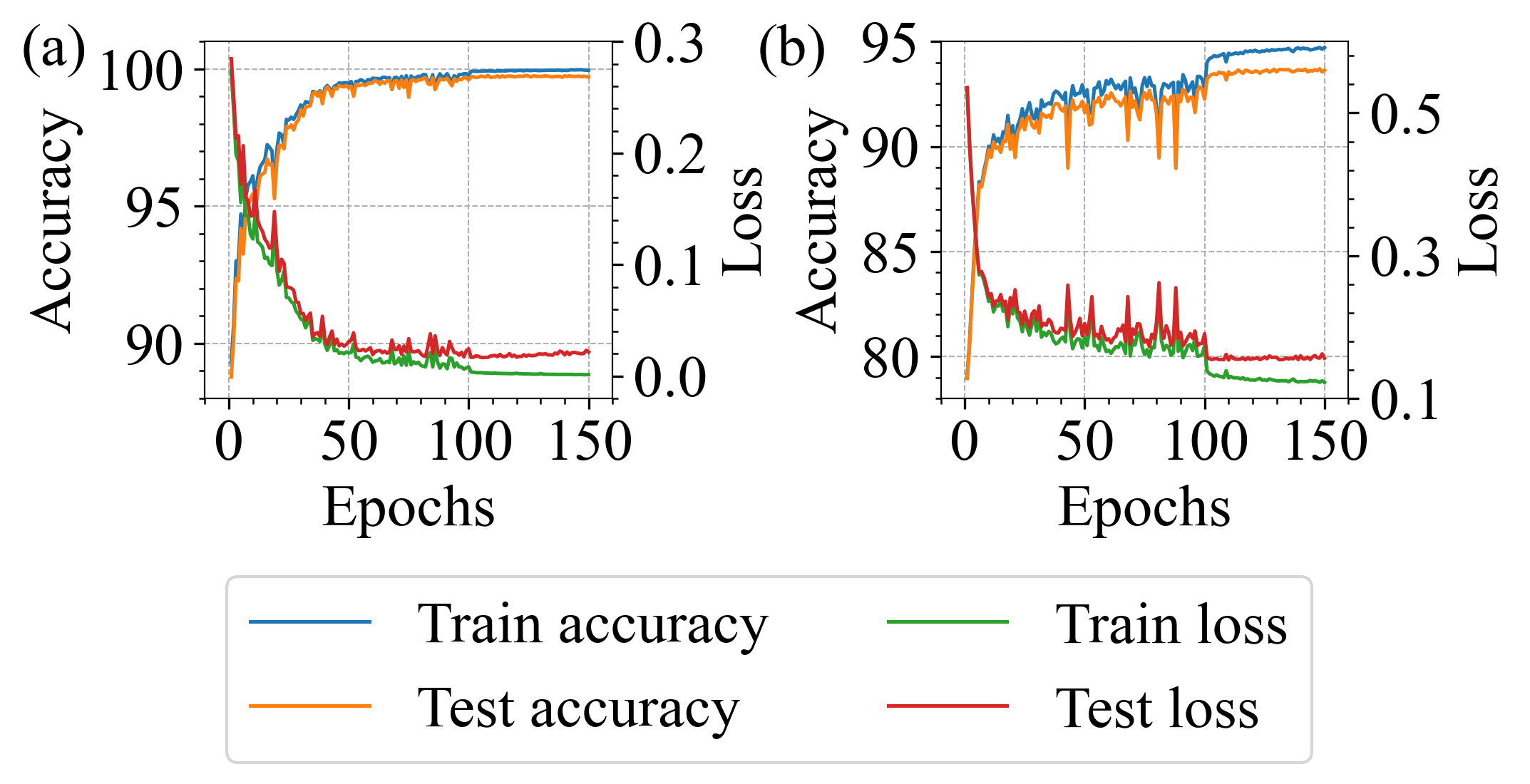}
    \caption{Results of the training process for the optimized CNN in the 2 and 3 qubit cases. (a) Accuracy and loss for the final two-qubit CNN. The architecture has $4$ convolutional layers with $(4, 8, 8, 8)$ filters respectively, and two dense layers with $(64, 2)$ neurons. The dataset consisted of $E_c = 5 \times 10^4$ states per class, yielding a total training set size of $E = 10^5$. (b) Accuracy and loss for the final three-qubit CNN. The architecture has $5$ convolutional layers with $(32, 16, 8, 8, 8)$ filters respectively, and $3$ dense layers with $(128, 64, 2)$ neurons. The dataset consisted of $E_c = 10^5$ states per class, yielding a total training set size of $E = 2 \cdot 10^5$.}
    \label{fig.:pure_model}
\end{figure}

While the DNN achieves a marginally higher training accuracy, it also presents significant overfitting, shown by the divergence of the training and test losses. The reason the test accuracy does not collapse entirely is that classification depends only on the ranking of the output neuron activations (i.e., the argmax), not their precise values. Thus the class remains correct although the output vector diverges from the theoretical one. In contrast, the CNN shows stable training and test loss, signifying robust generalization. Moreover, advantage in performance between the DNN and the CNN allows to reduce the gap between the DNN and CNN training accuracies by increasing the dataset sizes (See Fig.~\ref{fig.:pure_model}(b)).

Figures~\ref{fig.:pure_model}(a) and~\ref{fig.:pure_model}(b) show the final results for the two- and three-qubit system. In the two-qubit case, model achieved near-perfect performance with a training accuracy of $100 \, \%$ and a test accuracy of $99.7 \, \%$. For the three-qubit system, the final accuracies were $94.72 \, \%$ for the training set and $93.81 \, \%$ for the test set. Higher accuracy are likely achievable with a larger network architecture and/or more training data. However, given our computational resources, the chosen model sizes and dataset volumes represent a practical limit for the subsequent stages of this research. The achieved accuracies are more than sufficient for proving the validity of the method.

\subsection{Mixed states: architecture optimization}\label{app.:mixed_arch._opt.}

\subsubsection{Two qubits}\label{app.:two_qubits}

As detailed in App.~\ref{app.:rand._2dm}, we considered two distinct methods for generating random density matrices. Each inducing a different prior distribution over the state space. We first test which distribution is more amenable to learning by a neural network. The results are summarized in Table~\ref{tab.:dataset_dist}. The rapid saturation of accuracy metrics indicates efficient learning. Minor variations in the accuracy can be attributed to stochastic processes in the initialization and training procedure, rather than over-training. The data suggests that the uniform distribution yields slightly superior performance across both training and test metrics. Consequently, the uniform distribution was selected for generating all two-qubit mixed state datasets.

\begin{table}[b!]
    \centering
    \begin{tabular}{|c|c|c|c|}
    \hline
    Epochs & Dist. & Tr. acc. ($\%$) & Tst. acc ($\%$)\\
    \hline
    \hline
    \multirow{2}{*}{$30$ epochs} & normal & $99,3$ & $95$ \\
    
                             & uniform & $97,4$ & $95,5$ \\
    \hline
    \multirow{2}{*}{$50$ epochs} & normal & $96,6$ & $94,1$ \\
                             & uniform & $97,9$ & $95,3$ \\
    \hline
    \multirow{2}{*}{$100$ epochs} & normal & $97,4$ & $93,8$ \\
                             & uniform & $98,7$ & $95,3$ \\
    \hline
    \end{tabular}
    \caption{Training and test accuracies for a CNN comparing distribution of the states in the dataset. Baseline architecture for all tests consists on a 1D convolutional layer with $64$ filters, a kernel size of $9$, padding of $4$, and a stride of $1$. A subsequent dense section with $5$ layers containing $(128, 64, 16, 8, 2)$ neurons, respectively. Standard hyperparameters were used across all tests: $\mathrm{bs} = 64$ and a $\mathrm{lr} = 10^{-3}$.}
    \label{tab.:dataset_dist}
\end{table}

\begin{figure}
    \includegraphics[width=\linewidth]{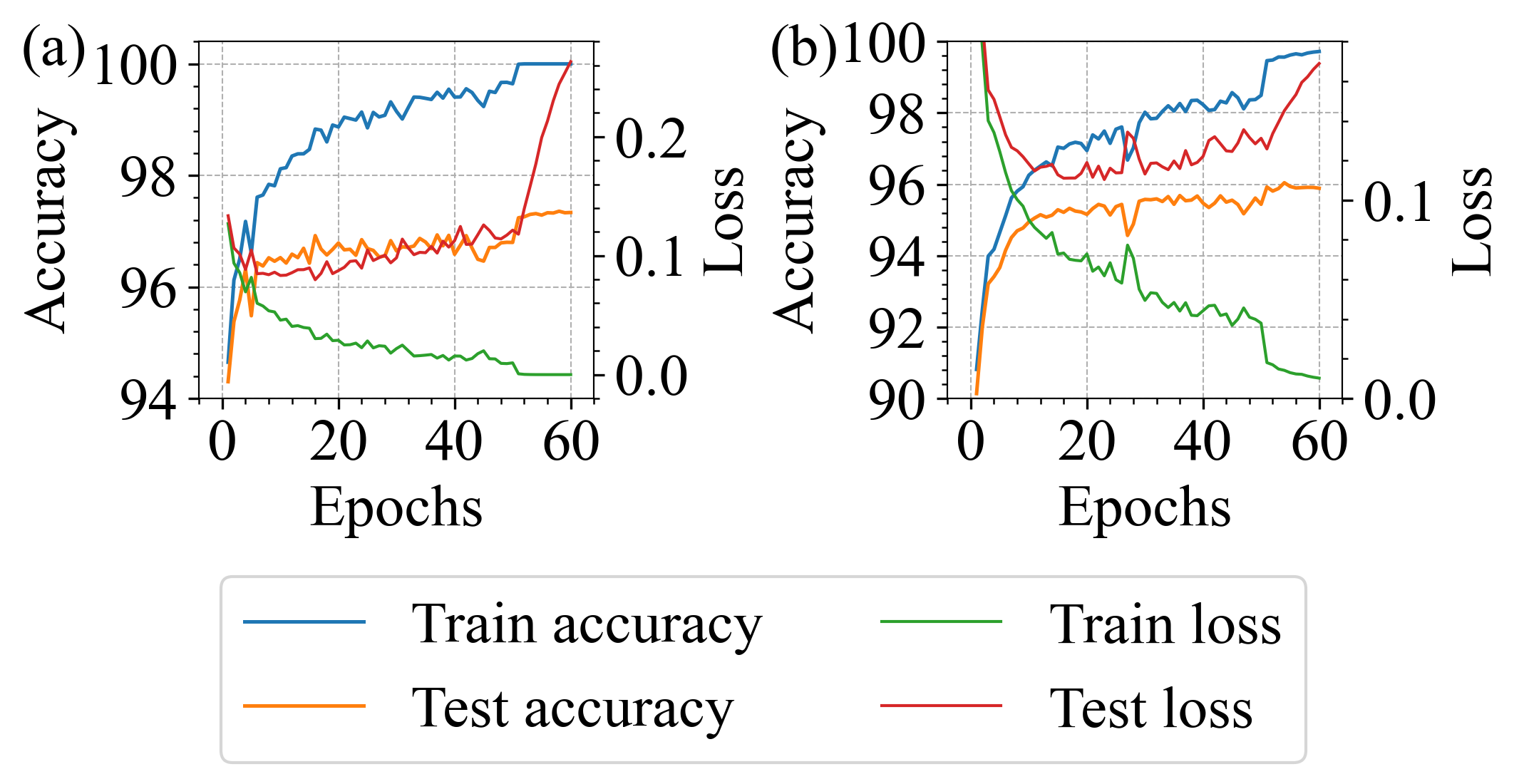}
    \caption{Comparative analysis of DNN and CNN performance on two-qubit mixed state entanglement classification. Both models utilize the ReLU activation function. The models were trained with $10^{-3}$ for $50$ epochs, followed by a reduced rate of $10^{-4}$ for $10$ epochs. Both used $\mathrm{bs} = 64$. The dataset consisted of $E_c = 2 \times 10^6$ states per class for a total $E = 4 \times 10^6$ states. The plot shows the training and test accuracies and the corresponding cross-entropy losses over the course of training. (a) DNN classifier with three hidden layers of $(256, 64, 2)$ neurons. (b) 1D-CNN classifier with a single convolutional layer with $64$ filters, followed by a dense section with $(256, 64, 2)$ neurons. The model was trained with a learning rate of $10^{-3}$ for 50 epochs, followed by $10^{-4}$ for 30 epochs.}
    \label{fig.:mixed_2q_dnn_vs._cnn}
\end{figure}

Figure~\ref{fig.:mixed_2q_dnn_vs._cnn} compares the training processes for a DNN (Fig.~\ref{fig.:mixed_2q_dnn_vs._cnn}(a)) and a CNN (Fig.~\ref{fig.:mixed_2q_dnn_vs._cnn}(b)). Both architectures achieve similar training accuracy. The DNN attains a slightly higher test accuracy. The level of overfitting, as indicated by the final loss values, is comparable between the two. A decisive advantage for the DNN is its faster convergence, requiring fewer epochs to finish training despite having a simpler architecture. The argmax-based classification scheme again ensures that the test accuracy remains stable even if the precise output neuron values diverge. These results justify the selection of the DNN over the CNN. The dataset size is four times larger than the dataset used for pure states, reflecting the increased complexity of the learning task. Higher accuracy are likely achievable with a larger network architecture and/or more training data. However, given our computational resources, the chosen model sizes and dataset volumes represent a practical limit for the subsequent stages of this research. The achieved accuracies are more than sufficient for proving the validity of the method.

\subsubsection{Three qubits}\label{app.:three_qubits}

\begin{figure}
    \centering
    \includegraphics[width=\linewidth]{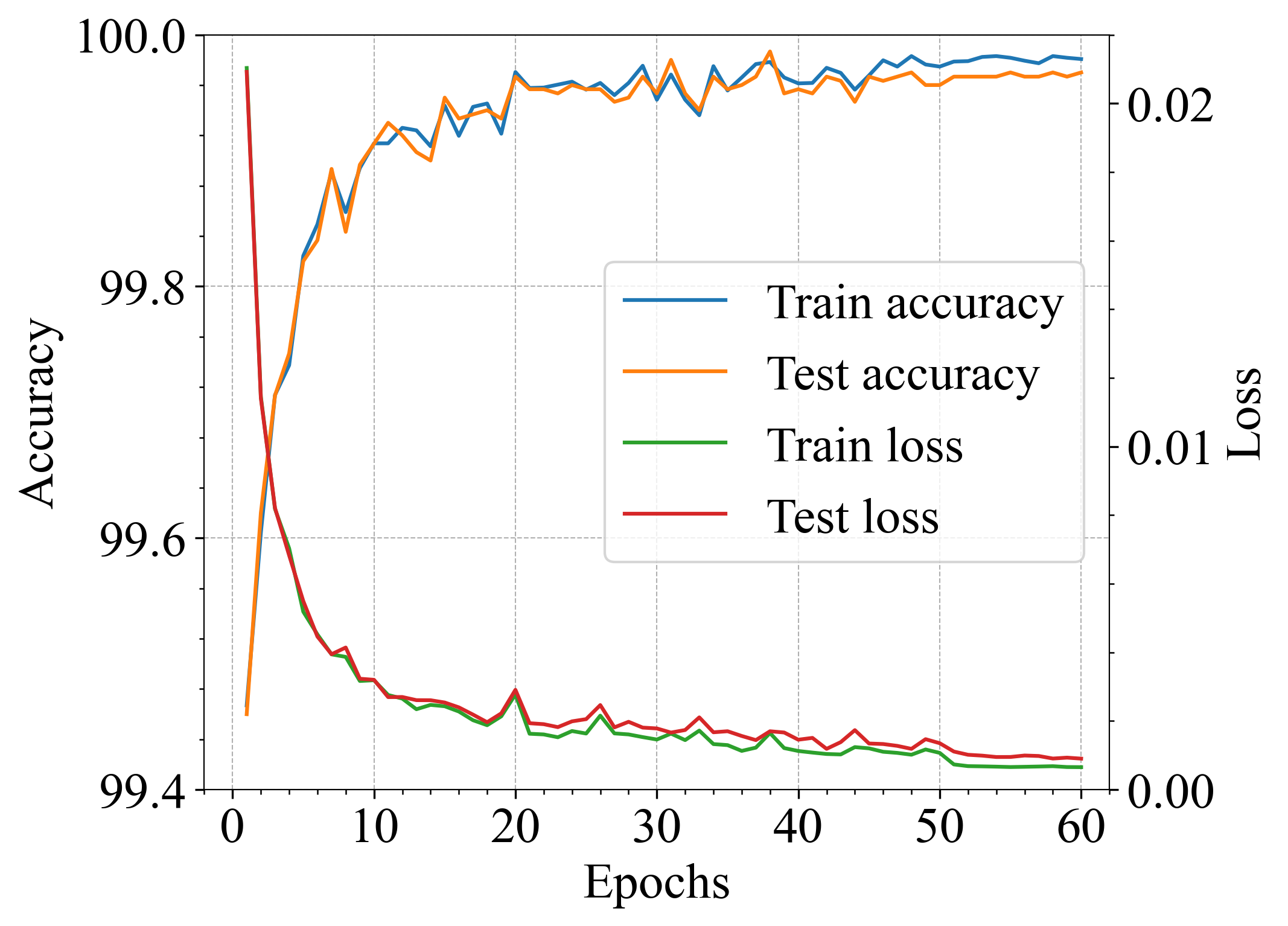}
    \caption{Test and train accuracies and losses during the training process of a DNN. The architecture consists of $(82, 2)$ neurons. The dataset for this model consisted of $E_c = 2.5 \times 10^4$ states per class, resulting in a total training set size of $E = 5 \times 10^4$ states. The learning rate was $10^{-3}$ for $50$ epochs and $10^{-4}$ for $10$ additional epochs.}
    \label{fig.:mixed_3q_model}
\end{figure}

The network depicted in Fig.~\ref{fig.:mixed_3q_model} represents the final selected architecture for the three-qubit mixed state system. The DNN architecture is notably more compact, comprising only two hidden layers with $(8, 2)$ neurons, respectively. The model achieved exceptional performance. Our primary goal was extreme computational efficiency for the subsequent explainability analysis. However, these results indicate that virtually perfect accuracy is attainable, potentially by further increasing either the network size or the dataset volume. Nevertheless, the achieved accuracies are vastly more than sufficient for our proof-of-concept purposes, demonstrating that the simple DNN model has successfully learned the intended classification rule.

For the training set, we are sampling the square overlap $|\langle\phi|\psi\rangle|^2$ from a beta distribution (see App.~\ref{app.:rand._3dm}). This method concentrates the sampled states towards the region where the classification occurs. One of the advantages of this is that helps the training of the network, providing more training examples on the region where the decision boundary has to be established. On the other hand, to make sure that the network is not being biased by the distribution of the dataset, we have also tested it against a dataset constructed by sampling $|\langle\phi|\psi\rangle|^2$ uniformly. The results are that the training accuracies are better for all models independently from the number of measurement used on the training. Thus, the network is able to properly generalized the trained results to unknown distributions. Moreover, the majority of the errors occur at states close to the decision boundary. By moving the distribution away, this states are classified with no problem, and thus the accuracy increases.

\subsection{Input order}
\label{app.:input_order}

The order in which measurement settings are presented as input to a neural network can significantly impact the performance of CNNs. CNNs process information using filters that aggregate data within a finite, localized window. Consequently, these architectures are inherently sensitive to local patterns but may be less effective at detecting long-range correlations between features that are distantly positioned in the input vector.

From an implementation perspective, the input features can be arranged in an arbitrary order, provided the chosen convention is applied consistently. The lexicographic order (Eq.~\eqref{eq.:lex_order}, is a natural and straightforward choice. However, from a physical standpoint, no particular ordering is inherently superior for the task of entanglement detection.

The locality constraint of CNNs can potentially affect the methodology in two critical ways. First, if the correlations essential for classification exist between measurement settings that are not adjacent in the input vector, the CNN's finite receptive field may fail to detect them, potentially compromising classification performance. This supposes a problem specially on shallow networks. Second, the measurement-importance ranking provided by the computation of Shapley values can be biased by the input order. Since a CNN's processing is dependent on local neighborhoods, the specific arrangement of features may artificially inflate or deflate the perceived importance of certain measurements based solely on their position relative to others, rather than their genuine physical relevance.

\begin{figure*}[t]
    \centering
    \includegraphics[width=\textwidth]{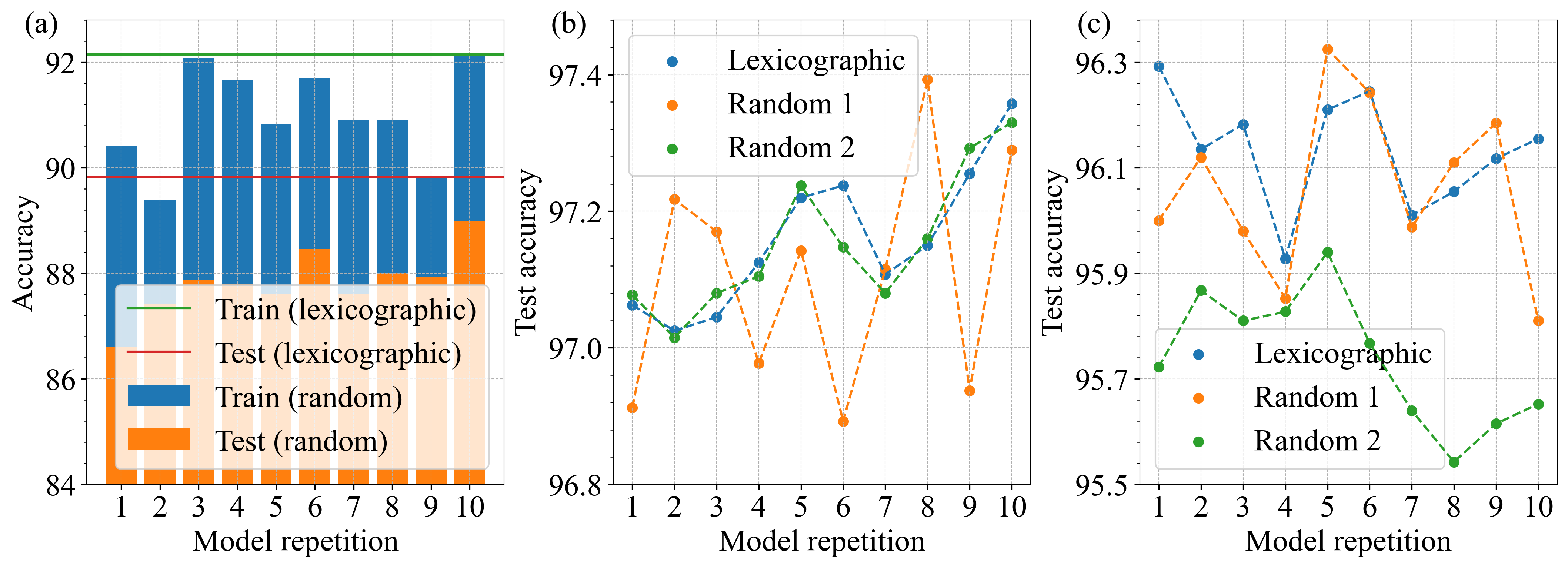}
    \caption{Impact of input feature order on model performance. Test accuracies are shown for models trained with the standard lexicographic order compared to models trained with random permutations of the input features. (a) Results for a CNN classifying pure three-qubit states. For each model, a different random order is chosen. (b) Results for a CNN classifying mixed two-qubit states. Two random orders were compared with the lexicographic one. (c) Results for a DNN classifying mixed two-qubit states, included to demonstrate the architecture's inherent invariance to input order. Two random orders were compared with the lexicographic one.}
    \label{fig.:model_order}
\end{figure*}

To investigate the first issue, we trained models using $10$ different random permutations of the measurement indices. We compared the final accuracies of these models against that of a model trained with the standard lexicographic order. This experiment was conducted for three scenarios: pure two-qubit states (CNN), mixed two-qubit states (CNN), and mixed two-qubit states (DNN). The results are presented in Fig.~\ref{fig.:model_order}. Models trained with random input orders exhibited similar or worse performance than the model using the lexicographic order, although the differences were not substantial. This suggests that while the input order does have a measurable impact on final accuracy, the lexicographic order performs at or above the average of random permutations. Based on this finding, we adopted the lexicographic order for all subsequent experiments, for both CNNs and DNNs, to maintain consistency.

To address the second issue, we compared the resulting measurement-importance rankings for both CNN and DNN architectures. This analysis was conducted exclusively for mixed states. This is because we have tested that the high information redundancy in pure states masks the effect of the CNN's finite filter. The results confirm that for CNNs, the computed Shapley values and the resulting measurement rankings are indeed sensitive to the input order of the measurements. In contrast, for DNNs, which possess full connectivity between all input and hidden layers, the Shapley value rankings remain consistent and invariant to the permutation of input features. This finding provides a strong motivation for employing DNNs rather than CNNs for mixed-state entanglement classification, as it ensures that the derived measurement importance is a genuine reflection of the physical properties of the measurements and is not an artifact of an arbitrary input representation.

\subsection{Non-ideal measurements} \label{app.:non-ideal_meas}

Recall that Eq.~\eqref{eq.:meas.} assumes an ideal scenario where obtaining complete knowledge of the probability distribution $p_A$ is possible, i.e. under an infinite number of shots. Here we illustrate the impact of a finite number of shots per measurement.

\begin{figure}
    \centering
    \includegraphics[width=\linewidth]{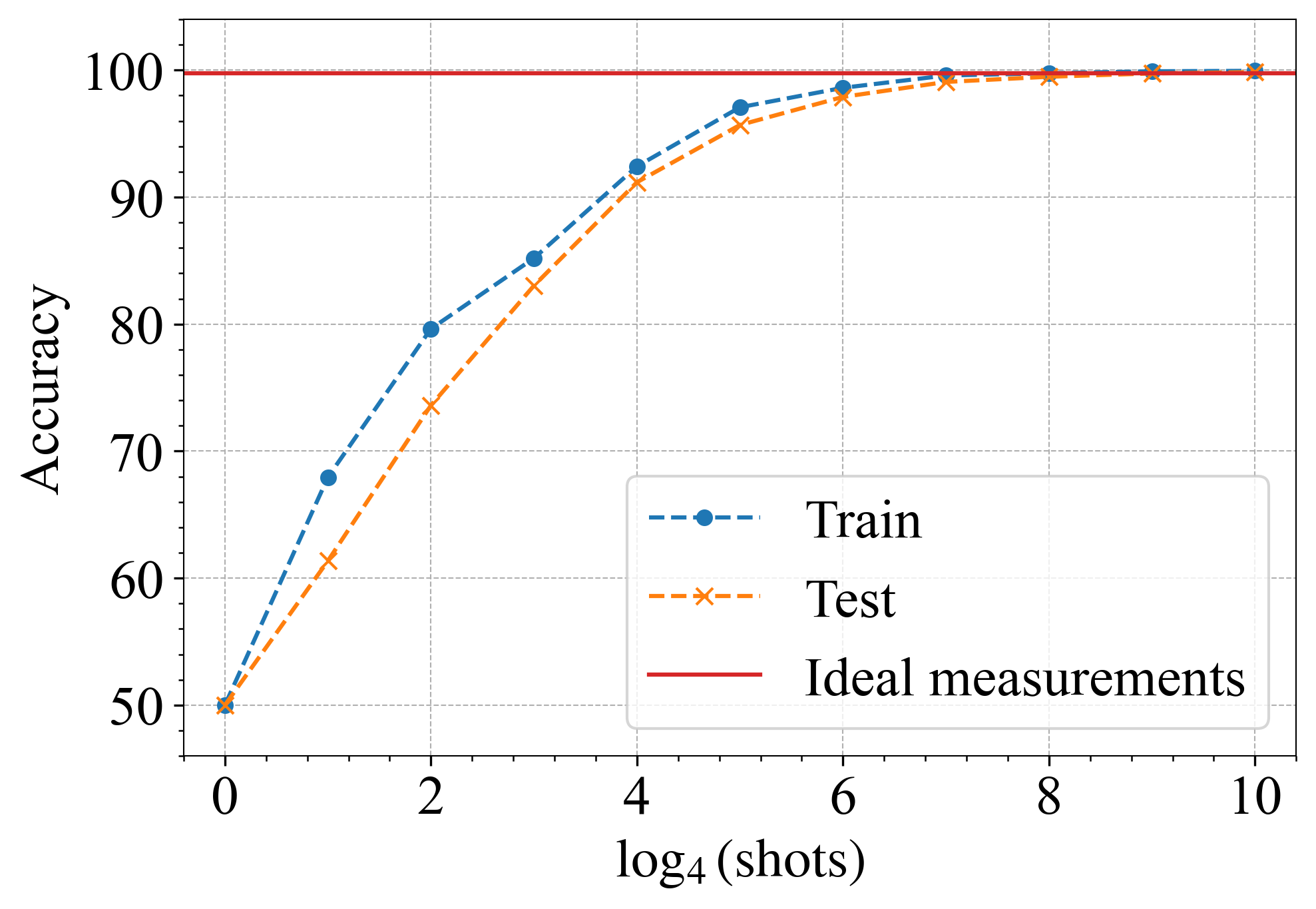}
    \caption{Network classification accuracy for two-qubit pure states in the non-ideal measurement scenario as a function of the number of shots used for estimation of the correlation tensor elements.}
    \label{fig.:non-ideal_meass}
\end{figure}

Assume we perform $M$ shots, labeled by $m = 1, 2 \ldots, M$. Each yields a result corresponding to the eigenvalue $a_{i_m}$, where $i_m$ indexes the eigenvalues of $A$ for every $m$. The measurement outcomes $\{a_{i_1}, \ldots, a_{i_M}\}$ can be used to estimate the probabilities $p_A(i \ | \ \varrho)$
\begin{equation}
    \hat p_A(i \ | \ \varrho) = \frac{m_i}{M} \, ,
\end{equation}
where $m_i$ is the number of times the eigenvalue $a_i$ was measured, and $\hat p_A$ is the empirical estimate of $p_A$. Using the estimated probabilities, the expected value can be approximated as  $\langle A, \varrho\rangle \approx \sum_i a_i \hat p_A (i \ | \ \varrho)$, so that
\begin{equation}
    \langle A, \varrho\rangle \approx \frac{1}{M} \sum_{m = 1}^M a_{i_m}.
\end{equation}
We then benchmark the network under these non-ideal scenario. Figure~\ref{fig.:non-ideal_meass} shows the classification accuracy for two-qubit pure states as a function of the number of shots used for estimation of the correlation tensor elements. All the design choices, hyperparameters and other arbitrary decisions are kept fixed. As we can see, the network converges to the ideal result as the number of shots is increased. This effect occurs both for the complete model, and for the reduced two-measurements model that uses $\sigma_{0i}$ operators.

\section{Shapley values}

\subsection{Approximation of Shapley values using DeepExplainer}
\label{app.:deepexplainer}

The core challenge in approximating Eq.~\eqref{eq.:shap} is evaluating the expectation that defines the model output when only a subset $S$ of features is known. SHAP requires to compute
\begin{equation}
    f(S) = \mathbb{E}[f(X) \, | \, X_S = x_S] \, ,
\end{equation}

but computing the true conditional expectation can be difficult in practice when features are dependent. \texttt{DeepExplainer} approximates the required marginalization by using a background (reference) dataset $X_{\text{back}} = \{x^{(1)}_{\text{back}}, x^{(2)}_{\text{back}}, \dots, x^{(K)}_{\text{back}}\}$. It replaces missing features by values drawn from the background dataset and approximates the expectation by the empirical average
\begin{equation}
f(S) \approx \frac{1}{K} \sum_{k=1}^{K} f\left( x_S, \, x^{(k)}_{\text{back}, \setminus S} \right) \, ,
\end{equation}

were $\left( x_S, \, x^{(k)}_{\text{back}, \setminus S} \right)$ is the hybrid input that uses the instance's values for features in $S$ and the $k$-th background sample for the remaining features. Note this is an interventional/marginal treatment of missingness: it marginalizes the missing features according to the chosen reference distribution rather than conditioning on them, so the choice of background samples materially affects the resulting attributions.

\texttt{DeepExplainer} builds on the DeepLIFT backpropagation rules (Rescale and RevealCancel) to efficiently compute attributions for deep networks. As shown by~\cite{Lundberg_2017}, DeepLIFT-style propagation can be combined with background sampling to produce approximate SHAP attributions. For each background reference $x_{\text{back}}$ a single backward pass yields contributions $\Phi_j [f, x, x_{\text{back}}]$, and averaging over $K$ references gives
\begin{equation}
\chi_j \approx \frac{1}{K} \sum_{k=1}^{K} \Phi_j\left[f, x, x^{(k)}_{\text{ref}}\right] \, .
\end{equation}

These estimates obey the additive (local accuracy) property with respect to the implemented marginalization. The $\chi_j$ sum to the difference between the model output on $x$ and the average model output over the background,
\begin{equation}
\sum_{i=1}^{|F|} \chi_j = f(x) - \frac{1}{K} \sum_{k=1}^{K} f\left[x^{(k)}_{\text{ref}}\right]
\end{equation}

However, because \texttt{DeepExplainer} uses DeepLIFT propagation rules and background sampling, the result is an approximation of the exact Shapley values for arbitrary non-linear dependencies. It is exact for linear models and accurate in many practical settings, but practitioners should be aware that (i) the choice of background samples affects semantics and magnitude of $\chi_j$ (ii) correlations among features mean marginal (interventional) and conditional expectations differ.

\subsection{Procedure for Shapley values approximation}
\label{app.:shap_approx.}

The \texttt{DeepExplainer} algorithm approximates the Shapley values for a subset of the dataset, called the sample dataset, using a background (or reference) dataset. This background dataset provides a baseline for evaluating the marginal contribution of each input feature. We constructed both datasets by randomly selecting $K_b$ and $K_s$ instances, respectively, from the entire training set. The \texttt{DeepExplainer} was then applied to the trained deep neural network to generate Shapley value explanations for each instance in the sample set.

The output of the algorithm is a tensor of Shapley values with dimensions $(K_s, 4^N, L)$. Each element $\chi_{k_s, j, c}$ quantifies the importance of the input feature $j = 0, \ldots, 4^N-1$ for classifying sample $k_s = 0, \ldots, K_s-1$ into class $c \in 0, \ldots, C-1\}$.

Moreover, Shapley values carry a sign, indicating whether a feature's contribution towards classification into a specific class is positive (supportive) or negative (antagonistic). For our analysis, which focuses on global measurement importance irrespective of directional effect, we work with the absolute values of the Shapley values, $|\chi_{k_s, j, c}|$.

To extract a meaningful global importance metric from this data, we perform two successive averaging operations. First, since we are interested in general classification performance rather than class-specific behavior, we average over all output classes. This yields a class-agnostic importance for each measurement and sample: $\frac{1}{C} \sum_{c=0}^{C-1} | \chi_{k_s, j, c} |$.

Second, to eliminate state-specific information and obtain a general state-agnostic metric, we average across all samples in the explanation set. This gives the final global importance for each measurement setting $j$: $\frac{1}{K_s \cdot C} \sum_{k_s=0}^{K_s-1} \sum_{c=0}^{C-1} | \chi_{k_s, j, c} |$.

This setup presents two possibilities for a rigorous analysis. If the sampling is uniform and the sample sizes $K_b$ and $K_s$ are sufficiently large, the background and sample datasets are expected to be representative of the entire training distribution. In this ideal scenario, the approximated Shapley values can be assumed to accurately reflect the true values over the complete dataset. However, for large models and complex datasets like ours, this approach is computationally prohibitive, as the computation time grows super-linearly with $K_b$. The tractable values for $K_b$ and $K_s$ are often too small to guarantee this level of representativeness.

In Table~\ref{tab.:order_per_sampling} we present the measurement settings for a system of pure two-qubit states, ordered using the approximated Shapley values for ten different samplings from the same model. Each value represents the lexicographic index $j = 4i_1 + i_2$, where $(i_1, i_2)$ are the local Pauli operator indices $\sigma_{i_1} \otimes \sigma_{i_2}$ being measured ($i_1, i_2 = 0, 1, 2, 3$). The values used were $K_b = 10^4$ and $K_s = 10^3$. As one can see, although a certain structure is preserved, there are variations in the middle ranks that are attributed to random sampling. The same effect appears for pure states of three qubits and mixed states of two and three qubits.

\begin{table}[b!]
    \centering
    \begin{tabular}{|c|c|}
    \hline
    & Measurement settings order \\
    \hline
    \hline
    Sampling $1$ & $0$  $1$  $3$  $8$  $4$  $2$ $11$ $ 7$ $12$ $10$ $5$  $9$  $6$ $15$ $14$ $13$ \\
    \hline
    Sampling $2$ & $0$  $1$  $3$  $8$  $4$  $2$ $7$ $11$ $12$ $10$ $9$  $5$  $6$ $15$ $14$ $13$ \\
    \hline
    Sampling $3$ & $0$  $1$  $3$  $8$  $4$  $2$ $7$ $11$ $12$ $10$ $9$  $5$  $6$ $15$ $14$ $13$ \\
    \hline
    Sampling $4$ & $0$  $1$  $3$  $8$  $4$  $2$ $7$ $11$ $12$ $10$ $9$  $5$  $6$ $15$ $14$ $13$ \\
    \hline
    Sampling $5$ & $0$  $1$  $3$  $4$  $8$  $2$ $7$ $11$ $12$ $10$ $5$  $9$  $6$ $15$ $14$ $13$ \\
    \hline
    Sampling $6$ & $0$  $1$  $3$  $8$  $4$  $2$ $7$ $11$ $12$ $10$ $9$  $5$  $6$ $15$ $14$ $13$ \\
    \hline
    Sampling $7$ & $0$  $1$  $3$  $8$  $4$  $2$ $11$ $ 7$ $12$ $ 5$ $10$  $6$  $9$ $15$ $14$ $13$ \\
    \hline
    Sampling $8$ & $0$  $1$  $3$  $8$  $4$  $2$ $7$ $12$ $11$ $ 9$ $10$  $5$  $6$ $15$ $14$ $13$ \\
    \hline
    Sampling $9$ & $0$  $1$  $3$  $8$  $4$  $2$ $7$ $11$ $12$ $10$ $5$  $6$  $9$ $15$ $14$ $13$ \\
    \hline
    Sampling $10$ & $0$  $1$  $3$  $8$  $4$  $2$ $11$ $ 7$ $12$ $10$ $9$  $5$  $6$ $15$ $14$ $13$ \\
    \hline
    \end{tabular}
    \caption{Measurement settings identified by lexicographic index $j$ and ordered by Shapley value order $\tilde \jmath$ for pure two-qubit states, from ten different samplings using the same model. Parameters: $K_b = 10^4$, $K_s = 10^3$.}
    \label{tab.:order_per_sampling}
\end{table}

To mitigate the randomness inherent in the sampling process, we perform multiple independent samplings. For $L$ independent trials, we draw different background and sample sets of sizes $K_b$ and $K_s$, compute the approximate Shapley values for each trial, and aggregate the results. After repeating this process for $L$ independent samplings, we obtain a tensor of Shapley values with shape $(L, K_s, 4^N, L)$. To derive a stable, aggregate estimate of the global importance for each input measurement $j$, we compute the mean absolute Shapley value across all samples, trials, and classes:
\begin{equation} \label{eq:mean_shapley}
    \hat \chi_j = \frac{1}{L \cdot K_s \cdot C} \sum_{l=0}^{L-1} \sum_{k_s=0}^{K_s-1} \sum_{c=0}^{C-1} | \chi_{l, k_s, j, c} |.
\end{equation}

The primary objective of this project is not to characterize a specific neural network instance but to identify the most important measurement settings for entanglement classification in a model-agnostic manner. This goal introduces an additional source of variability: the random initialization of the neural network weights. Different initializations can lead the model to converge to distinct local minima of the loss function. While these minima may yield similar predictive performance, they might utilize the input features in different ways, leading to divergent Shapley value profiles.

In Table~\ref{tab.:order_per_model} we present the measurement settings for pure two-qubit states, now ordered by averaging over $L = 10$ samplings for each of ten different models. While certain structure is preserved, there are significant variations in the rankings that are attributed to random model initialization. The same effect appears for all system under study.

\begin{table}[b!]
    \centering
    \begin{tabular}{|c|c|}
    \hline
    & Measurement settings order \\
    \hline
    \hline
    Model $1$ & 0  1  4  2  8  3  5 12  7 11  9  6 15 10 13 14 \\
    \hline
    Model $2$ & 0  1  3  8  4  2  7 11 12 10  9  5  6 15 14 13 \\
    \hline
    Model $3$ & 0  2  1  4  8  3 12 11  7 10  6 15  5  9 14 13 \\
    \hline
    Model $4$ & 0  1  2  4  8 12  3  5  7  9 10 13  6 15 14 11 \\
    \hline
    Model $5$ & 0  1 15  8  4 12  3  2 11  7  9  5  6 13 10 14 \\
    \hline
    Model $6$ & 0 15  3  2  4 12  8  1  7 11 10  6 14  5 13  9 \\
    \hline
    Model $7$ & 0  1  2 12  4  8  3  5  9  7  6 10 11 13 14 15 \\
    \hline
    Model $8$ & 0 15  1  2  3 12 11 14 13  7  6  4  5  9  8 10 \\
    \hline
    Model $9$ & 0 15  2  3  1  4 11  7  8 12  6 10 14  5  9 13 \\
    \hline
    Model $10$ & 0  1  8  2 12  4  3 15  7 14  6 10 11 13  9  5 \\
    \hline
    \end{tabular}
    \caption{Measurement settings identified by lexicographic index $j$ and ordered by Shapley value order $\tilde \jmath$ averaged over $L=10$ samplings for each of ten different models.}
    \label{tab.:order_per_model}
\end{table}

To address this model-specific variability, we employ an ensemble-based ranking method. The procedure involves training $M$ models with different random initializations. For each model $m$, we compute the aggregate importance scores $\hat \chi_j^{(m)}$ using Eq.~\eqref{eq:mean_shapley}. Instead of averaging these scores directly—which can be unstable if the distributions are inconsistent—we use a rank-based aggregation method. For each model $m$, we rank the $4^N$ measurement settings based on their $\hat\chi_j^{(m)}$ values (the highest-ranked setting receives a score of $4^N - 1$, the second-highest $4^N - 2$, and so on, down to $0$ for the lowest-ranked). The final score for each measurement setting $j$ is the sum of its ranks across all $M$ models. The final model-agnostic importance order, indexed by $\tilde \jmath$ is obtained by ranking measurement settings by these aggregated scores.

Table~\ref{tab.:Shapley_params.} shows the parameter choices $(M, L, K_b, K_s)$ for the four cases under study. These values were determined through numerical testing to provide stable results within our computational constraints. The parameters were optimized to satisfy three conditions simultaneously. (i) For fixed $(K_b, K_s)$, the similarity between Shapley value approximations $\hat \chi_{l, j} = \frac{1}{K_s \cdot C} \sum_{k_s = 0}^{K_s-1} \sum_{c = 0}^{C-1} |\chi_{l, k_s, j, c}|$ across different samplings $j$ is maximized. (ii) For fixed $(L, K_b, K_s)$, the similarity between aggregated Shapley values $\hat \chi_{j} = \frac{1}{L} \sum_{l=0}^{L-1} \hat \chi_{l, j}$ across different model initializations is maximized. (iii) The value of $M$ is sufficiently large to ensure the aggregated rankings of measurement settings converge to a stable order.

\begin{table}[hbtp]
    \centering
    \begin{tabular}{|c|c|c|}
    \hline
    & Pure & Mixed \\
    \hline
    \hline
    two qubits & $(10, 10, 10^4, 10^3)$ & $(10, 75, 3600, 400)$ \\
    \hline
    thee qubits & $(20, 60, 1800, 180)$ & $(20, 25, 5000, 1000)$ \\
    \hline
    \end{tabular}
    \caption{Values of the parameters $(M, J, K_b, K_s)$ used for the method of approximating the Shapley values on the cases under study.}
    \label{tab.:Shapley_params.}
\end{table}

\begin{figure}
    \centering
    \includegraphics[width=\linewidth]{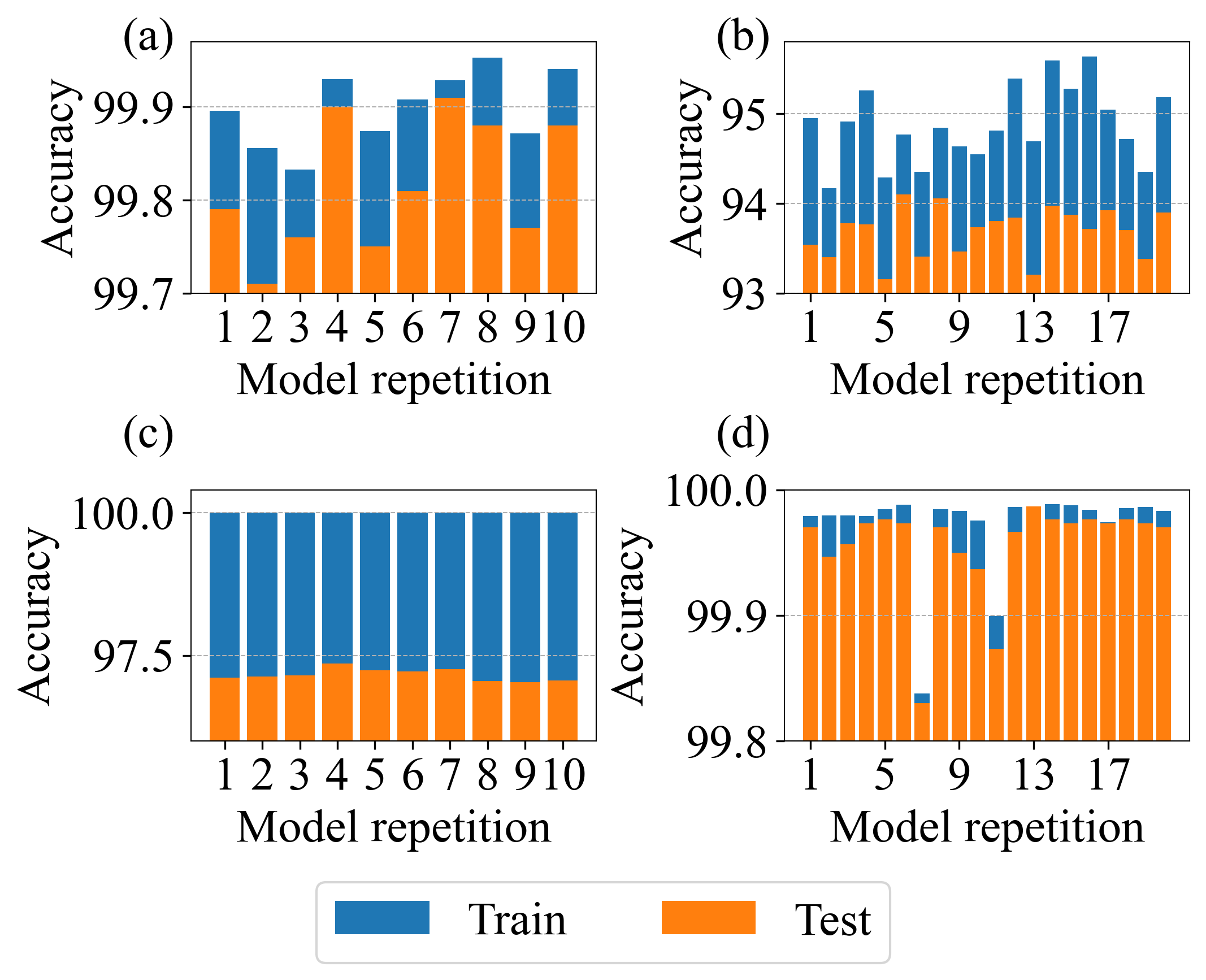}
    \caption{Training and test accuracies for all model instances. Panels (a) and (b) correspond to pure states of $2$ and $3$ qubits, respectively. The models were repeated $M=10$ times. Panels (c) and (d) correspond to mixed states of $2$ and $3$ qubits, respectively $M=20$ repetitions were performed.}
    \label{fig.:model_repeat}
\end{figure}

Figure~\ref{fig.:model_repeat} presents the final training and test accuracies achieved across these multiple runs for all four systems and their corresponding architectures. The observed performance metrics—including the gap between training and test accuracy—are influenced by several factors: the specific network architecture, the size of the dataset, and the intrinsic complexity of the classification task for each system.

\subsection{Masking measurement reduction}\label{app.:ablation_mask}

\begin{figure}
    \centering
    \includegraphics[width=\linewidth]{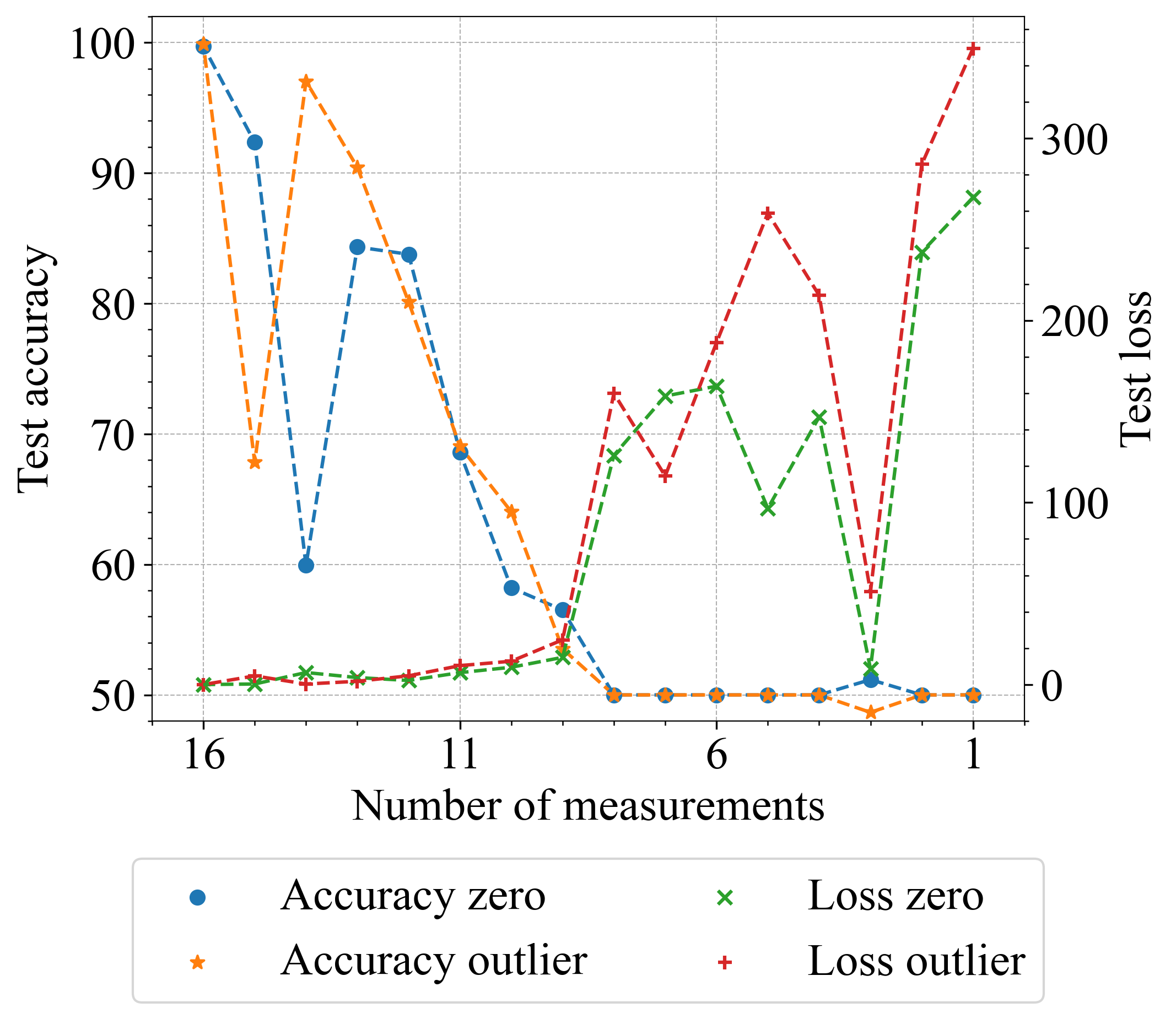}
    \caption{Accuracy of a pre-trained two-qubit neural network evaluated with progressively masked measurement settings.}
    \label{fig.:pure_2q_mask}
\end{figure}

The first approach to measurement reduction was to evaluate the performance of the trained neural networks while progressively masking measurement settings in order determined by the derived importance orders. Masking them with a value of zero carries a risk; the network might interpret this value as a valid input, since lies in the valid range, i.e. $[-1, 1]$. Masking them using an outlier value risks pushing the neural network values into regions outside the training distribution. The experiment was conducted for the two-qubit case, due to their uniformly negative outcome. The results are shown in Fig.~\ref{fig.:pure_2q_mask}, where the model architecture and hyperparameters are consistent with those defined in App.~\ref{app.:pure_arch._opt.}. The classification accuracy drops precipitously to near-random guessing after removing the first few measurements. Thus, the post-hoc masking is not a viable method for assessing measurement importance. Instead, since the network has been trained to use determined information collectively, it is required to train a new NN each time we wish to change the set of inputs.

\section{Analysis of the mixed three-qubit dataset}

\subsection{Solving the witness inequality}\label{app.:witness_inequality}

Several entanglement witnesses are being used, to detect different types of entanglement (Sect.~\ref{sect.:SLOCC_mixed}). A general expression for the entanglement witnesses is given by Eq.~\eqref{eq.:witness}. In the same manner, the form of the states that we will be generating is given by Eq.~\eqref{eq.3q_states}.

The objective is to generate states that are detected by the witness, therefore that satisfy $\text{tr}(\mathcal W_\phi \varrho_\psi) < 0$. By developing this expression, one gets
\begin{gather*}
    \tr(\mathcal W_\phi \varrho_\psi) = \alpha - \tr(P_\phi \varrho_\psi) = \alpha - \frac{\beta}{8} - (1 - \beta) \, |\langle \phi | \psi \rangle|^2 \, , \\
    \tr(\mathcal W_\phi \varrho_\psi) < 0 \iff \beta \left(\frac{1}{8} - \gamma\right) > \alpha - \gamma \, ,
\end{gather*}

where $\gamma := |\langle \phi | \psi \rangle|^2$, satisfying $\gamma \in [0, 1]$. A weaker assumptions fulfilled for this case is that $\alpha < 1/d$. With these conditions, the equation discerns three important regimes.

\begin{enumerate}
    \item If $\gamma \leq 1 / 8$, then $1 / 8 - \gamma \leq \alpha - \gamma$ with both sides positive. Thus $\beta \, (1 / 8 - \gamma) \leq \alpha - \gamma$, and the inequality is not fulfilled for all $\beta$.
    \item If $1 / 8 < \gamma \leq \alpha$, then $1 / 8 - \gamma \leq \alpha - \gamma$ with the left hand side negative and the right hand side positive. Thus $\beta \, (1 / 8 - \gamma) \leq \alpha - \gamma$, and the inequality is not fulfilled for all $\beta$.
    \item If $\alpha < \gamma$, then $1 / 8 - \gamma \leq \alpha - \gamma$ with both sides negative. Thus, depending on the value of $\beta$, $\beta(1 / 8 - \gamma)$ will be greater, smaller or equal than $\alpha - \gamma$. If we want to impose $\beta (1 / 8 - \gamma) \geq \alpha - \gamma$, then $\beta \leq \frac{\alpha - \gamma}{1 / 8 - \gamma}$ where the inequality is changed of direction since $1 / 8 - \gamma$ is negative.
\end{enumerate}

\begin{figure}
    \centering
    \includegraphics[width=\linewidth]{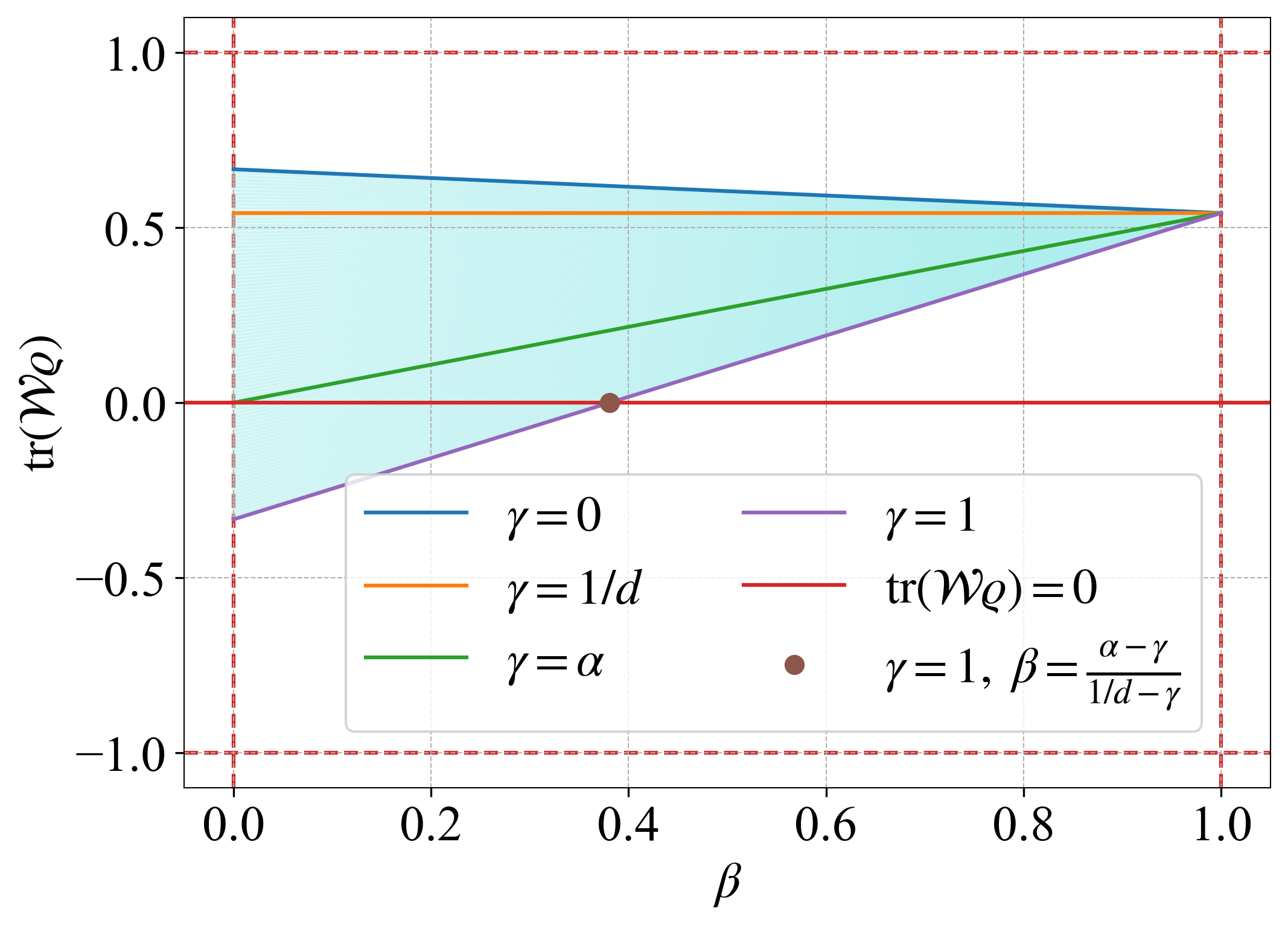}
    \caption{Value of the expected value of the witness for all values of $\beta$ and $\gamma$. The different lines represent different values for $\gamma$, which divide the the different regimes. The detected region lies below the $y = 0$ line.}
    \label{fig.:witness_vs._beta}
\end{figure}

For the state to be detected by the witness, the overlap $\gamma$ must be strictly bigger than the constant $\alpha$. Moreover, the degree of mixing has to be sufficiently small so that the state contains enough entanglement. Figure~\ref{fig.:witness_vs._beta} shows the value of the witness as a function $\beta$, and shows the separation between these three regions determined by the relation between $\alpha$ ($= 2/3$) and $\gamma$.

\subsection{Witnesses redundancy}\label{app.:W_redundancy}

Consider two entanglement witnesses $\mathcal W_\alpha = \alpha \mathbb I - P_\phi$ and $\mathcal W_{\alpha'} = \alpha' \mathbb I - P_\phi$ with respect to the same reference state $\ket{\phi}$ but different values of $\alpha$. Then, assume that $\alpha \ge \alpha'$. We want to prove that if $\mathcal W_\alpha$ detects an state $\varrho$, then that state is also detected by $\mathcal W_{\alpha'}$. If this is the case, $\mathcal W_{\alpha'}$ is at least as general as $\mathcal W_\alpha$ since it detects at least the same amount of states. If the inequality is not strict, i.e. $\alpha > \alpha'$, then is $\mathcal W_{\alpha'}$ more general than $\mathcal W_\alpha$, detecting strictly more states.

This scenario generalizes the case of the witnesses $(3/4, \ket{\rm GHZ})$ and $(1/2, \ket{\rm GHZ})$ with $\alpha = 3/4$, $\alpha' = 1/2$ and $\ket{\phi} = \ket{\rm GHZ}$. That a quantum state $\varrho$ is detected by $\mathcal W_\alpha$ is equivalent to
\begin{equation*}
    \tr(\mathcal W_\alpha \varrho) = \alpha - \langle \phi | \, \varrho \, | \phi \rangle \iff \alpha < \langle \phi | \, \varrho \, | \phi \rangle \, .
\end{equation*}

Since $\alpha > \alpha'$, it follows that $\alpha' < \alpha < \langle \phi | \, \varrho \, | \phi \rangle$, so that $\alpha' - \langle \phi | \, \varrho \, | \phi \rangle < 0$. And thus, $\mathcal W_{\alpha'}$ also detects $\varrho$. Notice that the implication does not follow in the oposite direction.

\subsection{Witness disjointness}\label{app.:W_disjointness}

Consider two entanglement witnesses $\mathcal W = \alpha \mathbb I - P_\phi$ and $\mathcal W' = \alpha' \mathbb I - P_{\phi'}$ with respect to different reference states $\ket{\phi}$, $\ket{\phi'}$ and values of $\alpha$, $\alpha'$. Then, assume that $\langle \phi | \phi' \rangle = 0$ and $\alpha + \alpha' \geq 1$. We want to prove that no state $\varrho_\psi = \beta/d \, \mathbb I + (1 - \beta) P_\psi$ detected by $\mathcal W$ is detected by $\mathcal W'$ and vice versa.

First, notice that since $\langle \phi | \phi' \rangle = 0$, there exists an orthonormal basis of $\mathbb C^d$ that contains both states, that is $\{\ket{\phi}, \ket{\phi'}, \ket{\phi_3}, \ldots, \ket{\phi_d}\}$. Take any pure state $\ket{\psi}$, then it can be decomposed as a linear combination of the elements in this basis
\begin{equation*}
    \ket{\psi} = u \ket{\phi} + u' |\phi'\rangle + u_3 \ket{\phi_3} + \cdots + u_d \ket{\phi_d} \, ;
\end{equation*}

and furthermore satisfies
\begin{equation*}
    \langle \psi | \psi \rangle =|u|^2 + |u'|^2 + |u_3|^2 + \cdots + |u_d|^2 = 1 \, .
\end{equation*}

Let $\gamma := |u|^2$, $\gamma' := |u'|^2$ and $\gamma^\perp := |u_3|^2 + \cdots + |u_d|^2$, then $\gamma + \gamma' + \gamma^\perp = 1$. If both witnesses detect the state $\varrho_\psi$,
\begin{eqnarray*}
    \tr(\mathcal W \varrho_\psi) &=& \alpha - \gamma < 0 \\
    \tr(\mathcal W' \varrho_\psi) &=& \alpha' - \gamma' < 0
\end{eqnarray*}

But there is a contradiction, since $\gamma + \gamma' + \gamma^\perp = 1$ but $\gamma + \gamma' > 1$. Thus, by contradiction, the assumption must be wrong and one of the witnesses must not detect the state.

\subsection{Decomposing the witness in the Pauli basis}
\label{app.:wit._decomposition}

In section~\ref{sect.:dataset_mixed} by Eqs.~\eqref{eq.:witness} and~\eqref{eq.3q_states}, it is established how the states and witnesses we are working with are parametrized. The witness $\mathcal W_\phi$ will detect state $\varrho_\psi$ if
\begin{equation*}
    \alpha - \frac{\beta}{8} - (1 - \beta) \, \tr(P_\phi P_\psi) < 0 \, .
\end{equation*}

Indeed, since $P_\phi, P_\psi$ represent density matrices of pure states, one could write them according to Eq.~\eqref{eq.:Pauli} as
\begin{equation*}
    P_\phi = \sum_{i, j, k = 0}^3 p_{ijk} \sigma_{ijk} \, , \quad P_\psi = \sum_{i, j, k = 0}^3 r_{ijk} \sigma_{ijk} \, ;
\end{equation*}

with $p_{000} = r_{000} = 1$, and $\frac{1}{8} \sum_{i, j, k = 0}^3 p_{ijk}^2 = \frac{1}{8} \sum_{i, j, k = 0}^3 r_{ijk}^2 = 1$. The value of the inequality is thus dependent on this overlap $\tr(P_\psi P_\phi)$, which can be extended as
\begin{eqnarray*}
    \tr(P_\psi P_\phi) &=& \frac{1}{64} \sum_{i, j, k = 0}^3 \sum_{l, p, q = 0}^3  p_{ijk}  r_{lpq} \tr(\sigma_{ijk} \sigma_{lpq}) \\
    &=& \frac{1}{8} \sum_{i, j, k = 0}^3 p_{ijk} r_{ijk} \, .
\end{eqnarray*}

For the Shapley method, we are not interested in the particular values that the inequality takes, but only in which measurement settings are the important to evaluate it. In general, we are assuming the states $\ket{\psi}$ to be general, and thus no relevant conditions can be imposed on the values of $r_{ijk}$. Nevertheless, the states $\ket{\phi}$ are fixed for some witnesses chosen, and thus their values may dominate the previous equation. In particular, we are choosing states $\ket{\phi} = \ket{\rm GHZ}, \ket{\rm W}$, which decompose as follows: For the $\ket{\rm GHZ}$, 
\begin{align}
    p_{033} &= 1 \, , \ p_{303} = 1 \, , \ p_{330} = 1 \, ,\nonumber \\
         p_{111} &= 1 \, , \ p_{112} = -1 \, , \ p_{212} = - 1 \, , \ p_{221} = -1 \, \nonumber,
\end{align}
while for the $\ket{\rm W}$ state, 
 \begin{align*}
        p_{003} &= 1/3 \, , \ p_{030} = 1/3 \, , \ p_{300} = 1/3 \, , \\
        p_{011} &= 2/3 \, , \ p_{022} = 2/3 \, , \ p_{033} = -1/3 \, , \\
        p_{101} &= 2/3 \, , \ p_{202} = 2/3 \, , \ p_{303} = -1/3 \, , \\
        p_{110} &= 2/3 \, , \ p_{220} = 2/3 \, , \ p_{330} = -1/3 \, , \ \\
        p_{113} &= 2/3 \, , \ p_{131} = 2/3 \, , \ p_{311} = 2/3 \, , \\
        p_{223} &= 2/3 \, , \ p_{232} = 2/3 \, , \ p_{322} = 2/3 \, , \ p_{333} = -1 \, .
    \end{align*}
The rest of the components are equal to zero.

\subsection{Detection as a function of relative phases}
\label{app.:rel._phases}

\begin{figure}
    \centering
    \includegraphics[width=\linewidth]{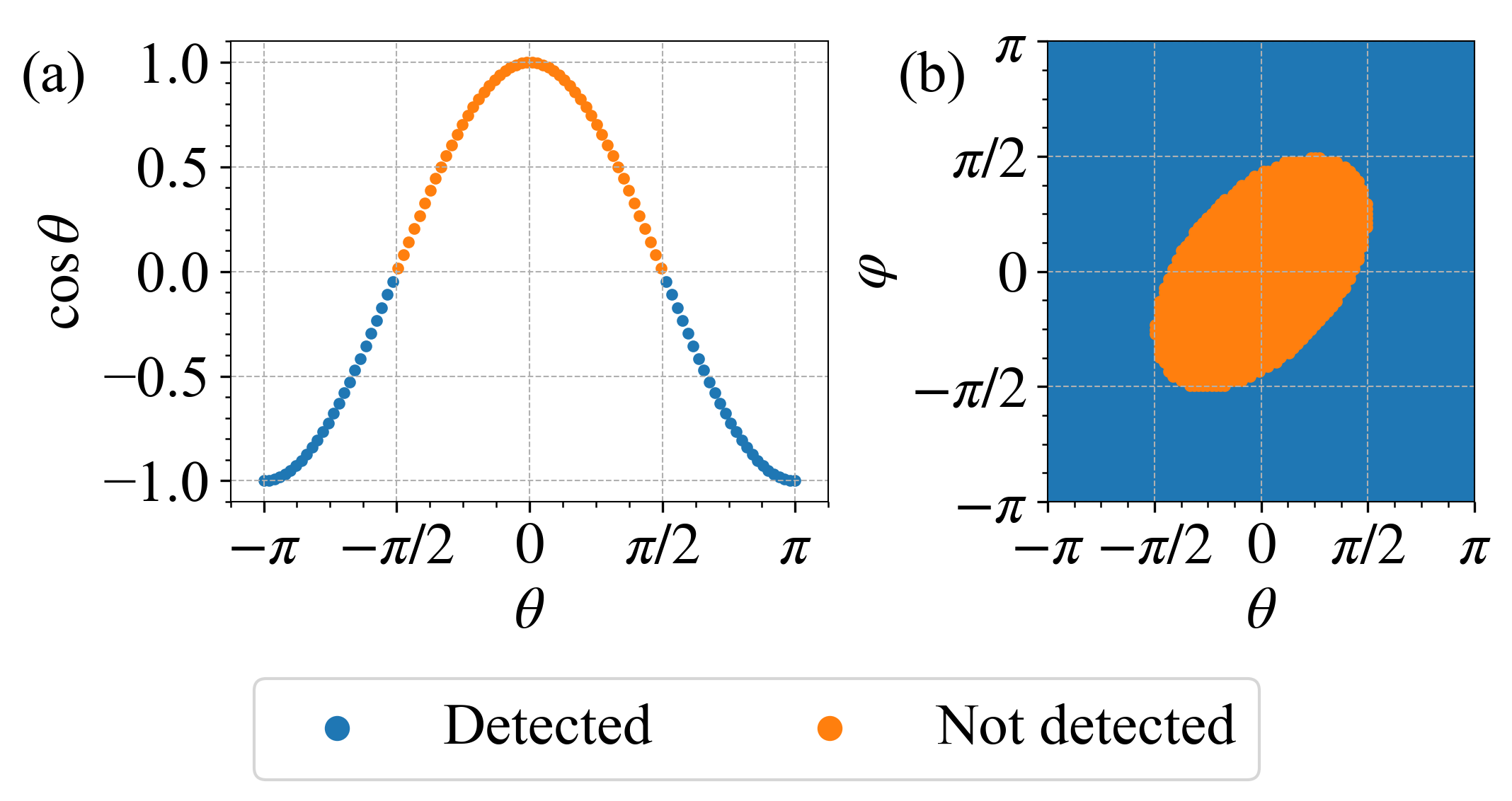}
    \caption{Neural network classification behavior for states with relative phases. (a) Classification probability for the GHZ-class as a function of the relative phase $\theta$. (b) Classification probability for the W-class as a function of the relative phases $\theta_1$ and $\theta_2$.}
    \label{fig.:mixed_3q_phases}
\end{figure}

In Sect.~\ref{sect.:dataset_mixed} by Eqs.~\eqref{eq.:witness} and~\eqref{eq.3q_states}, it is established how the considered states and witnesses are parametrized. This appendix analyzes the behavior of entanglement witness inequalities when applied to pure states $\ket{\psi}$ that differ from the witness state $\ket{\phi}$ by relative phases. Assume $\ket{\psi} = \ket{\phi (\theta_1, \theta_2, \ldots)}$ is parametrized by relative phases $\theta_1, \theta_2, \ldots \in [-\pi, \pi)$. The witness condition $\tr(\mathcal W_\phi \varrho_\psi) < 0$ simplifies to
\begin{equation*}
    \tr(\mathcal W_\phi \varrho_\psi) = \alpha - |\langle \phi | \phi (\theta_1, \theta_2, \ldots) \rangle |^2 < 0 \, .
\end{equation*}

We analyze this condition for the two specific witnesses used in our study.
\begin{enumerate}
    \item GHZ Witness: For the witness determined by $\ket{\phi} \equiv \ket{\rm GHZ} = \tfrac{1}{\sqrt{2}} (\ket{000} + \ket{111})$, we consider the one-parameter family of states $\ket{\phi (\theta)} := \tfrac{1}{\sqrt{2}} (\ket{000} + e^{i \theta} \ket{111})$. The overlap and witness value are computed as follows:
    \begin{gather*}
        \langle\phi|\phi(\theta)\rangle = \frac{1}{2} (1 + e^{i\theta}) \, , \\
        |\langle\phi|\phi(\theta)\rangle|^2 = \frac{1}{4} (1 + e^{i\theta})(1 + e^{-i\theta}) = \frac{1}{2} (1 + \cos \theta) \, ; \\
        \tr(\mathcal W_\phi \varrho_{\phi(\theta)}) = \alpha - \frac{1}{2} (1 + \cos \theta) \, .
    \end{gather*}
    
    Consider the used value $\alpha = \tfrac{1}{2}$ to get the witness condition $\cos \theta > 0$. Restricting the domain to $\theta \in [-\pi, \pi)$, this inequality is satisfied for $\theta \in \left(-\tfrac{\pi}{2}, \tfrac{\pi}{2}\right)$.

    \item W Witness: For the witness determined by $\ket{\phi} \equiv \ket{\rm W} = \tfrac{1}{\sqrt{3}} (\ket{001} + \ket{010} + \ket{100})$, we consider the two-parameter family $\ket{\phi (\theta_1, \theta_2)} := \tfrac{1}{\sqrt{3}} (\ket{001} + e^{i \theta_1} \ket{010} + e^{i \theta_2} \ket{100})$. The overlap is given by:    
    \begin{gather*}
        \langle \phi | \phi(\theta_1, \theta_2) \rangle = \frac{1}{3} (1 + e^{i\theta_1} + e^{i\theta_2}) \, , \\
        |\langle \phi | \phi(\theta_1, \theta_2) \rangle|^2 = \frac{1}{9}[3 + 2 \cos \theta_1 + 2 \cos \theta_2 + 2 \cos(\theta_1 - \theta_2)].
    \end{gather*}
    
    With the used value $\alpha = \frac{2}{3}$, the witness condition becomes:   
    \begin{equation*}
        \cos \theta_1 + \cos \theta_2 + \cos(\theta_1 - \theta_2) > \frac{3}{2}.
    \end{equation*}

    This inequality defines the detection region in the $(\theta_1, \theta_2)$ phase space for the W witness.
\end{enumerate}

We benchmark the neural network's classification behavior against these theoretical predictions. Figure~\ref{fig.:mixed_3q_phases} shows the network's output probabilities for states in these phase-parameterized families. The results demonstrate that the neural network successfully reproduces the theoretical detection boundaries derived above, validating its ability to learn the underlying witness-based classification rules.

\bibliography{paper_bib}

\end{document}